\documentclass[%
 aip,
 amsmath,amssymb,
 reprint,
 onecolumn
]{revtex4-1}

\usepackage{graphicx}
\usepackage{array}[=2016-10-06] 
\usepackage{bm}

\usepackage[utf8]{inputenc}
\usepackage[T1]{fontenc}
\usepackage{etoolbox}

\usepackage{siunitx}
\usepackage{mathtools}
\usepackage{url}
\usepackage{tikz}
\usepackage{booktabs}
\usepackage{mathrsfs}
\usepackage[]{hyperref}

\newcommand{\bmnu}{\bm{\nu}}
\newcommand{\dispE}{\mathscr{E}}

\newcommand{\calN}{\mathcal{N}}

\DeclareSymbolFont{TimesMathOps}{OT1}{ztmcm}{m}{n}
\DeclareMathSymbol{\cem}{\mathord}{TimesMathOps}{1}

\makeatletter
\def\@email#1#2{%
 \endgroup
 \patchcmd{\titleblock@produce}
  {\frontmatter@RRAPformat}
  {\frontmatter@RRAPformat{\produce@RRAP{*#1\href{mailto:#2}{#2}}}\frontmatter@RRAPformat}
  {}{}
}%
\makeatother
\begin{document}


\title[Scattering Measures and Information Geometry]{Scattering Measures and Information Geometry}
\author{Tatsuro Oda}
 \affiliation{Institute for Solid State Physics, The University of Tokyo}
 \email{oda@issp.u-tokyo.ac.jp}

\date{\today}

\begin{abstract}
We formulate scattering experiments in terms of parameterized 
finite measures on outcome spaces and develop the associated information
geometry defined by the Fisher metric. 
The decomposition of a scattering measure into total mass and normalized
shape separates Fisher information into contributions from changes in
overall scattering strength and redistribution among outcomes. 
Measurement processes are represented by probability
or sub-probability kernels acting on the scattering measure. 
Such kernels contract Fisher information, placing a broad class of 
experimental operations within a common information-theoretic framework. 
This formulation connects scattering experiments to the theory
of optimal experimental design through the Fisher geometry of scattering
measures.  The same finite-measure structure arises naturally in quantum 
measurements with incomplete detection, clarifying its place in the 
hierarchy from quantum measurement to classical outcome statistics.
\end{abstract}

\maketitle

\section{Introduction}
\label{sec:introduction}

A scattering process assigns scattering strength to physically
distinguishable outcomes.  These outcomes may be continuous variables
such as momentum, discrete quantities such as spin state, 
or combinations of both.  
The natural mathematical object is therefore a finite measure 
on the outcome space. A differential cross section is then a density
representation of this measure with respect to a chosen reference measure, 
rather than the primary object itself.

This viewpoint has appeared previously in mathematical scattering
theory.  Babbitt developed probabilistic formulations of
both classical and quantum scattering cross sections in terms of
measures on scattering outcome spaces, with differential cross sections
appearing as Radon--Nikodym derivatives~\cite{Babbitt1971, Babbitt1972}.  
Related measure-theoretic treatments of cross sections have
also appeared in the classical \(n\)-centre scattering problem
\cite{Knauf2002}.

Measure-theoretic descriptions of scattering and statistical theories of
parameter distinguishability are thus well established in their
respective settings.  What is less explicit in the conventional
description of a scattering experiment is how the physical scattering
distribution, instrumental response, counting statistics, and data
reduction fit into a single mathematical structure. 
In particular, the information carried by the total scattering strength 
and by the distribution of outcomes is often treated separately, as are
information loss through instrumental transformations and the resulting
tradeoffs in experimental design.

In this work, we connect these elements by taking the scattering measure
as the starting point for the full experimental chain.  We call the
finite measure describing the physical distribution of scattering
strength the \emph{scattering measure}.  Instrumental response and data
reduction transform this measure into the recorded measure:
deterministic operations act by pushforwards of measures, whereas
stochastic processes are represented by probability or sub-probability
kernels.  The physical scattering process, detector response, and data
reduction can therefore be treated within a common sequence of
transformations between measures on different outcome spaces.

For a parameterized family of scattering measures
\(\{\sigma_\lambda\}\), Fisher information then provides a local measure
of distinguishability between neighboring physical models
\cite{RN830,ay2018parametrized}.  It can be defined intrinsically from
the measure and its parameter derivative, without choosing a density
representation, consistently with the information-geometric
characterization of the Fisher metric
\cite{Chentsov1982,AmariNagaoka2000,Le2017}.

The finite, rather than normalized, nature of the scattering measure is
essential.  Its intrinsic Fisher information decomposes exactly into
\emph{mass} and \emph{shape} contributions, associated respectively
with changes in the total scattering strength and in the normalized
distribution of outcomes.  This decomposition is intrinsic to the
geometry of finite measures and does not depend on a particular
counting model.  Under Poisson counting statistics, the same intrinsic
metric is realized as the Fisher information of the observed point
process per unit exposure.  The decomposition therefore separates two
fundamentally distinct sources of parameter distinguishability:
the absolute scale of the scattering intensity and normalized distribution.

The kernel formulation also gives a common description of information
loss under measurement and data reduction.  For a
parameter-independent probability kernel, the transformed score is the
conditional expectation of the original score given the recorded
outcome, so that Fisher information contracts and the loss is determined
by the unresolved conditional variation of the score.  
Together, the finite-measure Fisher geometry and its contraction under
measurement transformations lead naturally to experimental design.
Instrumental settings and data reduction jointly determine the observed
measure and hence the information available about the parameters of
interest.

We further connect the scattering-measure framework to quantum
measurements with incomplete detection, where the Born rule generates
the outcome probabilities, thereby clarifying its place in the hierarchy
from quantum measurement to classical outcome statistics.

In Sec.~\ref{sec:measure}, we formulate scattering measures, reference
measures, and their density representations. Section~\ref{sec:kernel} 
then describes how measurement processes transform the scattering 
measure and thereby induce recorded measures through pushforwards 
and stochastic kernels. In Sec.~\ref{sec:fisher_scattering_measure}, we
develop the intrinsic Fisher geometry of parameterized finite scattering
measures, including the mass--shape decomposition and its realization
in Poisson counting experiments. Information contraction under
measurement kernels is examined in Sec.~\ref{sec:information_contraction}.
Section~\ref{sec:design} uses these structures to formulate measurement
design as an optimization problem on the observed measure. Finally,
Sec.~\ref{sec:quantum_origin} connects the classical hierarchy to quantum
measurement, where the Born rule generates outcome probabilities before
subsequent instrumental transformations. 
Frequently used notation is summarized in Appendix~\ref{app:notation}.


\section{Scattering measures, reference measures, and densities}
\label{sec:measure}

\subsection{Outcome spaces and scattering measures}
\label{subsec:event_space_scattering_measure}

Let \(X\) denote the space of physically distinguishable
outcomes, and let \(\mathcal X\) be a \(\sigma\)-algebra of subsets of
\(X\). Depending on the physical setting of a scattering experiment, 
\(X\) may be a space of outgoing directions, final momenta, energy, 
spin outcomes, or a product of several such spaces.

The specification of \(X\) is part of the physical modeling. 
It determines which variables are used to describe scattering outcomes 
and against which variables the scattering strength is resolved. 
For example, an experiment that resolves both scattering direction 
and final energy may use an outcome space of the form
\(X=S^2 \times \mathbb{R}_{+}\), 
whereas an energy-integrating measurement describes outcomes only on 
the space of scattering directions. 
These spaces represent different sets of physically distinguishable 
outcomes.

A scattering process assigns a nonnegative scattering strength to measurable
sets of outcomes. We represent this assignment by a finite positive measure
\(\sigma\) on \((X,\mathcal X)\), which we call the
\emph{scattering measure}. Specifically,
\begin{equation}
\sigma:\mathcal X\to[0,\infty),
\label{eq:scattering_measure_definition}
\end{equation}
 and for any countable collection of
pairwise disjoint sets \(A_n\in\mathcal X\),
\begin{equation}
\sigma\left(\bigsqcup_{n=1}^{\infty}A_n\right)
=
\sum_{n=1}^{\infty}\sigma(A_n).
\end{equation}
Its total mass is denoted by
\begin{equation}
r \coloneqq
\sigma(X) < \infty .
\label{eq:scattering_measure_total_mass}
\end{equation}
For any measurable set \(A\in\mathcal X\), the value \(\sigma(A)\geq 0 \)
gives the scattering strength associated with outcomes in \(A\). 

The physical interpretation and normalization of \(\sigma\) depend on 
the physical quantity that it is taken to represent.
In a common physical description of scattering, 
\(\sigma(A)\) is taken to represent a cross section for scattering into the set of outcomes \(A\). 
Its values carry the physical dimension of area. 
A cross section quantifies the scattering strength that determines the expected scattering rate into \(A\), together with the incident flux and the number of scattering centers.

\subsection{Reaction channels and changes in particle multiplicity}

The advantage of describing the physical outcomes by an abstract
measurable space becomes particularly apparent when different reaction channels have different final-state
structures. 
For a process with several channels, the outcome space may be
represented as the measurable disjoint union
\begin{equation}
X
=
\bigsqcup_{c\in\mathcal C} X_c ,
\label{eq:reaction_channel_disjoint_union}
\end{equation}
where \(X_c\) is the outcome space associated with reaction channel \(c\).
The individual spaces \(X_c\) need not share the same coordinates or even
the same dimensionality.  A scattering or reaction measure \(\sigma\) is
defined on the full space \(X\), while
\(\sigma(X_c)\)
gives the total weight associated with channel \(c\).

The same construction naturally accommodates changes in particle
multiplicity.  For example, electron--positron annihilation may 
populate sectors with different
photon multiplicities, 
\(
X = X_{2\gamma} \sqcup X_{3\gamma} \sqcup \cdots ,
\) 
where \(X_{n\gamma}\) denotes the space of final states containing
\(n\) photons.

\subsection{Density representations and reference measures}
\label{subsec:reference_measure_density}

While the scattering measure \(\sigma\) intrinsically assigns scattering strength 
to measurable outcome sets, scattering experiments often seek 
a local description of how this strength is distributed over the outcome space.  
Such a description makes the scattering directly visible as a function of 
outcome variables, for example as an angular distribution or an energy spectrum, 
which reflect the structure and dynamics of the scattering system. 

To construct such a local representation, one must first specify how 
size is measured on the outcome space. This is provided by a reference measure
 \(\mu\) on \((X,\mathcal X)\). If \(\sigma \ll \mu\), that is, if \(\sigma\) is
absolutely continuous with respect to \(\mu\), the Radon--Nikodym
theorem~\cite{Klenke2008} guarantees the existence of a nonnegative
measurable function \(s\) such that
\begin{equation}
\sigma(A)
=
\int_A s(x)\,d\mu(x)
\end{equation}
for every measurable set \(A\in\mathcal X\). Here \(x\in X\) denotes
an individual outcome. The function
\begin{equation}
s
\coloneqq
\frac{d\sigma}{d\mu}
\label{eq:RN-density_sigma}
\end{equation}
is the density of the scattering measure with respect to \(\mu\).
In the usual terminology of scattering experiment, differential cross sections are precisely such density representations. 
The density \(s=d\sigma /d\mu\) depends on the chosen reference measure and on its coordinate representation, whereas the underlying measure \(\sigma\) does not. 
Therefore, apparently different differential cross sections may represent the
same underlying scattering measure.

\subsubsection{Example: Angular differential cross section}
\label{subsubsec:angular_scattering_example}

As a simple example, consider elastic angular scattering, for which the outgoing direction is the quantity of interest. The outcome space may then be taken as \(X=S^2\). For instance, the forward hemisphere defines the measurable region
\(A=
\left\{
\Omega=(\theta,\phi)\in S^2
\;\middle|\;
0\leq\theta\leq\frac{\pi}{2}
\right\}
\).  Using the solid-angle measure
\(
d\Omega=\sin\theta\,d\theta\,d\phi
\) as a reference measure on \(S^2\), the cross section assigned to \(A\) is
\begin{align}
\sigma(A)
=
\int_A
\frac{d\sigma}{d\Omega}(\Omega)\,d\Omega
=
\int_0^{2\pi}\int_0^{\pi/2}
\frac{d\sigma}{d\Omega}(\theta,\phi)
\sin\theta\,d\theta\,d\phi.
\end{align}
Thus, \(d\sigma/d\Omega\) is the Radon--Nikodym density of the scattering measure with respect to the solid-angle measure. 
The total mass of the scattering measure,
\(
r = \sigma(S^2)
=
\int_{S^2}
\frac{d\sigma}{d\Omega}\,d\Omega,
\)
is the total cross section of angular scattering.  
The solid-angle measure is particularly natural for angular scattering because it is invariant under the action of \(SO(3)\).

\subsection{Kinematic construction of energy–angle outcome measures}
\label{subsec:outcome-space-reference-measure}

Differential cross sections provide the standard density-based
description used throughout scattering physics. 
They describe how the scattering cross
section is distributed over experimentally distinguishable
final-state outcomes. Their measure-theoretic formulation involves
three conceptually distinct ingredients.
The first is the specification of the final-state outcome space. 
The second is a reference measure determined by the kinematics 
of the outgoing particle. 
The third is the scattering measure generated by 
the probe--target interaction for the specified initial state. 
We focus here on the first two, which provide the measure-theoretic basis 
for the subsequent analysis of scattering measures and their transformations, 
while the third determines how the mass of the scattering measure is 
distributed over the allowed final states.

Starting from the asymptotic momentum space of the outgoing particle, 
we show how its dispersion relation naturally leads to the
energy--angle variables conventionally used to express the
double-differential cross section. 
We then take Compton scattering as a physical example showing 
how energy--momentum conservation restricts the allowed final states 
within the chosen outcome space, while the interaction determines how 
the scattering cross section is distributed among them.

\subsubsection{From momentum space to energy--angle outcomes}
\label{subsubsec:energy_angle_outcome}

In a scattering experiment, the relevant final outcomes need not be described in the
full six-dimensional phase space. Once the outgoing particle has left the
localized interaction region, its position along the asymptotic trajectory
does not distinguish a new scattering outcome. The final state may therefore
be specified by the outgoing momentum \(\mathbf{p}_f\), or equivalently by the wave vector
\(\mathbf{k}_f=\mathbf{p}_f/\hbar\), where \(\hbar\) is the reduced Planck constant. We thus take the asymptotic final-state
outcome space to be \(X=\mathbb{R}^3_{\mathbf{k}_f} \) 
with reference measure \(d\mu_{f}=d^3k_f\).

The outgoing momentum and energy are related by the dispersion
relation \(
E_f
=
E(\mathbf k_f)
\label{eq:final_state_dispersion}
\).  Since
the target may exchange energy with the probe, the allowed \(E_f\) depends on the initial state and on the final states of the target. 
For each possible final energy \(E_f\), define the constant-energy
surface
\begin{equation}
\Sigma_{E_f}
=
\left\{
\mathbf k_f\in\mathbb R^3
\mathrel{}\middle|\mathrel{}
E(\mathbf k_f)=E_f
\right\}.
\end{equation}
The coarea formula~\cite{RN826} resolves the momentum-space measure
into a continuous family of such surfaces:
\begin{equation}
d^3k_f
=
\frac{d\Sigma_{E_f}}
{|\nabla_{\mathbf k_f}E(\mathbf k_f)|}
\,dE_f.
\label{eq:momentum_measure_coarea}
\end{equation}
Here, \(d\Sigma_{E_f}\) denotes the surface measure on
\(\Sigma_{E_f}\). The coarea Jacobian can be written in terms of the group speed as
\(
v_g(\mathbf k_f) = |\nabla_{\mathbf k_f}E(\mathbf k_f)| /\hbar \).
This representation does not reduce the dimension of the final-state
space; it simply replaces the three momentum coordinates by the energy
\(E_f\) and two coordinates on the corresponding constant-energy surface.

Suppose now the dispersion relation is isotropic,
\(
E(\mathbf k_f)
=
E(k_f)\) with \(
k_f
\equiv
|\mathbf k_f|\). 
Each constant-energy surface \(\Sigma_{E_f}\) is then a sphere of radius \(k_f\). Parametrizing
this sphere by the outgoing direction \(\Omega\in S^2\), its surface
measure is
\( d\Sigma_{E_f}
=
k_f^2\,d\Omega.
\label{eq:energy_shell_surface_element}
\) 
For an isotropic dispersion relation,
\begin{equation}
 d^3k_f
 = \frac{1}{\left| \frac{dE}{dk_{f}}\right|} d\Sigma_{E_f} dE_f
 = \frac{1}{\left| \frac{dE}{dk_{f}}\right|} k_{f}^{2} \,d\Omega\,dE_f.
\label{eq:isotropic_final_state_measure}
\end{equation}
Here \(1/ |\frac{dE}{dk_{f}}|\) is the Jacobian associated with the
change of radial variable from \(k_f\) to \(E_f\). 
For a freely propagating nonrelativistic particle of mass \(m\), the corresponding relation is
\( d^3k_f 
= m k_f/\hbar^2 \, d\Omega\,dE_f
\). 
For a photon with \(E_f=\hbar c k_f\), where \(c\) is the speed of light, 
\(d^3k_f
=k_f^2 / (\hbar c) \, d\Omega\,dE_f \).

The conventional double-differential cross section is defined using
the coordinate product measure \(d\mu_{\Omega E}= d\Omega\,dE_f\) 
on the energy--angle outcome space. If \(\sigma \ll \mu_{\Omega E}\), there exists a density \(s(\Omega,E_f)\) such that
\begin{equation}
\sigma(A)
=
\int_A
s(\Omega,E_f)\,d\mu_{\Omega E}
\end{equation}
for every measurable region \(A\) of outgoing directions and final
energies. This density is conventionally denoted by
\begin{equation}
\frac{d^2\sigma}{d\Omega\,dE_f}
\equiv  s(\Omega,E_f)
= \frac{d\sigma}{d\mu_{\Omega E}}.
\label{eq:double_differential_cross_section_rn}
\end{equation}
Thus, despite its notation, the double-differential cross section is more fundamentally a 
Radon--Nikodym density of the scattering measure with respect to
the product measure \(d\Omega\,dE_f\).

\subsubsection{Compton scattering and the Klein--Nishina formula}
\label{subsec:Compton_scattering_example}

Compton scattering from a free electron provides a physical example in which the constraints imposed by energy--momentum conservation and the distribution of scattering-measure mass can be seen explicitly. 
Let the incident photon energy \(E_i\) be fixed. If only
the outgoing photon direction is taken as the outcome, the scattering
measure on \(S^2\) is
\begin{equation}
\sigma_{E_i}(d\Omega)
=
\frac{d\sigma_{\mathrm{KN}}}{d\Omega}
\,d\Omega,
\label{eq:klein_nishina_angular_measure}
\end{equation}
where, for unpolarized photons, the angular differential cross section
is given by the Klein--Nishina formula,
\begin{equation}
\frac{d\sigma_{\mathrm{KN}}}{d\Omega}
=
\frac{r_e^2}{2}
\left(\frac{E_f}{E_i}\right)^2
\left[
\frac{E_f}{E_i}
+
\frac{E_i}{E_f}
-
\sin^2\theta
\right].
\label{eq:klein_nishina_formula}
\end{equation}
Here, \(r_e\) is the classical electron radius and \(\theta\) is the
photon scattering angle.

The photon dispersion relation, \(E_f=\hbar c k_f\), 
fixes the geometry and reference measure of the outgoing-photon state space, whereas the relation between \(E_f\) and the scattering angle follows from four-momentum conservation for the photon--electron system. 
For a stationary initial electron, one obtains the standard Compton
energy--angle relation, 
\begin{equation}
E_f
=
\dispE(\theta;E_i)
\coloneqq
\frac{E_i}
{1+\dfrac{E_i}{m_ec^2}(1-\cos\theta)}.
\label{eq:compton_kinematic_relation}
\end{equation}

If both the outgoing direction and energy are retained, the outcome space is
\(
X_{\Omega E}=S^2\times\mathbb R_+
\).  
For fixed \(E_i\), the kinematically allowed final states lie on the
two-dimensional graph
\begin{equation}
\mathcal G_{E_i}
=
\left\{
(\Omega,E_f)\in S^2\times\mathbb R_+
\;\middle|\;
E_f=\dispE(\theta;E_i)
\right\}.
\label{eq:compton_kinematic_graph}
\end{equation}
The scattering measure is concentrated on this graph, with its mass
distributed according to the Klein--Nishina differential cross section.
More precisely, for any bounded measurable function \(f\),
\begin{equation}
\int_{S^2\times\mathbb R_+}
f(\Omega,E_f)\,
\sigma_{E_i}(d\Omega\,dE_f)
=
\int_{S^2}
f\!\left(\Omega,\dispE(\theta;E_i)\right)
\frac{d\sigma_{\mathrm{KN}}}{d\Omega}\,d\Omega .
\label{eq:compton_measure_integral}
\end{equation}
This relation characterizes the measure through its action on arbitrary
observables \(f\). The same measure may be written in the customary Dirac-delta notation as
\begin{equation}
\sigma_{E_i}(d\Omega\,dE_f)
=
\frac{d\sigma_{\mathrm{KN}}}{d\Omega}
\delta\!\left(E_f-\dispE(\theta;E_i)\right)
\,d\Omega\,dE_f .
\label{eq:compton_delta_measure}
\end{equation}
Indeed, integrating this expression against an arbitrary function
\(f\) gives
\begin{align}
&\int_{S^2}\int_{\mathbb R_+}
f(\Omega,E_f)
\frac{d\sigma_{\mathrm{KN}}}{d\Omega}
\delta\!\left(E_f-\dispE(\theta;E_i)\right)
\,d\Omega\,dE_f =
\int_{S^2}
f\!\left(\Omega,\dispE(\theta;E_i)\right)
\frac{d\sigma_{\mathrm{KN}}}{d\Omega}\,d\Omega,
\end{align}
which is exactly Eq.~\eqref{eq:compton_measure_integral}. Thus, the
Dirac-delta expression is an equivalent representation of the same measure because
the two expressions give identical integrals for every function \(f\).

The scattering measure is supported on the two-dimensional graph
\(\mathcal G_{E_i}\), which is embedded in the three-dimensional outcome space
\(S^2\times\mathbb R_+\).
As a two-dimensional subset of this three-dimensional space,
\(\mathcal G_{E_i}\) has zero measure with respect to the product measure
\(d\Omega\,dE_f\).
The scattering measure is therefore singular with respect to
\(d\Omega\,dE_f\) and has no ordinary density with respect to
this product measure, although it can be represented
formally using a Dirac delta function as in
Eq.~\eqref{eq:compton_delta_measure}. 
Its angular marginal, however, is absolutely
continuous with respect to \(d\Omega\), with density
\(d\sigma_{\mathrm{KN}}/d\Omega\).
This illustrates why the scattering measure is more fundamental than
any particular differential cross section representation.


\section{Transformations of scattering measures}
\label{sec:kernel}

The scattering measure describes the physical distribution of outcomes
before recording, whereas experimental data are apparatus-mediated
records or reduced representations.  We describe the transformations
between these stages at the level of finite measures, using measurable
maps for deterministic operations and stochastic kernels for
instrumental response and data reduction.  Sub-probability kernels are
used when events may be lost.

\subsection{Recorded outcome spaces}
\label{subsec:recorded_outcome_spaces}

Let \((X,\mathcal X)\) denote the
measurable space of physical scattering outcomes and \((Y,\mathcal Y)\)
the measurable space of recorded or reduced outcomes retained 
for analysis. 
A recorded outcome may include a
detected particle position, time stamp, spin-channel
label, or a product of several such variables, whereas a reduced outcome
may be a reconstructed physical coordinate, 
such as energy, momentum, or a histogram-bin index. 
Specifying these spaces determines which 
distinctions remain available at each stage of the experiment.

\subsection{Deterministic transformations and pushforward measures}
\label{subsec:deterministic_transformations}

\begin{figure}[t]
\centering
\begin{tikzpicture}[>=stealth]


\draw[thick] (-3,0) ellipse (1.3 and 1.6);
\draw[thick] ( 3,0) ellipse (1.3 and 1.6);

\draw[thick, dashed] (-3,0) ellipse (1.05 and 1.0);
\draw[thick, dashed] ( 3,0) ellipse (0.75 and 1.3);

\node at (-3, 2.1) {Physical outcome \((X,\mathcal X)\)};
\node at ( 3, 2.1) {Recorded outcome \((Y,\mathcal Y)\)};

\node at (-3,-0.4) {\(T^{-1}(B)\)};
\node at ( 3,-0.4) {\(B\)};

\draw[->, thick] (-1.1,1.0) to[bend left=15]
node[above] {\(T:X\to Y\)} (1.1,1.0);

\draw[->, thick] (1.1,-1.0) to[bend left=15]
node[below] {\(B\mapsto T^{-1}(B)\)} (-1.1,-1.0);

\fill (-2.7,0.35) circle (1.5pt);
\node[left] at (-2.7,0.35) {\(x\)};

\fill (2.7,0.35) circle (1.5pt);
\node[right] at (2.7,0.35) {\(T(x)\)};

\draw[->, thin] (-2.62,0.35) to[bend left=8] (2.62,0.35);


\node at (-3,-3.0)
{\(\mathcal M_+(X,\mathcal X)\)};

\node at ( 3,-3.0)
{\(\mathcal M_+(Y,\mathcal Y)\)};

\draw[->, very thick] (-1.65,-3.0) -- (1.65,-3.0)
node[midway, above] {\(T_\#\)};

\node at (-3,-3.55) {\(\sigma\)};
\node at ( 3,-3.55) {\(\nu=T_\#\sigma\)};

\node at (0,-4.25) {
\(
\bigl(T_\#\sigma\bigr)(B)
=
\sigma\!\left(T^{-1}(B)\right)
\)
};

\node[anchor=west, font=\small\itshape] at (-5.2,-1.9)
{\(\uparrow\) outcome level};

\draw[densely dashed, gray] (-5.0,-2.2) -- (5.0,-2.2);

\node[anchor=west, font=\small\itshape] at (-5.2,-2.45)
{\(\downarrow\) measure level};

\end{tikzpicture}

\caption{
A deterministic measurement acts at two related levels.
At the outcome level, the measurable map \(T:X\to Y\) sends
\(x\) to \(T(x)\), while a measurable recorded set
\(B\in\mathcal Y\) is pulled back to the physical event
\(T^{-1}(B)\in\mathcal X\).
The same map induces the pushforward
\(T_\#:\mathcal M_+(X,\mathcal X)\to
\mathcal M_+(Y,\mathcal Y)\) on finite positive measures,
which maps the scattering measure \(\sigma\) to the
recorded measure \(\nu=T_\#\sigma\). 
Measurable sets and observables are pulled back, whereas measures are pushed forward. 
}
\label{fig:pushforward_preimage}
\end{figure}
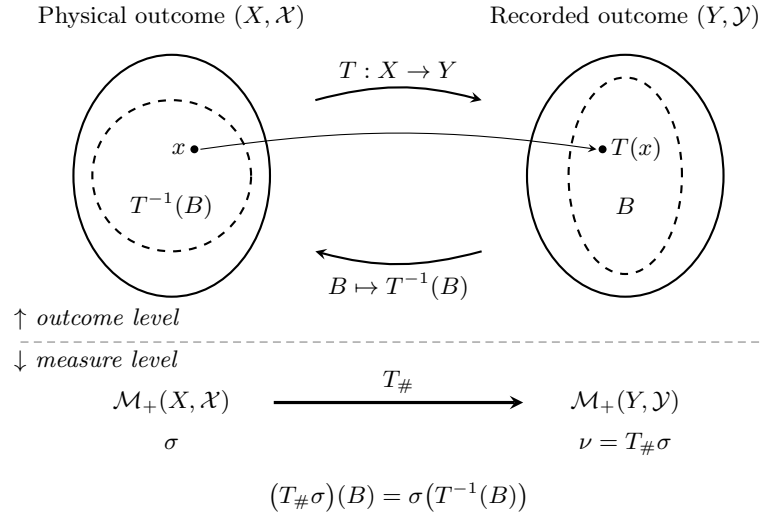

A measurable map
\( 
T:X\longrightarrow Y
\) 
describes a deterministic transformation for which the recorded outcome
\(y=T(x)\) is uniquely determined by the physical outcome \(x\).
Here, measurability means that \(
T^{-1}(B)\in\mathcal X\) for every \(B\in\mathcal Y\), 
so that the set of physical outcomes mapped into any measurable
recorded set \(B\) is itself measurable. 
The map \(T\) defines the zero--one-valued kernel
\(
K_T(B\mid x)
=
\mathbf 1_B\!\left(T(x)\right)
\), 
where \(\mathbf 1_B(y)=1\) if \(y\in B\), and \(0\) otherwise.

Since 
\(
\mathbf 1_B(T(x))=\mathbf 1_{T^{-1}(B)}(x)
\), 
the measure induced on \(Y\) by \(T\) is 
\begin{align}
\nu(B)
&=
\int_X
\mathbf 1_B(T(x))\,
\sigma(dx)
\nonumber\\
&=
\int_X
\mathbf 1_{T^{-1}(B)}(x)\,
\sigma(dx) = \sigma\!\left(T^{-1}(B)\right).
\label{eq:deterministic_output_measure}
\end{align}
The pushforward of \(\sigma\) by \(T\), denoted by
\(T_\#\sigma\), is the measure on \((Y,\mathcal Y)\) defined by
\begin{equation}
\bigl(T_\#\sigma\bigr)(B)
:=
\sigma\!\left(T^{-1}(B)\right),
\qquad
B\in\mathcal Y.
\label{eq:pushforward_definition}
\end{equation}
The symbol \(T_\#\) thus denotes the operation that maps a measure
on \(X\) to a measure on \(Y\) through \(T\).  
Hence, the induced measure in Eq.~\eqref{eq:deterministic_output_measure} 
is precisely the pushforward of
\(\sigma\) by \(T\):
\(
\nu
=
T_\#\sigma.
\)

\subsubsection{Induced maps and pushforward--pullback duality}

Figure~\ref{fig:pushforward_preimage} illustrates the two levels of
mapping involved in a deterministic transformation. At the outcome
level, the measurable map \(T:X\to Y\) sends an individual physical
outcome \(x\) to the recorded outcome \(T(x)\), while measurable
recorded events are pulled back through \(B\mapsto T^{-1}(B)\).
At the measure level, the same transformation induces a map between
spaces of finite positive measures,
\begin{equation}
T_\# :
\mathcal M_+(X,\mathcal X)
\longrightarrow
\mathcal M_+(Y,\mathcal Y),
\qquad
\sigma \longmapsto T_\#\sigma ,
\label{eq:measure_level_map_deterministic}
\end{equation}
where \(\mathcal M_+(X,\mathcal X)\) and \(\mathcal M_+(Y,\mathcal Y)\) denote the spaces of finite
positive measures on the respective measurable spaces.

The measure-level viewpoint also reveals a simple duality between the
forward transformation of measures and the backward transformation of
functions defined on the outcome spaces. Let
\(f:Y\to\mathbb R\) be a bounded measurable function of the recorded
outcome. Such a function assigns a numerical value or weight to each
recorded outcome \(y\in Y\). When this assignment has a physical or
experimental interpretation, we refer to \(f\) as an observable on the
outcome space \(Y\). For example, \(f\) may represent a measured
quantity, a weighting function, or the indicator of a measurable
region.

The pushforward relation can be expressed in integral form as
\begin{equation}
\int_Y
f(y)\,
\bigl(T_\#\sigma\bigr)(dy)
=
\int_X
f(T(x))\,
\sigma(dx).
\label{eq:pushforward_integral_identity}
\end{equation}
Thus, an observable defined on the recorded outcome space may be
integrated against the pushforward measure, or equivalently pulled
back to the physical outcome space. The resulting
function
\begin{equation}
(T^*f)(x)
\coloneq 
(f\circ T)(x)
=
f(T(x))
\end{equation}
is the pullback of \(f\) by \(T\). 
The physical meaning of this pullback is straightforward. For a
physical outcome \(x\in X\), \(T^*f(x)\) is the value of the recorded
observable \(f\) that would be assigned to the recorded outcome
\(T(x)\). Thus, \(T^*f\) may be viewed as the observable \(f\) expressed
back on the physical outcome space \(X\).

Schematically, at the outcome level,
\begin{equation*}
\begin{array}{ccccc}
X
& \xrightarrow{\qquad T\qquad} &
Y
\\[1.0ex]
T^*f:X\to\mathbb R
& \mathrel{\reflectbox{\ensuremath{\longmapsto}}} &
f:Y\to\mathbb R
\\[0.2ex]
\mathrel{\rotatebox{90}{$\in$}}
&&
\mathrel{\rotatebox{90}{$\in$}}
\\[-0.8ex]
\mathcal F(X)
& \xleftarrow{\qquad T^*\qquad} &
\mathcal F(Y)
\end{array}
\end{equation*}
where \(\mathcal F(X)\) and \(\mathcal F(Y)\) denote the spaces of
bounded measurable real-valued functions on \(X\) and \(Y\), 
respectively. The schematic makes the reversal explicit: while \(T\)
maps outcomes from \(X\) to \(Y\), the induced pullback \(T^*\) maps
functions in the opposite direction, from \(\mathcal F(Y)\) to
\(\mathcal F(X)\).

Introducing the natural duality pairing between a bounded measurable function
and a finite measure~\cite{ay2017information},
\begin{equation}
\langle f,\sigma\rangle
:=
\int f\,d\sigma,
\label{eq:duality_pairing}
\end{equation}
Eq.~\eqref{eq:pushforward_integral_identity} can be written compactly as
\begin{equation}
\langle f,T_\#\sigma\rangle
=
\langle T^*f,\sigma\rangle .
\label{eq:pushforward_pullback_duality}
\end{equation}
Thus, \(T_\#\) acts forward on measures, while \(T^*\) acts backward
on functions of the outcome, and the two operations are dual with
respect to integration.  
The corresponding relation for measurable events is recovered by taking \(f=
\mathbf 1_B\), for which \( T^{*}\mathbf 1_B=\mathbf 1_{T^{-1}(B)}\).

The deterministic operations most commonly encountered in scattering
analysis---including coordinate transformation, finite binning,
projection or marginalization, and radial reduction---are all instances
of this same pushforward construction. Their explicit realizations are
collected in Appendix~\ref{app:concrete_reductions}; here
we retain the measure-level structure that will be used below.

\subsection{Stochastic transformations and probability kernels}
\label{subsec:physical_to_recorded_outcomes}

A deterministic transformation assigns a unique recorded outcome
\(T(x)\) to each physical outcome \(x\). More generally, an experimental
operation is described by a probability distribution over recorded outcomes. 
A probability kernel describes the conditional
distribution of the recorded outcome given the input outcome. Intuitively,
\begin{equation}
K(B\mid x)
=
\Pr\!\left(\text{output}\in B \mid \text{input}=x\right), \nonumber
\end{equation}
that is, the probability that the recorded outcome lies in a measurable
region \(B\in\mathcal Y\), given the input outcome \(x\).

Such a probability kernel \(K\) from \((X,\mathcal X)\) to
\((Y,\mathcal Y)\) is required to satisfy two conditions in terms of the different inputs to \(K(B\mid x)\). 
First, for each fixed input outcome \(x\in X\),
\begin{equation}
K(\,\cdot\mid x):
\mathcal Y\longrightarrow[0,1],
\qquad
B\longmapsto K(B\mid x), \nonumber
\end{equation}
takes a measurable recorded event \(B\in\mathcal Y\) as its input and
returns its probability. For fixed \(x\), the mapping 
\(B\mapsto K(B\mid x)\) is a probability measure on \((Y,\mathcal Y)\). 
Second, for each fixed recorded event \(B\in\mathcal Y\),
\begin{equation}
K(B\mid\cdot):
X\longrightarrow[0,1],
\qquad
x\longmapsto K(B\mid x), \nonumber
\end{equation}
takes an outcome \(x\in X\) as its input and returns the
probability that the recorded outcome lies in \(B\). As a function of
\(x\), it is required to be measurable with respect to \(\mathcal X\),
so that it can be integrated over the input measure.

\subsubsection{Kernel-induced measurement maps}

Let \((X,\mathcal X)\) be an input outcome space carrying a finite
measure \(\sigma\), and let \((Y,\mathcal Y)\) be a recorded outcome
space. A probability kernel \(K\) from \(X\) to \(Y\) assigns to each
\(x\in X\) a probability measure \(K(\,\cdot\mid x)\) on \(Y\).
It induces a finite measure \(\nu\) on \(Y\) according to
\begin{equation}
\nu(B)
=
\int_X K(B\mid x)\,\sigma(dx),
\qquad
B\in\mathcal Y .
\label{eq:detector_output_measure2}
\end{equation}
By analogy with the deterministic pushforward
Eq.~\eqref{eq:pushforward_definition}, we write
\(
\nu = K_\#\sigma .
\)

At the level of finite measures,  a probability kernel induces a map
\begin{equation}
K_\# :
\mathcal M_+(X,\mathcal X)
\longrightarrow
\mathcal M_+(Y,\mathcal Y),
\qquad
\sigma\longmapsto K_\#\sigma .
\label{eq:measure_level_map_kernel}
\end{equation}
The generalization from deterministic to stochastic transformations 
preserves the measure-level structure: an input finite measure is
mapped forward to an output finite measure. 

The corresponding backward action on observables is defined as follows.
For a bounded measurable function \(f:Y\to \mathbb R\), let
\begin{equation}
(K^*f)(x)
:=
\int_Y f(y)\,K(dy\mid x).
\label{eq:kernel_backward_action}
\end{equation}
For a fixed input outcome \(x\), \(K^*f(x)\) is the conditional average
of the observable \(f\) over the possible recorded outcomes described
by \(K(\,\cdot\mid x)\).  Using the pairing introduced in 
Eq.~\eqref{eq:duality_pairing}, these actions satisfy 
\footnote{Since \(f\) is bounded, \(K(\,\cdot\mid x)\) is a probability
measure, and \(\sigma\) is finite, the relevant integral is absolutely
convergent; hence Fubini's theorem applies.}
\begin{align}
\langle f,K_\#\sigma\rangle
&=
\int_Y f(y)\,(K_\#\sigma)(dy)
\nonumber\\
&=
\int_Y f(y)
\left[
\int_X K(dy\mid x)\,\sigma(dx)
\right]
=
\int_X
\left[
\int_Y f(y)\,K(dy\mid x)
\right]
\sigma(dx)
\nonumber\\
&=
\int_X (K^*f)(x)\,\sigma(dx)
= \langle K^*f,\sigma\rangle .
\label{eq:kernel_pushforward_backward_duality}
\end{align}
Thus, \(K_\#\) acts forward on finite measures, whereas \(K^*\) acts
backward on observables. This is the stochastic counterpart of the deterministic 
pushforward--pullback duality in Eq.~\eqref{eq:pushforward_pullback_duality}.

A deterministic transformation is recovered as a special case.
For a measurable map \(T:X\to Y\), define the kernel
\(K_T(B\mid x) =\mathbf 1_B(T(x))\). The backward action then becomes
\(
(K_T^*f)(x)
=
\int_Y f(y)\,K_T(dy\mid x)
=
f(T(x))
=
(T^*f)(x).
\)

The essential extension from deterministic to stochastic
transformations therefore occurs at the outcome level: a unique output outcome \(T(x)\) is replaced by a conditional distribution of possible output outcomes \(K(\cdot \mid x)\). At the level of measures
and observables, however, the same forward--backward structure is
retained:
\begin{equation*}
\begin{array}{ccccc}
\mathcal M_+(X,\mathcal X)
&
\xrightarrow{\qquad K_\#\qquad}
&
\mathcal M_+(Y,\mathcal Y)
\\[0.6ex]
\sigma
&
\longmapsto
&
K_\#\sigma
\\[1.0ex]
K^*f: X\to \mathbb R
&
\mathrel{\reflectbox{\ensuremath{\longmapsto}}}
&
f: Y \to \mathbb R
\\[0.1ex]
\mathrel{\rotatebox{90}{$\in$}}
&&
\mathrel{\rotatebox{90}{$\in$}}
\\[-0.7ex]
\mathcal F(X)
&
\xleftarrow{\qquad K^*\qquad}
&
\mathcal F(Y)
\\[1.1ex]
\multicolumn{3}{c}{
\displaystyle
\langle f,K_\#\sigma\rangle
=
\langle K^*f,\sigma\rangle
}
\end{array}
\label{eq:kernel_forward_backward_structure}
\end{equation*}

This framework provides a common mathematical structure 
for a broad class of measurement and data-reduction operations 
represented by measurable maps, probability kernels, 
or sub-probability kernels. 
Successive measurement processes may be combined into a single effective
kernel.  The composition rule and the corresponding backward propagation
of observables are given in Appendix~\ref{app:kernel_composition}.

\subsubsection{Mass preservation and event loss}

When \(K(Y\mid x)=1\) for every \(x\), the kernel preserves the total
mass: \(\nu(Y)=\sigma(X)\). 
More generally, a sub-probability kernel satisfies
\(K(Y\mid x)\leq 1\), 
so that \(\nu(Y)\leq \sigma(X)\). 
The missing mass then represents input events that do not produce a
retained output, as may occur through finite detection efficiency, acceptance or
event rejection.

By contrast, consider a mass-preserving Gaussian detector-resolution
kernel. For a scalar physical coordinate \(x\in\mathbb R\), recorded
coordinate \(y\in\mathbb R\), and a measurable set
\(B\subseteq\mathbb R\), let
\begin{equation}
K_{\gamma}(B\mid x)
=
\int_B k_\gamma(y\mid x)\,dy,
\qquad
k_\gamma(y\mid x)
=
\frac{1}{\sqrt{2\pi}\gamma}
\exp\left[
-\frac{(y-x)^2}{2\gamma^2}
\right].
\label{eq:Gaussian_detector_resolution_kernel}
\end{equation}
Since \(\int_{\mathbb R}k_\gamma(y\mid x)\,dy=1\), 
this kernel preserves the total mass of the scattering measure while
spreading each physical outcome over a Gaussian distribution of
recorded outcomes with standard deviation \(\gamma\).
Thus, preservation of total mass does not imply preservation of
statistical information. The corresponding information loss is
quantified in terms of Fisher information in 
Sec.~\ref{subsubsec:gaussian_blur}.

\paragraph*{Event selection as a sub-probability kernel.} 
Event selection provides a simple realization of the complementary,
mass-losing case. Let \(S\in\mathcal Y\) denote the set of accepted
recorded outcomes. If accepted outcomes retain their original
coordinates, the corresponding kernel is
\(
K_S(B\mid y)
=
\mathbf 1_S(y)\mathbf 1_B(y)\) for  \(B\in\mathcal Y\). 
Because \(
K_S(Y\mid y)
=
\mathbf 1_S(y)
\leq 1
\),  outcomes outside \(S\) are not retained. The transformed measure is
\(
\bigl((K_S)_\#\nu\bigr)(B)
=
\nu(S\cap B)
\)
whose total mass is 
\(
\bigl((K_S)_\#\nu\bigr)(Y) 
=
\nu(S) \leq \nu(Y)
\).  
The missing mass is therefore
\(
\nu(Y)
-
\bigl((K_S)_\#\nu\bigr)(Y)
=
\nu(Y\setminus S)
\), 
corresponding precisely to the recorded outcomes excluded by the
selection rule. 
A deterministic map assigns every
input outcome to an output and therefore preserves total mass, whereas
a sub-probability kernel can represent the
loss of outcomes from the retained set \(S\) without explicitly 
retaining the discarded region \(Y\setminus S\).

This distinction will reappear in the quantum-measurement setting, 
where unrecorded outcomes may likewise be represented explicitly 
by completing the outcome space, 
as discussed in Sec.~\ref{subsec:quantum_outcome_measures}.

\subsubsection{Discrete channel mixing}
\label{subsubsec:discrete_channel_kernel}

For discrete outcome spaces, the preceding kernel structure reduces to 
an ordinary mixing matrix. Such discrete outcomes arise naturally in 
many measurements, such as energy levels, spin, polarization, or qubit states. 
In practice, however, these outcomes may not be perfectly discriminated, 
so an event originating in one input channel can be recorded in another.

Let \(X=\{1,\ldots,N\}\) and \(Y=\{1,\ldots,M\}\), and define 
\(K_{ji}=K(\{j\}\mid i)\), 
the conditional probability that an event in input channel \(i\) is
recorded in output channel \(j\).
Writing the measures for singletons as 
\(\sigma_i=\sigma(\{i\})\) and
\(\nu_j=\nu(\{j\})\), the kernel transformation
\(\nu=K_\#\sigma\) becomes
\(
\nu_j
=
\sum_{i=1}^{N}K_{ji}\sigma_i\),  or, in vector form, 
\(
\bmnu=K\bm{\sigma}
\), where \(\bm{\sigma}=(\sigma_1, \ldots, \sigma_N)^{\mathsf T}\) and \(\bmnu= (\nu_1, \ldots, \nu_M)^{\mathsf T}\). Thus, the general kernel integral reduces to matrix multiplication for
discrete channels. The kernel is mass preserving when
\(\sum_j K_{ji}=1\) for every input channel \(i\); column sums smaller
than unity describe event loss.

As the simplest example, consider symmetric mixing between two channels, 
\begin{equation}
K_\varepsilon
=
\begin{pmatrix}
1-\varepsilon & \varepsilon \\
\varepsilon & 1-\varepsilon
\end{pmatrix},
\qquad
0\leq\varepsilon\leq\frac{1}{2}.
\label{eq:two_channel_kernel}
\end{equation}
This kernel preserves the total mass: \(\nu_1 + \nu_2 = \sigma_1 +\sigma_2 \)
, but progressively reduces the
distinguishability of the two input channels as \(\varepsilon\)
increases. The corresponding loss of Fisher information is evaluated
explicitly in Sec.~\ref{subsubsec:discrete_channel_mixing}.

\subsection{Observability and information loss under measurement transformations}
\label{subsec:observability_information_loss}

The fact that successive measurement transformations can be represented 
by a single composed kernel in the forward model does not imply 
reversibility. Through a sequence of operations
between outcome spaces of different dimensions or types, the intermediate or
original event record generally cannot be reconstructed from the final
one. 

The dual action on observables provides a natural way to express which
of these distinctions remain accessible after the transformation.  For
a deterministic reduction \(T:X\to Y\), an observable on the reduced
space is represented on \(X\) by its pullback \(T^*f=f\circ T\).
Consequently, only observables that are constant on the fibers of \(T\)
can be constructed from the reduced record.  For a stochastic kernel
\(K\), the corresponding backward action of Eq.~\eqref{eq:kernel_backward_action} 
plays the analogous role: it identifies the functions of the incoming
outcome that can be represented through observables of the transformed
record.

Whether the loss of such distinctions is relevant to inference depends
on the parameterized family under consideration.  A transformation may
erase details of the original event record without removing any
distinction relevant to a particular parameter, whereas another
transformation may merge outcomes whose relative weights change with
that parameter.  Event-level irreversibility and loss of
parameter-relevant information are therefore distinct notions.

This distinction will be made precise after introducing the score and
Fisher information in the next Sec.~\ref{sec:fisher_scattering_measure}.  
The score associated with the transformed record will be shown to
correspond to the component of the original score that remains
resolvable from that record, expressed through conditional expectation.
The unresolved component then determines the Fisher information lost
under the transformation.

Thus the backward action on observables identifies which distinctions
remain accessible after measurement or data reduction, while the
information-geometric analysis developed below determines which of
those distinctions matter for inference and quantifies the information
that is lost.  In particular, a reduction may be irreversible at the
level of individual events while remaining information preserving for
a specified parameter.


\section{Fisher geometry of parameterized scattering measures}
\label{sec:fisher_scattering_measure}

We now consider a parameterized family of finite scattering measures
\(\{\sigma_{\lambda}\}_{\lambda\in\Theta}\) and introduce a local geometry
that quantifies the distinguishability of neighboring measures.

We first define the tangent measure and intrinsic score, leading to a
Fisher metric independent of any particular density representation or
counting model. This metric admits an orthogonal decomposition into
total-mass and normalized-shape contributions. We then introduce
Poisson point process and relate the resulting experimental Fisher
information to the intrinsic metric through the exposure.

\subsection{Parameterized families and tangent measures}
\label{subsec:parameterized_scattering_measures}

Consider a one-parameter family of finite positive scattering measures
\(\{\sigma_\lambda\}_{\lambda\in\Theta}\), 
\(\Theta\subseteq\mathbb R\) 
on a measurable outcome space \((X,\mathcal X)\).
Equivalently, the family defines a map
\begin{equation}
\sigma:
\Theta
\longrightarrow
\mathcal M_+(X,\mathcal X),
\qquad
\lambda
\longmapsto
\sigma_\lambda ,
\label{eq:parameterized_measure_model}
\end{equation}
where \(\mathcal M_+(X,\mathcal X)\) denotes the space of finite
positive measures on \((X,\mathcal X)\). 
We assume that the family is differentiable in the sense that there
exists a finite signed measure \(\dot{\sigma}_\lambda\) satisfying
\begin{equation}
\frac{d}{d\lambda}
\int_X f(x)\,d\sigma_\lambda(x)
=
\int_X f(x)\,d\dot{\sigma}_\lambda(x)
\label{eq:tangent_measure_definition}
\end{equation}
for every bounded measurable function \(f\).
Hereafter, an overdot denotes differentiation with respect to
\(\lambda\).
The signed measure \(\dot{\sigma}_\lambda\) is the tangent measure of
the family at \(\lambda\).
In particular, for \(A\in\mathcal X\),
\(\dot{\sigma}_\lambda(A)
= d \sigma_\lambda(A)/d\lambda\). 
Thus, the tangent measure collects the infinitesimal changes of the
scattering strength assigned to all measurable subsets of the outcome
space.

We further assume
\(
\dot{\sigma}_\lambda\ll\sigma_\lambda
\).
The Radon--Nikodym derivative
\begin{equation}
u_\lambda(x)
\coloneqq
\frac{d\dot{\sigma}_\lambda}
     {d\sigma_\lambda}(x)
\label{eq:intrinsic_measure_score}
\end{equation}
will be called the \emph{intrinsic score} of the scattering-measure
family. It describes the infinitesimal change of the measure relative
to the measure itself, without introducing a reference measure or a
particular density representation.

\subsection{Relative entropy and intrinsic Fisher metric}
\label{subsec:intrinsic_fisher_scattering_measure}

The intrinsic score describes the first-order relative variation of a
scattering measure.  To connect this local variation to the
distinguishability of two finite measures at finite separation, it is
useful to start with a relative-entropy divergence.

For finite positive measures \(\sigma\) and \(\tau\), with
\(\sigma\ll\tau\), we use the generalized Kullback--Leibler (KL) divergence~\cite{Leskela2024}
\begin{equation}
\mathcal D(\sigma\Vert\tau)
\coloneqq
\int_X
\ln\!\left(
\frac{d\sigma}{d\tau}
\right)
d\sigma
-
\sigma(X)
+
\tau(X).
\label{eq:extended_KL_measure}
\end{equation}
The additional mass terms are required because the measures need not be
normalized.  If \(\sigma\) and \(\tau\) are probability measures,
their total masses are both unity and
Eq.~\eqref{eq:extended_KL_measure} reduces to the usual
KL divergence.

We now consider two neighboring members of the parameterized family,
\(\sigma_\lambda\) and \(\sigma_{\lambda+\delta\lambda}\), where
\(\delta\lambda\) denotes a sufficiently small parameter increment.
We assume that the two measures are mutually absolutely continuous, i.e., 
\( \sigma_{\lambda+\delta\lambda} \ll \sigma_\lambda \) and \(  \sigma_\lambda \ll \sigma_{\lambda+\delta\lambda}\), and further assume sufficient smoothness 
such that their derivative admits the expansion 
\begin{equation}
\frac{d\sigma_{\lambda+\delta\lambda}}
     {d\sigma_\lambda}
=
1
+
\delta\lambda\, u_\lambda
+
\frac{(\delta\lambda)^2}{2}v_\lambda
+
o\!\left((\delta\lambda)^2\right),
\label{eq:local_measure_ratio_expansion}
\end{equation}
where
\(u_\lambda=d\dot{\sigma}_\lambda/d\sigma_\lambda\)
 is the intrinsic score and \(v_\lambda = d\ddot{\sigma}_\lambda/d\sigma_\lambda \) denotes the second-order relative variation of the measure.

Setting \(z = \delta\lambda\, u_\lambda
+ (\delta\lambda)^2 v_\lambda /2 
+ o\!\left((\delta\lambda)^2\right) \) and  using
\(-\ln(1+z) = -z+z^2/2+o(z^2)\),
the logarithmic term in the generalized KL divergence
becomes
\begin{align}
\int_X
\ln
\frac{d\sigma_\lambda}
     {d\sigma_{\lambda+\delta\lambda}}
\,d\sigma_\lambda
=
-\delta\lambda
\int_X u_\lambda\,d\sigma_\lambda
-\frac{(\delta\lambda)^2}{2}
\int_X v_\lambda\,d\sigma_\lambda
+\frac{(\delta\lambda)^2}{2}
\int_X u_\lambda^2\,d\sigma_\lambda
+
o\!\left((\delta\lambda)^2\right).
\label{eq:KL_log_expansion}
\end{align}
For the mass terms,
\begin{align}
\sigma_{\lambda+\delta\lambda}(X)
-
\sigma_\lambda(X)
&=
\delta\lambda
\int_X u_\lambda\,d\sigma_\lambda
+
\frac{(\delta\lambda)^2}{2}
\int_X v_\lambda\,d\sigma_\lambda
+
o\!\left((\delta\lambda)^2\right).
\label{eq:KL_mass_expansion}
\end{align}
The first-order terms and the terms involving \(v_\lambda\) therefore
cancel in Eq.~\eqref{eq:extended_KL_measure}, leaving
\begin{equation}
\mathcal D
\left(
\sigma_\lambda
\Vert
\sigma_{\lambda+\delta\lambda}
\right)
=
\frac{(\delta\lambda)^2}{2}
\int_X
u_\lambda(x)^2\,d\sigma_\lambda(x)
+
o\!\left((\delta\lambda)^2\right).
\label{eq:KL_local_Fisher}
\end{equation}
The leading nonvanishing term of the generalized
KL divergence is quadratic in \(\delta\lambda\).
Consequently, although the divergence is generally asymmetric at finite separation, the two directions agree to
quadratic order:
\(
\mathcal D(
\sigma_\lambda\Vert\sigma_{\lambda+\delta\lambda})
=
\mathcal D(
\sigma_{\lambda+\delta\lambda}\Vert\sigma_\lambda)
+
o((\delta\lambda)^2).
\)
The directional asymmetry appears only beyond the quadratic
order. In particular, its leading third-order structure is associated
with the Amari--Chentsov cubic tensor~\cite{ay2017information}.

The local expansion above identifies
\(
\int_X u_\lambda(x)^2\, d\sigma_\lambda(x)
\) 
as the squared length of the tangent
\(\dot{\sigma}_\lambda\) to the trajectory
\(\lambda \mapsto \sigma_\lambda\).
Bilinearizing this squared length to compare arbitrary tangent
directions \(\tau_1\) and \(\tau_2\) leads to
\begin{equation}
g_\sigma(\tau_1,\tau_2)
=
\int_X
\frac{d\tau_1}{d\sigma}
\frac{d\tau_2}{d\sigma}
\, d\sigma,
\end{equation}
which is the Fisher--Rao metric on the space of finite positive
measures. 
Here, \(\tau_1\) and \(\tau_2\) are tangent signed measures at a finite
positive measure \(\sigma\), satisfying
\(\tau_1,\tau_2\ll\sigma\) and \(
d\tau_1 / d\sigma,\  d\tau_2 / d\sigma \in L^2(\sigma)\), 
where
\(L^2(\sigma)
=\left\{ f:
\int_X |f(x)|^2\,d\sigma(x)<\infty
\right\}\).
Thus, \(g_\sigma\) assigns an inner product to tangent directions,
providing the Fisher--Rao geometry directly for the parameterized
measure models~\cite{ay2018parametrized}.

For the parameterized scattering measure family, the relevant tangent
direction is \(\dot\sigma_\lambda\), so we set \(\tau_1=\tau_2=\dot\sigma_\lambda\). Since
\(u_\lambda=
d\dot\sigma_\lambda / d\sigma_\lambda\), 
its squared Fisher norm is
\begin{align}
\mathcal I^{(\sigma)}(\lambda)
&\coloneqq
g_{\sigma_\lambda}
\left(
\dot\sigma_\lambda,
\dot\sigma_\lambda
\right)
\nonumber\\
&=
\int_X
u_\lambda(x)^2\,d\sigma_\lambda(x).
\label{eq:intrinsic_fisher_scattering_measure}
\end{align}
We refer to
\(\mathcal I^{(\sigma)}(\lambda)\) as the
\emph{intrinsic Fisher information}. 
Equation~\eqref{eq:KL_local_Fisher} can therefore be written as
\begin{equation}
\mathcal D
\left(
\sigma_\lambda
\Vert
\sigma_{\lambda+\delta\lambda}
\right)
=
\frac{1}{2}
\mathcal I^{(\sigma)}(\lambda)
(\delta\lambda)^2
+
o\!\left((\delta\lambda)^2\right).
\label{eq:KL_local_Fisher_metric}
\end{equation}
This relation reveals the geometric meaning of the intrinsic Fisher information: 
the finite distinguishability of two scattering measures, as quantified by the generalized KL divergence, reduces locally to the Fisher metric.  This special role is also reflected in Chentsov's theorem, which
characterizes the Fisher metric, up to an overall constant, by its invariance
under information-preserving statistical transformations
\cite{Chentsov1982, Le2017}.

Along the one-parameter family, an infinitesimal parameter change \(d\lambda\) induces the tangent displacement \(d\sigma_\lambda=\dot\sigma_\lambda\,d\lambda\). 
Here \(d \lambda \) denotes the infinitesimal parameter displacement,  whereas \(\delta\lambda \) above denotes a small
finite parameter increment used to compare neighboring measures. 
The corresponding infinitesimal line element is
\begin{equation}
ds_{\mathrm F}^{\,2}
= g_{\sigma_{\lambda}}(d\sigma_\lambda, d\sigma_\lambda)=  g_{\sigma_{\lambda}}( \dot\sigma_\lambda,  \dot\sigma_\lambda) d\lambda^{2}
=
\mathcal I^{(\sigma)}(\lambda)\,d\lambda^2,
\label{eq:fisher_line_element_single}
\end{equation}
and therefore 
\begin{equation}
\frac{ds_{\mathrm F}}{|d\lambda|}
=
\sqrt{\mathcal I^{(\sigma)}(\lambda)}
\label{eq:fisher_speed_single}
\end{equation}
gives the local speed of the scattering measure family in Fisher geometry.  It quantifies how rapidly the scattering measure changes,
in the Fisher-geometric sense, per unit change of the physical
parameter \(\lambda\). Thus, a larger value of
\(\sqrt{\mathcal I^{(\sigma)}(\lambda)}\) indicates a stronger local
sensitivity of the scattering measure to that parameter.

This construction does not require a particular density
representation.
If the family is dominated by a common parameter-independent reference
measure \(\mu\), then  
\( d\dot{\sigma}_\lambda(x)
=\partial_\lambda s_\lambda(x)\,d\mu(x),
\)
and wherever \(s_\lambda(x)>0\),
\(
u_\lambda(x)
= \partial_\lambda s_\lambda(x) / s_\lambda(x)
=
\partial_\lambda\ln s_\lambda(x)
\). Consequently,
\begin{equation}
\mathcal I^{(\sigma)}(\lambda)
= \int_X
\frac{
\left[\partial_\lambda s_\lambda(x)\right]^2
}{
s_\lambda(x)
}
\,d\mu(x)
=
\int_X
s_\lambda(x)
\left[\partial_\lambda \ln s_\lambda(x)\right]^2
\,d\mu(x).
\label{eq:intrinsic_fisher_density_representation}
\end{equation}
The usual density-based score
\(\partial_\lambda\ln s_\lambda\) is recovered as the density
representation of the intrinsic measure-level score
\(d\dot{\sigma}_\lambda/d\sigma_\lambda\).
The invariance of the generalized KL divergence and the intrinsic
Fisher metric under changes of density representation is shown
explicitly in Appendix~\ref{app:representation_invariance}.

For a vector parameter
\(
\bm{\lambda}=(\lambda^1,\ldots,\lambda^m)
\),
define
\begin{equation}
u_i(x;\bm{\lambda})
\coloneqq
\frac{
d(\partial_i\sigma_{\bm{\lambda}})
}{
d\sigma_{\bm{\lambda}}
}(x),
\qquad
\partial_i
\equiv
\frac{\partial}{\partial\lambda^i}.
\label{eq:intrinsic_score_multiparameter}
\end{equation}
The intrinsic Fisher matrix is
\begin{equation}
\mathcal I_{ij}^{(\sigma)} (\bm{\lambda})
=
\int_X
u_i(x;\bm{\lambda})
u_j(x;\bm{\lambda})
\,d\sigma_{\bm{\lambda}}(x),
\label{eq:intrinsic_fisher_matrix}
\end{equation}
and the associated Fisher line element is
\begin{equation}
ds_{\mathrm F}^2
=
\mathcal I_{ij}^{(\sigma)} (\bm{\lambda})\,
d\lambda^i d\lambda^j,
\label{eq:multiparam_fisher_line_element}
\end{equation}
where summation over repeated parameter indices is understood.

Using a path parameter \(t\), consider a path
\(t\mapsto\bm{\lambda}(t)\) in parameter space, and the derivative of
\(\lambda^i\) with respect \(t\). Along this path, the chain rule gives the induced tangent variation
of the scattering measure as
\begin{equation}
\frac{d}{dt}\sigma_{\bm{\lambda}(t)}
=
\frac{d \lambda^{i}}{dt}\,
\partial_i\sigma_{\bm{\lambda}}.
\label{eq:induced_tangent_variation}
\end{equation}
The corresponding speed in the Fisher geometry is
\begin{equation}
\frac{ds_{\mathrm F}}{dt}
=
\sqrt{
  \frac{d \lambda^{i}}{dt}
\mathcal I_{ij}^{(\sigma)}(\bm{\lambda})
\frac{d \lambda^{j}}{dt}
}.
\label{eq:multiparam_Fisher_speed}
\end{equation}

\subsection{Mass--shape decomposition of the intrinsic Fisher geometry}
\label{subsec:mass_shape_intrinsic}

A finite positive measure contains two kinds of variation that are not
distinguished in an ordinary probability model: its total mass may
change, and its normalized distribution over the outcome space may
change. These two components are already separated at the level of
finite relative entropy, and become orthogonal in the local Fisher
geometry.

Let \(
\sigma=r_{\sigma}P\) and \(\tau=r_{\tau}Q\), 
where
\(
r_{\sigma}=\sigma(X)
\),
\(
r_{\tau}=\tau(X)
\),
and \(P\) and \(Q\) are probability measures.
Assuming \(P\ll Q\), the generalized KL divergence decomposes as
\begin{equation}
\mathcal D(\sigma\Vert\tau)
=
r_{\sigma}
\ln\frac{r_{\sigma}}{r_{\tau}}
-r_{\sigma}
+r_{\tau}
+
r_{\sigma}
D_{\mathrm{KL}}(P\Vert Q),
\label{eq:KL_mass_shape_decomposition}
\end{equation}
where \(
D_{\mathrm{KL}}(P\Vert Q)
=
\int_X
\ln\!\left(\frac{dP}{dQ}\right)
\, dP
\) denotes the ordinary KL divergence.
A finite separation between two positive measures consists of
a difference in their total masses together with a difference in their
normalized shapes.

The corresponding measure-level structure follows by writing a positive
parameterized scattering measure as
\begin{equation}
\sigma_\lambda
=
r_\lambda P_\lambda.
\label{eq:measure_mass_shape_decomposition}
\end{equation}
Here \(r_\lambda=\sigma_\lambda(X)\) represents the total mass (scattering strength), whereas \(P_\lambda\) describes the distribution over the outcome space. It is normalized as \(P_\lambda(X)=1\). 
Differentiation of
Eq.~\eqref{eq:measure_mass_shape_decomposition} with respect to \(\lambda\) gives
\begin{equation}
\dot\sigma_\lambda
=
\dot r_\lambda P_\lambda
+
r_\lambda\dot P_\lambda .
\label{eq:mass_shape_tangent}
\end{equation}
The first term changes the total mass without changing the normalized
shape, while the second redistributes the measure at fixed total mass.
Since the measure elements satisfy \(d\sigma_\lambda/dP_\lambda=r_\lambda\), the intrinsic score separates as
\begin{align}
u_{\lambda}
=
\frac{d\dot{\sigma}_\lambda}{d\sigma_\lambda}
&=
\frac{\dot r_\lambda}{r_\lambda}
+
\frac{d\dot P_\lambda}{dP_\lambda} \\
&\equiv
u^{(r)}+u^{(P)}. \nonumber
\label{eq:rate_shape_score}
\end{align}

Because \(P_\lambda(X)=1\), the shape score has zero mean, 
\(
\int_X u^{(P)}\,dP_\lambda
= \dot P_\lambda(X) = 0
\). 
Hence the two score components are orthogonal in
\(L^2(\sigma_\lambda)\):
\begin{equation}
\left\langle
u^{(r)},u^{(P)}
\right\rangle_{L^2(\sigma_\lambda)}
=
\int_X
\frac{\dot r_\lambda}{r_\lambda}
u^{(P)}
\,d\sigma_\lambda
=
\dot r_\lambda
\int_X
u^{(P)}
\,dP_\lambda
=
0.
\label{eq:mass_shape_orthogonality}
\end{equation}
Hereafter, \(\|\cdot\|\) denotes the \(L^2(\sigma_\lambda)\) norm.
The Fisher norm therefore obeys the Pythagorean decomposition
\begin{align}
\mathcal I^{(\sigma)}(\lambda)
&=
g_{\sigma_\lambda}
\left(
\dot\sigma_\lambda,\dot\sigma_\lambda
\right)
=
\|u_\lambda\|^2
\nonumber\\
&=
\|u^{(r)}\|^2
+
\|u^{(P)}\|^2
\nonumber\\
&=
\frac{\dot r_\lambda^{\,2}}{r_\lambda}
+
r_\lambda\,\mathcal I^{(P)}(\lambda),
\label{eq:intrinsic_mass_shape_FI_decomposition}
\end{align}
where
\begin{equation}
\mathcal I^{(P)}(\lambda)
\coloneqq
\int_X
\left[
u_\lambda^{(P)}(x)
\right]^2
\,dP_\lambda(x)
\label{eq:shape_fisher_information}
\end{equation}
This shows that mass and shape variations form orthogonal directions 
in the Fisher geometry of finite measures.

\paragraph*{Multiparameter case}
For a vector parameter
\(
\bm{\lambda}
=
(\lambda^1,\ldots,\lambda^m)
\),
the score in each parameter direction has the analogous decomposition
\begin{equation}
u_i
=
u_i^{(r)}
+
u_i^{(P)},
\qquad
u_i^{(r)} = \frac{\partial_i r_{\bm{\lambda}}}{r_{\bm{\lambda}}},
\qquad
u_i^{(P)} =\frac{d (\partial_i P_{\bm{\lambda}})}{dP_{\bm{\lambda}}}.
\label{eq:mass_shape_score_multiparameter}
\end{equation}
Since
\(
\int_X u_i^{(P)}\,dP_{\bm{\lambda}}=0
\),
the intrinsic Fisher matrix becomes
\begin{equation}
\mathcal I_{ij}^{(\sigma)}
=
\frac{
(\partial_i r_{\bm{\lambda}})(\partial_j r_{\bm{\lambda}})
}{
r_{\bm{\lambda}}
}
+
r_{\bm{\lambda}} \, 
\mathcal I_{ij}^{(P)},
\label{eq:intrinsic_mass_shape_Fisher_matrix}
\end{equation}
with
\begin{equation}
\mathcal I_{ij}^{(P)}
=
\int_X
u_i^{(P)}(x)\,
u_j^{(P)}(x)
\,dP_{\bm{\lambda}}(x), \qquad i, j= 1,\ldots, m.
\end{equation}
The mass contribution is 
\(
(1 / r_{\bm{\lambda}})
(\nabla r_{\bm{\lambda}})(\nabla r_{\bm{\lambda}})^{\mathsf T}
\), where \(\nabla \equiv (\partial_1, \ldots , \partial_m)^{\mathsf T}\).
This is an outer product and hence has rank at most one. Locally, the
total mass therefore distinguishes only the parameter-space direction
along which \(r\) changes; all remaining distinguishability must come
from changes in the normalized shape.

Changes in total scattering strength and changes in the normalized scattering pattern constitute
orthogonal contributions to local distinguishability, independently of any
particular statistical observation model.

\subsubsection{Geometric interpretation of the mass--shape 
decomposition: orthogonality and cone structure}
\label{subsubsec:physical_rate_shape}
The mass--shape decomposition distinguishes how a
parameter variation acts on the finite measure: through its total mass,
through a redistribution of its normalized shape over the outcome
space, or through both. A physical parameter
may contribute to both components; a shift, broadening, or deformation
of a scattering feature may also change its integrated intensity.

The mass--shape decomposition admits a natural geometric interpretation. 
To make this structure explicit, we introduce coordinates adapted to this decomposition and write a positive finite measure as
\(
\sigma_{(r, \bm{\theta})} = r P_{\bm{\theta}}
\), 
where \(r\) is its total mass and \({\bm{\theta}}=(\theta^1,\ldots,\theta^m)\) parametrizes the normalized
statistical model \(P_{\bm{\theta}}\).

The meaning of coordinates adapted to the mass--shape decomposition is illustrated below with two simple examples. 
For a measure supported on two discrete outcomes \(x_1\) and \(x_2\), with weights \(\lambda^{1}\) and \(\lambda^{2}\), respectively, the adapted coordinates are 
\( 
r=\lambda^1+\lambda^2 \),  \(\theta=\lambda^1/(\lambda^1+\lambda^2)\) so that \(\sigma_{(r,\theta)}=r[\theta\delta_{x_1}+(1-\theta)\delta_{x_2}]\). 
As a continuous example, consider a finite measure on velocity space with a Maxwellian density, 
\(d\sigma_{(n,T)}(\mathbf v)=n\,p_T(\mathbf v) \,d^3v\), 
with \(\int_{\mathbb R^3}p_T(\mathbf v)\,d^3v=1\). 
The physical parameters are already adapted coordinates: the number
density \(n\) gives the total mass, while the temperature \(T\)
parametrizes the normalized shape over velocity space.

Returning to the general family
\(\sigma_{(r,\bm{\theta})}=rP_{\bm{\theta}}\), the Fisher metric
introduced above induces on the normalized family the components
\begin{equation}
g^{(P)}_{ij}(\bm{\theta})
\coloneqq
g_{P_{\bm{\theta}}}
\left(
\partial_i P_{\bm{\theta}},
\partial_j P_{\bm{\theta}}
\right)
=
\mathcal I_{ij}^{(P_{\bm{\theta}})}
=
\int_X
u_i^{(P)}(x;\bm{\theta})\,
u_j^{(P)}(x;\bm{\theta})\,
dP_{\bm{\theta}}(x),
\label{eq:probability_fisher_metric}
\end{equation}
where
\begin{equation}
u_i^{(P)}(x;\bm{\theta})
\equiv
\frac{d(\partial_i P_{\bm{\theta}})}
     {dP_{\bm{\theta}}}(x).
\label{eq:score_shape_dependent}
\end{equation}

In the adapted coordinates \((r, \boldsymbol \theta)= (r, \theta^{1}, \ldots , \theta^{m})\), the Fisher metric is an \((m+1)\times (m+1)\) matrix with coordinate components \(g_{ab}\), where \(a,b\in\{r,1,\ldots,m\}\), and takes a block-diagonal form:
\begin{itemize}
\item the radial component is \(g_{rr}=1/r\) (one dimensional), 
\item the mixed \(1 \times m\) and \(m \times 1\) blocks, \(g_{rj}\) and \(g_{ir}\), respectively, vanish owing to the mass--shape orthogonality, 
\item the remaining \(m \times m\) block \(g_{ij}= r\, g_{ij}^{(P)}(\boldsymbol \theta)\) is the Fisher metric of the normalized \(m\)-dimensional statistical model, scaled by the total mass \(r\). When the normalized model is locally identifiable, this block has full rank \(m\).
\end{itemize}
In matrix form, the metric components are
\begin{equation}
\begin{pmatrix}
g_{rr} & g_{rj} \\
g_{ir} & g_{ij}
\end{pmatrix}
=
\begin{pmatrix}
\dfrac{1}{r} & \mathbf 0^{\mathsf T} \\[2mm]
\mathbf 0 & \ r\,g^{(P)}_{ij}(\boldsymbol\theta)
\end{pmatrix}.
\label{eq:finite_measure_metric_blocks}
\end{equation}
This block-diagonal structure directly yields the associated quadratic form,
\begin{equation}
ds_{\mathrm F}^{2}
=
g_{rr}\,dr^{2}
+
2g_{ir}\,dr\,d\theta^{i}
+
g_{ij}\,d\theta^{i}d\theta^{j}.
\end{equation}
Since \(g_{ir}=0\), this reduces to
\begin{equation}
ds_{\mathrm F}^{2}
=\frac{dr^{2}}{r}
+r\,ds_{P}^{2},
\qquad
ds_{P}^{2} \equiv
g^{(P)}_{ij}(\bm{\theta})\,
d\theta^{i}d\theta^{j}.
\end{equation}

Introducing a new radial coordinate
\(
\varrho \coloneq  2\sqrt{r}
\), we have \(
\frac{dr^2}{r}=d\varrho^2
\). 
The Fisher line element becomes
\begin{equation}
ds_{\mathrm F}^{2}
=
d\varrho^{2}
+
\frac{\varrho^{2}}{4}\,ds_{P}^{2}.
\label{eq:finite_measure_cone_metric}
\end{equation}
This form is directly analogous to the Euclidean metric in polar coordinates,
\(
ds^{2}
=
d\varrho^{2}
+
\varrho^{2}d\Omega^{2}
\), 
with a correspondence 
\(
d\Omega^{2}
\longleftrightarrow
\frac{1}{4}ds_{P}^{2}
\). 
Thus, the Fisher metric of the normalized family \(\{P_{\boldsymbol\theta}\}\), 
which quantifies shape information, plays a role analogous to that of the angular metric.

\begin{figure}[t]
\centering
\includegraphics[width=6.5cm]{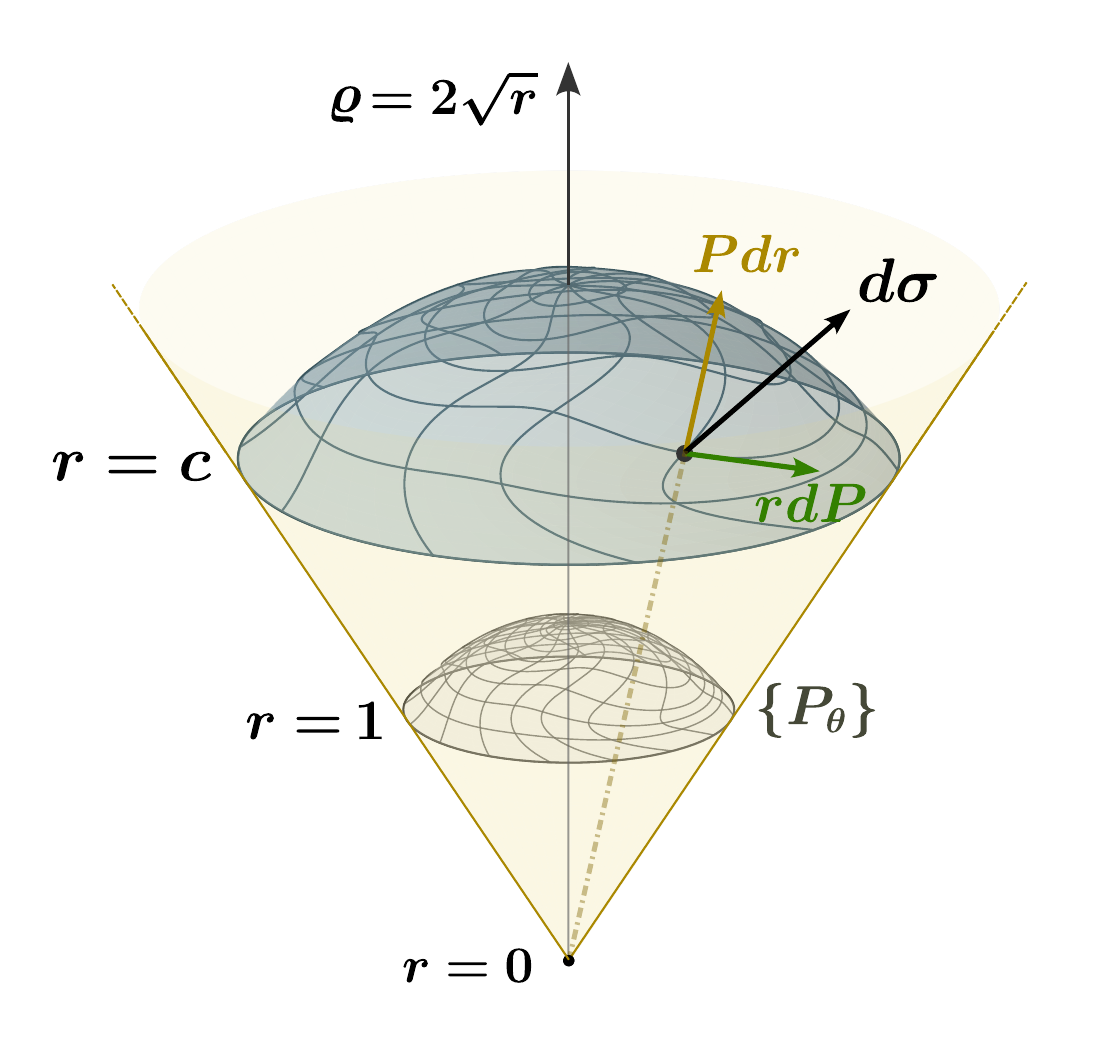}
\caption{
Schematic cone form \(ds_{\mathrm F}^{2}=d\varrho^{2}+(\varrho^{2}/4)ds_{P}^{2}\) of finite measures \(\sigma=rP_{\bm{\theta}}\) with the radial coordinate \(\varrho=2\sqrt r\). 
The apex \(r=0\) represents the zero measure, while the \(r=1\) level set contains the normalized statistical family \(\{P_{\boldsymbol\theta}\}\); the level set \(r=c\) is its radially scaled counterpart. At a point, an infinitesimal variation decomposes as \(d\sigma=P\,dr+r\,dP\) into a radial mass variation and a tangential shape variation.}
\label{fig:finite_measure_cone}
\end{figure}

Equation~\eqref{eq:finite_measure_cone_metric} gives a
geometric picture of the finite-measure family as a metric cone over
the normalized statistical model, as illustrated in
Fig.~\ref{fig:finite_measure_cone}. The cone structure reflects both
the orthogonality of the mass and shape directions and the specific
radial scaling of the shape metric. Variations of the total mass are
radial, whereas variations of the normalized outcome distribution are
tangential to constant-mass sections.

This representation makes several features of the mass--shape
decomposition geometrically transparent. Each constant-\(\rho\)
section is a scaled copy of the normalized statistical model, while
motion in the radial direction changes the total mass. 
The factor \(\varrho^{2}/ 4 =r \) shows that the distinguishability
of a given normalized-shape variation increases with the total mass. 
The zero measure appears formally as the cone apex,
where all shape directions collapse.

Such a cone structure of finite measures is closely related to the
Hellinger--Fisher--Rao geometry of nonnegative measures discussed by
Mielke, e.g., Sec.~3 of Ref.~\cite{Mielke2025Hellinger}.
A related decomposition also arises in quantum information geometry,
although there the state space is usually restricted to normalized
quantum states.  For a family of density operators
\(\rho\), an unfolding based on the spectral
decomposition represents each state in terms of its eigenvalues
\(\mathbf p\) and a diagonalizing unitary \(U\)\cite{DiNocera2022,Ciaglia2024}.
Under this construction, monotone quantum metrics separate into a
classical Fisher--Rao contribution associated with variations of the
eigenvalues and a genuinely quantum contribution associated with
variations of the eigenvectors~\cite{Petz1996}; the former is universal 
within this class of metrics, whereas the latter depends on the 
particular quantum metric.

The trace normalization of the density operator removes an independent radial
degree of freedom.  If the comparison is extended to positive,
not necessarily normalized, operators, the radial mass coordinate of the
finite-measure cone provides an additional direction.  This leads
naturally to the schematic decomposition
\[
\text{mass}
\;+\;
\text{classical spectral shape}
\;+\;
\text{genuinely quantum direction}.
\]
A systematic development of this quantum extension is beyond the scope
of the present work.

\subsection{Poisson point process and Fisher information}
\label{subsec:poisson_scattering_likelihood}
\subsubsection{Poisson point process}
The preceding construction is intrinsic to the family of finite
scattering measures and does not assume a particular statistical
observation model. We now consider its realization in an ideal
counting experiment in which individual scattering events are observed
according to a Poisson point process.

Let \(x_1,\ldots,x_N\in X\) denote the observed events, where the total
number of events \(N\) is itself random. The event list is represented
by the counting measure
\begin{equation}
\calN
=
\sum_{i=1}^{N}\delta_{x_i}, 
\label{eq:counting_measure}
\end{equation}
where \(\delta_{x}\) is the Dirac measure concentrated at \(x\). 
The sum is over the finite set of observed events, and the indexing \(i= 1,\ldots , N\) 
merely labels these events for convenience; the point process itself is unordered. 
The event locations \(x_i\) may take values in a continuous outcome space. 
 
For any measurable set \(A\in\mathcal X\),
\(\calN(A)=\sum_{i=1}^N \delta_{x_i}(A)\) is the number of observed events falling in \(A\), and \(\calN(X)=N\). 
Viewed as a random counting measure,
\(\calN\) is a point process~\cite{Kallenberg2017} on \(X\).
A histogram with bins \(B_1,\ldots,B_m\) is simply obtained  by
evaluating this counting measure on each bin, \(
N_j=\calN(B_j)\), 
\(j=1,\ldots,m \).

For a parameter-independent exposure \(\mathcal E\), define the
intensity measure
\begin{equation}
\Lambda_\lambda
\coloneqq
\mathcal E\,\sigma_\lambda.
\label{eq:ideal_count_intensity}
\end{equation}
The exposure may include incident fluence, acquisition time, sample
amount, and other parameter-independent factors that scale the expected
number of events. The counting measure is modeled as a Poisson point process,
denoted \(
\calN \sim \operatorname{PPP}(\Lambda_\lambda),
\)
with intensity measure \(\Lambda_\lambda\), so that
\begin{equation}
\mathbb E_\lambda[\calN(A)]
=
\Lambda_\lambda(A)= \mathcal E\,\sigma_\lambda (A),
\end{equation}
and counts in disjoint measurable sets are independent
\cite{DaleyVereJones2003,Kutoyants1998,Kallenberg2017}. 
The schematic chain from the physical scattering measure to the observed counting measure is
\begin{equation*}
\sigma_\lambda
\xrightarrow[\text{exposure}]{\times\,\mathcal E}
\Lambda_\lambda
\xrightarrow[\text{Poisson point process}]{\operatorname{PPP}}
\calN = \sum_{i=1}^{N}\delta_{x_i}.
\label{eq:scattering_to_point_process}
\end{equation*}
The Poisson point process with intensity measure
\(\Lambda_\lambda\) defines a probability law
\(\mathbb P_{\Lambda_\lambda}\) for the random counting measure
\(\calN\). 
Under \(\mathbb P_{\Lambda_\lambda}\), 
for any finite collection of pairwise disjoint measurable sets
\(A_1,\ldots,A_J\), the random counts
\(\calN(A_1),\ldots,\calN(A_J)\) are independent and satisfy
\begin{equation}
\calN(A_j)
\sim
\operatorname{Poisson}\!\left(\Lambda_\lambda(A_j)\right)
\qquad j=1,\ldots,J.
\end{equation}
Explicitly, for a measurable set \(A\in\mathcal X\), the probability law
\(\mathbb P_{\Lambda_\lambda}\) assigns the Poisson probability
\begin{equation}
\mathbb P_{\Lambda_\lambda}
\bigl(\calN(A)=k \bigr)
=
e^{-\Lambda_\lambda(A)}
\frac{\Lambda_\lambda(A)^k}{k!}, \qquad k=0,1,2\ldots\,.
\end{equation}

\paragraph*{Relative entropy of Poisson laws.}
For two Poisson point-process laws
\(\mathbb P_{\Lambda}\) and \(\mathbb P_{\Gamma}\) with finite
intensity measures \(\Lambda\) and \(\Gamma\), respectively,
 the ordinary KL divergence between the
process laws has exactly the same form as the generalized KL divergence
introduced above for their intensity measures: \(
D_{\mathrm{KL}}
\left(
\mathbb P_{\Lambda}
\Vert
\mathbb P_{\Gamma}
\right)
=
\mathcal D(\Lambda\Vert\Gamma)\). 
Thus, the generalized KL divergence of finite measures acquires a
direct statistical interpretation as the KL divergence between the
corresponding Poisson counting experiments.

\subsubsection{Likelihood and Fisher information}
\label{subsubsec:likelihood_S_l}
The likelihood describes the probability of the observed counting
measure \(\calN \) as a function of the parameter \(\lambda\).
It is convenient to work with the log-likelihood ratio relative to a
fixed reference parameter \(\lambda_0\):
\begin{equation}
\ell(\lambda;\calN)
-
\ell(\lambda_0;\calN)
\coloneqq
\ln 
\frac{
d\mathbb P_{\Lambda_\lambda}
}{
d\mathbb P_{\Lambda_{\lambda_0}}
}
(\calN),
\end{equation}
Under the same local absolute-continuity assumptions used in Sec.~\ref{subsec:intrinsic_fisher_scattering_measure}, the
log-likelihood ratio relative to a fixed reference intensity
\(\Lambda_{\lambda_0}\) can be written as
\begin{equation}
\ell(\lambda;\calN)-\ell(\lambda_0;\calN)
=
\int_X
\ln\!\left(
\frac{d\Lambda_\lambda}
     {d\Lambda_{\lambda_0}}
\right)
\calN(dx)
-
\Lambda_\lambda(X)
+
\Lambda_{\lambda_0}(X).
\label{eq:poisson_log_likelihood_ratio}
\end{equation}
Since the exposure is independent of \(\lambda\), multiplication by
\(\mathcal E\) does not change the intrinsic score:
\begin{equation}
\frac{d\dot\Lambda_\lambda}{d\Lambda_\lambda}
=
\frac{d\dot\sigma_\lambda}{d\sigma_\lambda}
=
u_\lambda.
\label{eq:poisson_intrinsic_score}
\end{equation}
Differentiating Eq.~\eqref{eq:poisson_log_likelihood_ratio} with
respect to \(\lambda\) gives the likelihood score
\begin{align}
S_\lambda(\calN)
\coloneqq
\frac{\partial\ell(\lambda;\calN)}{\partial\lambda}
=
\int_X
u_\lambda(x)\,\calN(dx)
-
\dot\Lambda_\lambda(X)
=
\int_X
u_\lambda(x)
\left[
\calN(dx)-\Lambda_\lambda(dx)
\right].
\label{eq:poisson_likelihood_score_intrinsic}
\end{align}
Thus, the likelihood score is obtained by integrating the intrinsic
score against the fluctuation of the observed counting measure around
its expectation.

For a Poisson point process, compensated Poisson integrals satisfy, 
for any test function \(f\in L^2(\Lambda_\lambda)\), 
\begin{equation}
\mathbb E_\lambda
\left[
\left(
\int_X f(x)
[\calN(dx)-\Lambda_\lambda(dx)]
\right)^2
\right]
=
\int_X f(x)^2\,d\Lambda_\lambda(x).
\label{eq:poisson_isometry}
\end{equation}
Applying this identity with \(f=u_\lambda\) gives the Fisher information of the Poisson counting experiment,
\begin{align}
F(\lambda)
&\coloneqq
\mathbb E_\lambda
\left[
S_\lambda(\calN)^2
\right]
\nonumber\\
&=
\int_X
u_\lambda(x)^2\,d\Lambda_\lambda(x)
=
\mathcal I^{(\Lambda)}(\lambda).
\label{eq:poisson_fisher_intrinsic}
\end{align}
The usual Fisher information obtained from the Poisson likelihood
therefore coincides exactly with the intrinsic Fisher information of
the intensity measure. Since
\(\Lambda_\lambda=\mathcal E\sigma_\lambda\),
\begin{equation}
F(\lambda)
=
\mathcal E\,\mathcal I^{(\sigma)}(\lambda).
\label{eq:poisson_fisher_exposure_scaling}
\end{equation}
Thus, Poisson observation does not introduce a different information
metric; it provides a statistical realization of the intrinsic Fisher
geometry of the finite measure.

For a vector parameter
\(\bm{\lambda}=(\lambda^1,\ldots,\lambda^m)\), the same relation holds
componentwise:
\begin{equation}
F_{ij}(\bm{\lambda})
=
\mathcal E\,
\mathcal I_{ij}^{(\sigma)}(\bm{\lambda}).
\label{eq:poisson_fisher_matrix_intrinsic}
\end{equation}
Thus, the Poisson counting experiment inherits the same local Fisher
geometry as the underlying scattering-measure family, scaled by the
exposure. 
If a common reference measure exists,
\(d\sigma_{\bm{\lambda}}=s_{\bm{\lambda}}\,d\mu\), this becomes
\begin{equation}
F_{ij}(\bm{\lambda})
=
\mathcal E
\int_X
\frac{
\partial_i s_{\bm{\lambda}}(x)\,
\partial_j s_{\bm{\lambda}}(x)
}{
s_{\bm{\lambda}}(x)
}
\,d\mu(x).
\label{eq:fisher_matrix_scattering_measure}
\end{equation}

\subsubsection{Mass--shape decomposition under Poisson statistics.}
The mass--shape decomposition derived above carries directly over to
the intensity measure. Defining
\begin{equation}
M_\lambda
\coloneqq
\Lambda_\lambda(X)
=
\mathcal E r_\lambda,
\end{equation}
which represents the expected total count,  we have
\begin{equation}
\Lambda_\lambda
=
M_\lambda P_\lambda,
\end{equation}
with the same normalized measure \(P_\lambda\) as for
\(\sigma_\lambda=r_\lambda P_\lambda\). Therefore, the mass-shape decomposition
Eq.~\eqref{eq:intrinsic_mass_shape_FI_decomposition} immediately gives the rate--shape decomposition, 
\begin{equation}
F(\lambda)
=
\frac{\dot M_\lambda^{\,2}}{M_\lambda}
+
M_\lambda\,\mathcal I^{(P)}(\lambda).
\label{eq:poisson_mass_shape_FI_decomposition}
\end{equation}
This decomposition has a direct statistical interpretation. The total
count satisfies \(N
\sim
\operatorname{Poisson}(M_\lambda)\), whereas, conditional on \(N\), the event locations are independently
distributed according to
\(
x_1,\ldots,x_N
\overset{\mathrm{i.i.d.}}{\sim}
P_\lambda.\) 
The first term in
Eq.~\eqref{eq:poisson_mass_shape_FI_decomposition} is therefore the
information carried by fluctuations of the total event count, while
the second is the information carried by the distribution of event
locations, accumulated over the expected number of events.

For a vector parameter \(\bm{\lambda}  = (\lambda^1,\ldots,\lambda^m) \), the corresponding Fisher
matrix is
\begin{equation}
F_{ij}
=
\frac{
(\partial_i M_{\bm{\lambda}})
(\partial_j M_{\bm{\lambda}})
}{
M_{\bm{\lambda}}
}
+
M_{\bm{\lambda}}\,
\mathcal I_{ij}^{(P)}, \qquad i,j=1,\ldots, m.
\label{eq:poisson_mass_shape_Fisher_matrix}
\end{equation}

\subsection{Response functions as parameterized scattering measures}
\label{subsec:response_measure_family}

\subsubsection{Score structure of the factorized response}
\label{subsubsec:factorized_response}

In many applications, the scattering density is further factorized into
known probe-dependent or kinematic factors and target-dependent quantities
such as form factors, structure factors, or response functions.
Such quantities should be distinguished from both the scattering measure
and its density: they arise from a model-dependent factorization of a
particular density representation.

The purpose here is not to derive the scattering law from the microscopic 
structure and dynamics of the target system, but rather to view such a 
factorization from the perspective of the underlying measure. 
This makes it possible to apply the mass--shape geometry developed above
directly to scattering models.

We consider a scattering density \(s_{\boldsymbol \lambda}(x)\), \(x\in X\), 
with respect to a reference measure \(\mu\) at a fixed experimental setting \(a\). 
The setting label \(a\) is suppressed below for notational simplicity. 
A commonly used factorization has the schematic form 
\begin{equation}
s_{\boldsymbol\lambda}(x)
=
\underbrace{g(x)}_{\substack{\text{known}\\ \text{prefactor}}}
\sum_{\alpha,\beta}
\underbrace{W_{\alpha\beta}(x)}_{\substack{\text{probe--target}\\ \text{weight}}}
\underbrace{R^{\alpha\beta}(x;\boldsymbol\lambda)}_{\substack{\text{response}\\ \text{function}}}.
\label{eq:parameterized_response_factorization}
\end{equation}
Here \(g(x)\) collects known kinematic or normalization factors, \(W_{\alpha\beta}(x)\) 
represents probe--target coupling and selection weights such as scattering length or 
form factor weights, \(R^{\alpha\beta}(x;\bm{\lambda})\) contains the
target-dependent response, and \(\alpha\) and \(\beta\) run over the response-component indices
relevant to the probe--target coupling. The parameter vector
\(\bm{\lambda}\) is written explicitly only in the response function, 
since the quantities to be inferred characterize the target system.
For example, in neutron or x-ray scattering, \(R\) may represent a dynamic structure factor \(S(\mathbf Q,\omega;\bm{\lambda})\).

Equation~\eqref{eq:parameterized_response_factorization} is a
factorization of a density, rather than of the scattering measure
itself.  Returning to the underlying measure, the scattering density is the
Radon--Nikodym density of \(\sigma_{\bm{\lambda}}\) with respect to
the reference measure \(\mu\):
\(
\sigma_{\bm{\lambda}}(dx)
=
s_{\bm{\lambda}}(x)\,\mu(dx)\) with
\(s_{\bm{\lambda}}
=
d \sigma_{\bm{\lambda}} /d\mu\). 
Thus, the parameter dependence of the target response induces a parameterized 
family of finite scattering measures through the response-function representation above.
By separating the known, parameter-independent probe weighting from the
parameter-dependent target response, the response-function factorization
expresses the intrinsic geometry directly in terms of the target response.

If the known prefactor \(g\) and the probe--target weights
\(W_{\alpha\beta}\) are independent of \(\boldsymbol\lambda\), write
\begin{equation}
s_{\boldsymbol\lambda}(x)
=
g(x)\,h_{\boldsymbol\lambda}(x),
\qquad
h_{\boldsymbol\lambda}(x)
\coloneqq
\sum_{\alpha,\beta}
W_{\alpha\beta}(x)
R^{\alpha\beta}(x;\boldsymbol\lambda).
\label{eq:response_effective_density}
\end{equation}
For a fixed reference measure, the intrinsic score introduced above
then takes the simple form
\begin{equation}
u_i(x)
=
\frac{\partial_i s_{\boldsymbol\lambda}(x)}
     {s_{\boldsymbol\lambda}(x)}
=
\frac{\partial_i h_{\boldsymbol\lambda}(x)}
     {h_{\boldsymbol\lambda}(x)}.
\label{eq:response_score}
\end{equation}
Thus the score  has a direct response-function interpretation:
it is the local relative parameter sensitivity of the probe-weighted
target response.  The parameter-independent prefactor \(g(x)\) does
not enter this local sensitivity, although it changes the normalized
scattering distribution over which the score is averaged.

The mass--shape decomposition derived in
Sec.~\ref{subsec:mass_shape_intrinsic} can therefore be read directly
in response-function terms.  We write
\(
\mathbb E_{P_{\boldsymbol\lambda}}[f]
\coloneqq
\int_X f(x)\,P_{\boldsymbol\lambda}(dx)
\), the mean response score determines the relative mass variation,
\begin{equation}
\frac{\partial_i r_{\boldsymbol\lambda}}
     {r_{\boldsymbol\lambda}}
=
\mathbb E_{P_{\boldsymbol\lambda}}[u_i].
\end{equation}
Since the density of \(P_{\boldsymbol\lambda}\) is
\(p_{\boldsymbol\lambda}=s_{\boldsymbol\lambda}/r_{\boldsymbol\lambda}\),
the shape score is
\begin{equation}
u_i^{(P)}(x)
=
\frac{\partial_i p_{\boldsymbol\lambda}(x)}
     {p_{\boldsymbol\lambda}(x)}
=
u_i(x)
-
\frac{\partial_i r_{\boldsymbol\lambda}}
     {r_{\boldsymbol\lambda}}
=
u_i(x)
-
\mathbb E_{P_{\boldsymbol\lambda}}[u_i].
\label{eq:response_shape_score}
\end{equation}
Its fluctuations determine the shape Fisher information,
\begin{align}
\mathcal I_{ij}^{(P)}
&=
\int_X
u_i^{(P)}(x)\,
u_j^{(P)}(x)\,
P_{\boldsymbol\lambda}(dx)
\nonumber\\
&=
\mathbb E_{P_{\boldsymbol\lambda}}
\left[
\bigl(
u_i^{(P)}(x)
\bigr)
\bigl(
u_j-\mathbb E_{P_{\boldsymbol\lambda}}[u_j]
\bigr)
\right]
=
\operatorname{Cov}_{P_{\boldsymbol\lambda}}
\left(
u_i,u_j
\right).
\label{eq:response_shape_score_covariance}
\end{align}
The response-function representation therefore identifies the physical
response underlying the intrinsic score. It shows how a known
outcome-dependent probe weighting can redistribute the same local
response sensitivity between the observed mass and shape contributions,
as illustrated by the Maxwellian example below.

\subsubsection{Maxwellian response and probe-dependent weighting}
\label{subsubsec:maxwell_response}

A simple analytically tractable example is provided by the Maxwellian
free-particle response.  Let \(\varepsilon\) denote the energy-transfer
outcome, \(T\) the temperature, and \(k_{\mathrm B}\) the Boltzmann
constant.  For a fixed recoil-energy scale \(E_R>0\), the normalized
response is
\begin{equation}
f_T(\varepsilon)
=
\frac{1}{\sqrt{4\pi E_R k_{\mathrm B}T}}
\exp\left[
-\frac{(\varepsilon-E_R)^2}
{4E_R k_{\mathrm B}T}
\right].
\label{eq:maxwell_response}
\end{equation}
Its mean and variance are
\(\langle\varepsilon\rangle=E_R\), 
\(\operatorname{Var}(\varepsilon)
=2E_R k_{\mathrm B}T\). 

For a target with number density \(n\) of scattering centers, 
the intrinsic target response is 
proportional to \(n f_T(\varepsilon)\).  Since \(f_T\) is normalized, its
integrated strength is proportional to \(n\), while temperature changes
only the normalized response shape.

The response score associated with temperature is
\begin{equation}
u_T(\varepsilon)
\equiv
\frac{
\partial_T f_T(\varepsilon)
}{
f_T(\varepsilon)
}
=
-\frac{1}{2T}
+
\frac{
(\varepsilon-E_R)^2
}{
4E_R k_{\mathrm B}T^2
}.
\label{eq:maxwell_response_score}
\end{equation}
Because \(f_T\) is normalized,
\(
\int
u_T(\varepsilon)f_T(\varepsilon)\,d\varepsilon
=
\int
\partial_T f_T(\varepsilon)\,d\varepsilon
=
\partial_T
\int
f_T(\varepsilon)\,d\varepsilon
=
0
\). 
Thus temperature is a pure shape direction at the level of the
intrinsic target response.

Now include an energy-dependent probe and kinematic weighting
\(g(\varepsilon)>0\).  The corresponding scattering measure is
\(
\sigma_{n,T}(d\varepsilon)
=
n\,g(\varepsilon)f_T(\varepsilon)\,d\varepsilon \). 
Its total mass and normalized scattering distribution are
\begin{equation}
r_{n,T}
=
n\,\overline{g}_T,
\qquad
P_T^{(\sigma)}(d\varepsilon)
=
\frac{
g(\varepsilon)f_T(\varepsilon)
}{
\overline{g}_T
}\,d\varepsilon, 
\qquad
\overline{g}_T
\equiv
\int
g(\varepsilon)f_T(\varepsilon)\,d\varepsilon. 
\label{eq:maxwell_scattering_distribution}
\end{equation}
Although the intrinsic response score has zero mean under \(f_T\), its
mean under the probe-weighted distribution need not vanish.  Indeed,
\begin{align}
\frac{
\partial_T r_{n,T}
}{
r_{n,T}
}
=
\frac{
\partial_T\overline{g}_T
}{
\overline{g}_T
}
=
\frac{1}{\overline{g}_T}
\int
g(\varepsilon)
\partial_T f_T(\varepsilon)\,
d\varepsilon
=
\int
u_T(\varepsilon)\,
P_T^{(\sigma)}(d\varepsilon).
\label{eq:maxwell_score_mean_mass}
\end{align}
Thus an energy-dependent probe weighting can give a nonzero mean to a
response score that was purely shape sensitive at the response 
level.  This mean is precisely the relative change of the total
scattering mass.

The score of the normalized scattering distribution is correspondingly
the centered response score,
\begin{align}
\frac{
\partial_T
\left[
g(\varepsilon)f_T(\varepsilon)/\overline{g}_T
\right]
}{
g(\varepsilon)f_T(\varepsilon)/\overline{g}_T
}
&=
u_T(\varepsilon)
-
\frac{
\partial_T\overline{g}_T
}{
\overline{g}_T
}
=
u_T(\varepsilon)
-
\int
u_T(\varepsilon')\,
P_T^{(\sigma)}(d\varepsilon').
\label{eq:maxwell_centered_score}
\end{align}
Its Fisher information is therefore
\begin{align}
\mathcal I_{TT}^{(P)}(T)
=
\operatorname{Var}_{P_T^{(\sigma)}}(u_T)
&=
\int
\left[
u_T(\varepsilon)
-
\int
u_T(\varepsilon')\,
P_T^{(\sigma)}(d\varepsilon')
\right]^2
P_T^{(\sigma)}(d\varepsilon)
\nonumber \\
&=\int
u_T(\varepsilon)^2\,
P_T^{(\sigma)}(d\varepsilon)
-
\left[
\int
u_T(\varepsilon)\,
P_T^{(\sigma)}(d\varepsilon)
\right]^2 
.
\label{eq:maxwell_shape_score_variance}
\end{align}
The Fisher information of the full scattering measure is
\begin{align}
\mathcal I_{TT}^{(\sigma)}(n,T)
&=
n
\int
g(\varepsilon)
\frac{
[\partial_T f_T(\varepsilon)]^2
}{
f_T(\varepsilon)
}\,
d\varepsilon
=
r_{n,T}
\int
u_T(\varepsilon)^2\,
P_T^{(\sigma)}(d\varepsilon).
\label{eq:maxwell_response_fisher}
\end{align}

On the other hand, applying the mass--shape decomposition in Eq.~\eqref{eq:intrinsic_mass_shape_Fisher_matrix} directly to
\(\sigma_{n,T}\) gives
\begin{equation}
\mathcal I_{TT}^{(\sigma)}(n,T)
=
\frac{
[\partial_T r_{n,T}]^2
}{
r_{n,T}
}
+
r_{n,T}
\mathcal I_{TT}^{(P)}(T).
\label{eq:maxwell_mass_shape_fisher}
\end{equation}
Comparing Eqs.~\eqref{eq:maxwell_response_fisher} and
\eqref{eq:maxwell_mass_shape_fisher}, together with Eqs.~\eqref{eq:maxwell_score_mean_mass} 
and \eqref{eq:maxwell_shape_score_variance} 
shows explicitly that the two terms in the mass--shape decomposition are
the squared mean and the variance of the same temperature-response score:
\begin{equation}
\mathcal I_{TT}^{(\sigma)}(n,T)
=
r_{n,T}
\left[
\int
u_T(\varepsilon)\,
P_T^{(\sigma)}(d\varepsilon)
\right]^2
+
r_{n,T}
\operatorname{Var}_{P_T^{(\sigma)}}(u_T).
\label{eq:maxwell_mean_variance_fisher}
\end{equation}
Thus the mean response score describes the mass sensitivity generated
after probe-dependent weighting, whereas its fluctuations about that
mean carry the normalized-shape sensitivity.

As a simple limiting case, if \(g(\varepsilon)\) is constant, then
\(P_T^{(\sigma)}=f_T(\varepsilon)\,d\varepsilon\) and
\(\partial_T r_{n,T}=0\).  The temperature dependence therefore remains
a pure shape variation.  For the Maxwellian response,
\(
\mathcal I_{TT}^{(P)}(T)
=1 / (2T^2)\) and 
\(\mathcal I_{TT}^{(\sigma)}(n,T)
=r_{n,T} / (2T^2)\).

The distinction becomes particularly transparent if the number density
is also unknown.  Introducing the logarithmic density coordinate
\(\xi \equiv \ln n\),  
the corresponding Fisher elements are
\( \mathcal I_{\xi\xi}^{(\sigma)}
=r_{n,T}\) and \(
\mathcal I_{\xi T}^{(\sigma)} = \partial_T r_{n,T}\).  
If the number density is not known independently and the temperature
\(T\) is the parameter of interest, \(\xi=\ln n\) acts as a nuisance
parameter controlling the overall scattering strength.  Eliminating
this nuisance direction gives
\begin{equation}
\mathcal I_{TT}^{\mathrm{eff}}
=
\mathcal I_{TT}^{(\sigma)}
-
\frac{
\left(\mathcal I_{\xi T}^{(\sigma)}\right)^2
}{
\mathcal I_{\xi\xi}^{(\sigma)}
}
=
r_{n,T}\mathcal I_{TT}^{(P)}(T). 
\end{equation}
The mass-sensitive part of the temperature response is locally
indistinguishable from a change in number density and is therefore
removed by nuisance-parameter elimination, leaving only the
shape-sensitive component independently identifiable.  More generally,
probe weighting \(g\) may create additional rate sensitivity to a physical
parameter, but this information is useful only when the overall
scattering scale is independently constrained; otherwise it is
confounded with an amplitude nuisance parameter.

This example makes the complementary roles of the response-function and
mass--shape representations explicit.  Temperature changes only the
shape of the intrinsic Maxwellian response, whereas an outcome-dependent
probe weighting can convert part of this shape sensitivity into a change
of total scattering mass.  Whether that additional mass sensitivity
provides usable information therefore depends on how well the overall
scattering scale is known.


\section{Information contraction under measurement kernels}
\label{sec:information_contraction}

A measurement kernel maps the parameterized family
\(\{\sigma_\lambda\}\) to another family
\(\{K_{\#}\sigma_\lambda\}\). The central information-theoretic question
is then not only how a measure is transformed, but which
parameter-dependent distinctions survive that transformation. For
parameter-independent kernels, Fisher information cannot increase.

The following analysis returns to the concrete transformations of
Sec.~\ref{sec:kernel} and quantifies how these operations reshape the
statistical model and its information geometry.

\subsection{Score projection and information loss}
\label{subsec:data_processing_FI}

Let \(\{\sigma_\lambda\}\) be a differentiable family of finite
measures on \((X,\mathcal X)\), and let
\(K(dy\mid x)\) be a parameter-independent probability kernel from
\((X,\mathcal X)\) to \((Y,\mathcal Y)\).
The transformed family is \(
\nu_\lambda = K_\#\sigma_\lambda\). 
Since \(K(Y\mid x)=1\), the total mass is preserved:
\(
\nu_\lambda(Y)=\sigma_\lambda(X)=r_\lambda
\). 
Let
\begin{equation}
u_\lambda(x)
=
\frac{d\dot{\sigma}_\lambda}{d\sigma_\lambda}(x),
\qquad
v_\lambda(y)
=
\frac{d\dot{\nu}_\lambda}{d\nu_\lambda}(y)
\end{equation}
denote the input and output scores, respectively. 
Introducing the normalized joint distribution
\begin{equation}
P_\lambda(dx,dy)
=
\frac{\sigma_\lambda(dx)}{r_\lambda}
K(dy\mid x),
\label{eq:joint_distribution_dxdy}
\end{equation}
the output score is the conditional expectation of the input score,
\begin{equation}
v_\lambda(Y)
=
\mathbb E_\lambda
\left[
u_\lambda(X)\mid Y
\right].
\label{eq:conditional_score}
\end{equation}
Throughout this section, \(\mathbb E_\lambda\) denotes expectation with
respect to the joint probability distribution \(P_\lambda(dx,dy)\)
defined in Eq.~\eqref{eq:joint_distribution_dxdy},
induced by the kernel. 
The difference
\begin{equation}
u_\lambda(X)-v_\lambda(Y)
=
u_\lambda(X)
-
\mathbb E_\lambda
\left[
u_\lambda(X)\mid Y
\right]
\end{equation}
is the part of the input score that cannot be predicted from the
retained outcome \(Y\). We refer to it as the residual score.
By construction, its conditional mean vanishes: \(
\mathbb E_\lambda
\left[
u_\lambda(X)-v_\lambda(Y)\mid Y
\right]
=
0
\).

\begin{figure}[tb]
    \centering
    \includegraphics[width=5cm]{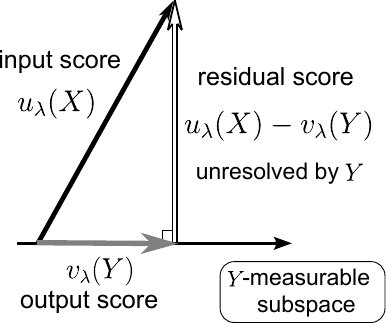}
    \caption{
    Geometric interpretation of the score contraction induced by a
    measurement kernel.
    The input score \(u_\lambda(X)\) is decomposed into the projected
    component
    \(v_\lambda(Y)=\mathbb E_\lambda[u_\lambda(X)\mid Y]\),
    which is determined by the recorded outcome \(Y\), and the
    orthogonal residual
    \(u_\lambda(X)-v_\lambda(Y)\).
    The horizontal direction schematically represents the subspace of
    \(Y\)-measurable functions in \(L^2(P_\lambda)\).
    The corresponding Pythagorean relation 
    \( \|u\|^2 = \|v\|^2 + \|u-v\|^2\) gives the contraction of
    Fisher information under the measurement.
    }
    \label{fig:score_projection}
\end{figure}

The measurement kernel therefore acts on the input score by conditional
averaging: the part predictable from the recorded outcome \(Y\) is
retained, whereas the remaining differences in the input score are
unresolved. Accordingly, the input score can be decomposed as
\begin{equation}
\begin{alignedat}{3}
u_\lambda(X)
&={}&
v_\lambda(Y)\ \quad 
&\;+\;&
\left[
u_\lambda(X)-v_\lambda(Y) 
\right]\qquad 
\\
&={}&
\underbrace{
\mathbb E_\lambda
\left[
u_\lambda(X)\mid Y
\right]
}_{\text{predictable from }Y}
&\;+\;&
\underbrace{
\left[
u_\lambda(X)
-
\mathbb E_\lambda
\left[
u_\lambda(X)\mid Y
\right]
\right]
}_{\text{unresolved by }Y\text{ (residual score)}}
\end{alignedat}
\label{eq:score_decomposition}
\end{equation}
By construction, the residual score has zero conditional mean given \(Y\),
\(
\mathbb E_\lambda
\left[
u_\lambda(X)-v_\lambda(Y)\mid Y
\right]
=0\), 
and is therefore orthogonal to every \(Y\)-measurable function,
including \(v_\lambda(Y)\).  
A geometric interpretation of this situation is illustrated in Fig.~\ref{fig:score_projection}.

Squaring Eq.~\eqref{eq:score_decomposition} and taking the expectation,
the cross term vanishes by orthogonality, giving the corresponding
Pythagorean relation, 
\begin{equation}
\mathbb E_\lambda
\left[
u_\lambda(X)^2
\right]
=
\mathbb E_\lambda
\left[
v_\lambda(Y)^2
\right]
+
\mathbb E_\lambda
\left[
\left\{
u_\lambda(X)-v_\lambda(Y)
\right\}^2
\right].
\end{equation}
The conditional variance
\(\operatorname{Var}_\lambda(u_\lambda(X)\mid Y)\)
quantifies the part of the input-score fluctuation that remains
undetermined by the recorded outcome \(Y\). 
Averaging this quantity over \(Y\) gives 
\(
\mathbb E_\lambda
\left[
\left\{
u_\lambda(X)-v_\lambda(Y)
\right\}^2
\right]
=
\mathbb E_\lambda
\left[
\operatorname{Var}_\lambda
\left(
u_\lambda(X)\mid Y
\right)
\right]
\), 
so that
\begin{equation}
\mathbb E_\lambda
\left[
u_\lambda(X)^2
\right]
=
\mathbb E_\lambda
\left[
v_\lambda(Y)^2
\right]
+
\mathbb E_\lambda
\left[
\operatorname{Var}_\lambda
\left(
u_\lambda(X)\mid Y
\right)
\right].
\label{eq:conditional_variance_identity}
\end{equation}
Since the \(X\)- and \(Y\)-marginals of \(P_\lambda\) are
\(\sigma_\lambda/r_\lambda\) and \(\nu_\lambda/r_\lambda\),
respectively, the two squared-score expectations are related to the
intrinsic Fisher information by
\begin{equation}
r_\lambda
\mathbb E_\lambda
\left[
u_\lambda(X)^2
\right]
=
\mathcal I^{(\sigma)}(\lambda),
\qquad
r_\lambda
\mathbb E_\lambda
\left[
v_\lambda(Y)^2
\right]
=
\mathcal I^{(\nu)}(\lambda).
\label{eq:Fisher_expectation_relation}
\end{equation}
Multiplying Eq.~\eqref{eq:conditional_variance_identity} by the preserved mass \(r_\lambda\) yields
\begin{equation}
\mathcal I^{(\sigma)}(\lambda)
-
\mathcal I^{(\nu)}(\lambda)
=
r_\lambda
\mathbb E_\lambda
\left[
\operatorname{Var}_\lambda
\left(
u_\lambda(X)\mid Y
\right)
\right]
\geq 0.
\label{eq:FI_loss_conditional_variance}
\end{equation}
Hence a parameter-independent probability kernel cannot increase the
intrinsic Fisher information,
\begin{equation}
\mathcal I^{(\nu)}(\lambda)
\leq
\mathcal I^{(\sigma)}(\lambda).
\label{eq:intrinsicFisherInfo_inequality}
\end{equation}

Equation~\eqref{eq:FI_loss_conditional_variance} gives more than the
inequality itself: the information loss is exactly the unresolved
conditional variance of the input score.
Equality holds if and only if
\begin{equation}
\operatorname{Var}_\lambda
\left(
u_\lambda(X)\mid Y
\right)
=0
\end{equation}
almost surely, that is, when the input score is determined by the
retained observation.
For a deterministic transformation \(T:X \to Y\), this requires
\(u_\lambda(x)\) to be constant almost everywhere on each fiber
\(T^{-1}(y)\).

Because a probability kernel preserves the total mass, the mass term
in the mass--shape decomposition is unchanged; the contraction occurs
entirely in the shape information.
Kernels that also remove events can modify both contributions and are
considered below. 
The assumption that \(K\) is parameter independent is essential.
If \(K=K_\lambda\), variations of the recorded measure contain
contributions from both the scattering measure and the measurement
kernel, and the contraction relation above does not apply in this
form.

\medskip
\paragraph*{Multipleparameter case.}
For a vector parameter
\(\boldsymbol{\lambda}=(\lambda^1,\ldots,\lambda^m)\),
the scalar score is replaced by the score vector
\begin{equation}
\mathbf u_{\boldsymbol{\lambda}}(x)
\equiv
\frac{
d\!\left(\nabla_{\boldsymbol{\lambda}}
\sigma_{\boldsymbol{\lambda}}\right)
}{
d\sigma_{\boldsymbol{\lambda}}
}(x),
\qquad
\nabla_{\boldsymbol{\lambda}}
\equiv
(\partial_1,\ldots,\partial_m)^{\mathsf T},
\label{eq:vector_intrinsic_score}
\end{equation}
where the Radon--Nikodym derivative is understood componentwise.
Likewise, for the output measure
\(\nu_{\boldsymbol{\lambda}}=K_{\#}\sigma_{\boldsymbol{\lambda}}\),
we define
\begin{equation}
\mathbf v_{\boldsymbol{\lambda}}(y)
\equiv
\frac{
d\!\left(\nabla_{\boldsymbol{\lambda}}
\nu_{\boldsymbol{\lambda}}\right)
}{
d\nu_{\boldsymbol{\lambda}}
}(y).
\end{equation}
As in the scalar case, the output score is the conditional mean of the
input score,
\begin{equation}
\mathbf v_{\boldsymbol{\lambda}}(Y)
=
\mathbb E_{\boldsymbol{\lambda}}
\left[
\mathbf u_{\boldsymbol{\lambda}}(X)
\mid Y
\right].
\end{equation}
Therefore, the conditional covariance decomposition gives
\begin{align}
\mathbb E_{\boldsymbol{\lambda}}
\left[
\mathbf u_{\boldsymbol{\lambda}}
\mathbf u_{\boldsymbol{\lambda}}^{\mathsf T}
\right]
&=
\mathbb E_{\boldsymbol{\lambda}}
\left[
\mathbf v_{\boldsymbol{\lambda}}
\mathbf v_{\boldsymbol{\lambda}}^{\mathsf T}
\right]
+
\mathbb E_{\boldsymbol{\lambda}}
\left[
\operatorname{Cov}_{\boldsymbol{\lambda}}
\left(
\mathbf u_{\boldsymbol{\lambda}}(X)
\mid Y
\right)
\right].
\end{align}
Multiplying by the preserved mass
\(r_{\boldsymbol{\lambda}}\) yields
\begin{equation}
\mathbf I^{(\sigma)}(\boldsymbol{\lambda})
-
\mathbf I^{(\nu)}(\boldsymbol{\lambda})
=
r_{\boldsymbol{\lambda}}
\mathbb E_{\boldsymbol{\lambda}}
\left[
\operatorname{Cov}_{\boldsymbol{\lambda}}
\left(
\mathbf u_{\boldsymbol{\lambda}}(X)
\mid Y
\right)
\right]
\succeq 0.
\label{eq:FI_loss_conditional_covariance}
\end{equation}
Here \(\succeq 0\) denotes positive semidefiniteness; equivalently,
the Fisher information cannot increase along any direction in parameter
space.

When a density representation
\(
\sigma_{\boldsymbol{\lambda}}(dx)
=
s_{\boldsymbol{\lambda}}(x)\,\mu(dx)
\)
is available, the score vector reduces to
\(
\mathbf u_{\boldsymbol{\lambda}}(x)
=
\nabla_{\boldsymbol{\lambda}}
\ln s_{\boldsymbol{\lambda}}(x)\). 
Writing
\(
s_{\boldsymbol{\lambda}}(x)
=
r_{\boldsymbol{\lambda}}
p_{\boldsymbol{\lambda}}(x)
\),
one obtains
\(
\nabla_{\boldsymbol{\lambda}}
\ln s_{\boldsymbol{\lambda}}(x)
=
\nabla_{\boldsymbol{\lambda}}
\ln r_{\boldsymbol{\lambda}}
+
\nabla_{\boldsymbol{\lambda}}
\ln p_{\boldsymbol{\lambda}}(x),
\) 
which is the multiparameter form of the mass--shape score decomposition.

\subsection{Physical consequences of information contraction}
\label{subsec:contraction_examples}

The contraction result has several direct consequences for common
experimental operations.  We summarize three simple examples here;
explicit calculations are given in
Appendix~\ref{app:contraction_examples}.

\subsubsection{Discrete channel mixing} 
\label{subsubsec:discrete_channel_mixing}
Consider the two-channel distribution 
\( \mathbf p_\lambda=(\lambda,1-\lambda)^{\mathsf T} \) subjected to 
the symmetric mixing kernel \(K_\varepsilon\) of Eq.~\eqref{eq:two_channel_kernel}. 
The recorded distribution is \(\mathbf q_\lambda = K_\varepsilon \mathbf p_\lambda 
= (q_\lambda,  1-q_\lambda )^{\mathsf T}\) with \(q_\lambda = \varepsilon+(1-2\varepsilon)\lambda\).  
The relative loss of per-event Fisher information is (see Sec.~\ref{app:discrete_channel_mixing})
\(
(\mathcal I^{(p)}(\lambda) - \mathcal I^{(q)}(\lambda))/ \mathcal I^{(p)}(\lambda)
= \varepsilon(1-\varepsilon) / q_\lambda(1-q_\lambda).
\) 
As \(\lambda\to0\) or \(1\), one has
\(q_\lambda\to\varepsilon\) or \(1-\varepsilon\), respectively, and
hence the recorded information vanishes. 
Thus even weak channel mixing can remove a large fraction of the
Fisher information near the boundary values of \(\lambda\).
Although \(K_\varepsilon\) is algebraically invertible for
\(\varepsilon\neq1/2\), this does not restore the lost statistical
information.  Reconstructing the underlying channel probabilities
requires applying \(K_\varepsilon^{-1}\), which amplifies finite-count
statistical fluctuations.  The inversion becomes increasingly
ill-conditioned as \(\varepsilon\to1/2\), since
\(\det K_\varepsilon=1-2\varepsilon\to0\).

\subsubsection{Resolution effect}
\label{subsubsec:gaussian_blur}   
A continuous resolution kernel of Eq.~\eqref{eq:Gaussian_detector_resolution_kernel} 
also gives a simple result. 
For a Gaussian location family of intrinsic width \(w\), the per-event Fisher 
information for the location parameter is 
\(
\mathcal I^{(p)}(\lambda) = 1 / w^2\). 
Convolution with a Gaussian instrumental resolution of width \(\gamma\) broadens 
the recorded distribution to width \(\sqrt{w^2+\gamma^2}\), giving 
\(\mathcal I^{(q)}(\lambda;\gamma) = 1 /(w^2+\gamma^2)\) (see Sec.~\ref{app:gaussian_blur}).  Hence, 
\(\mathcal I^{(q)}/\mathcal I^{(p)} = 1 / [1+(\gamma/w)^2].\) 
The information retained per event is therefore controlled directly
by the dimensionless ratio of instrumental to intrinsic width.

\subsubsection{Coarse graining.}  
Coarse graining can act selectively on different parameter directions. 
If the continuously recorded Gaussian outcome is reduced to a single
bit by a threshold centered at the nominal peak position, the 
 binary coarse-grained  local
Fisher information for the location parameter satisfies 
\( \mathcal I^{(b)} / \mathcal I^{(q)} = 2 /\pi ,\) 
whereas, under the same centered-threshold condition, 
the binary local Fisher information for the Gaussian width vanishes 
(see Sec.~\ref{app:binary_coarse_graining}). 
Thus coarse graining does not act uniformly on the statistical model:
it may retain substantial information along one parameter direction
while completely eliminating another.
In geometric terms, different tangent directions of the statistical
model can be contracted by different amounts.

\medskip

\subsubsection{Mass-preserving and lossy kernels} 
The contraction result also distinguishes two physically different
operations: loss of distinctions among retained events and loss of the
events themselves.
A parameter-independent mass-preserving kernel leaves the total
scattering mass \(r_\lambda\) unchanged.  Consequently, the mass
contribution to the Fisher information is preserved, while the shape
contribution may contract.

By contrast, uniform event loss with survival probability
\(0\leq\eta\leq1\), independent of outcome and of \(\lambda\), rescales
the scattering measure as \(
\widetilde{\sigma}_\lambda
=
\eta\,\sigma_\lambda,\) 
and therefore
\(
\mathcal I^{(\widetilde{\sigma})}(\lambda)
=
\eta\,
\mathcal I^{(\sigma)}(\lambda).\) 
Outcome-dependent acceptance may alter both the total mass and the
normalized shape.  If the kernel itself depends on \(\lambda\), its
parameter dependence must instead be included explicitly in the
statistical model.

\medskip

These examples show how the general contraction principle appears in
discrete mixing, finite resolution, coarse graining, and event loss.
The next section turns from information retained under a given
measurement to the design of measurements for a specified inference
task.


\section{Measurement design as optimization of observed measures}
\label{sec:design}

Once a parameterized scattering measure has been specified, each admissible
measurement design induces a corresponding family of observed finite measures.
Measurement design then amounts to choosing among these measures so as to
maximize the local statistical distinguishability relevant to the inference
task, subject to experimental constraints.

\subsection{Experimental configurations and observed intensity measures}
\label{subsec:design_kernels}

Let \(\mathsf A\) denote the set of measurement designs.
A design \(a\in\mathsf A\) may affect both the primary scattering measure
through the incident probe state and the subsequent measurement channel.
We therefore write the primary measure as
\(\sigma_{\boldsymbol\lambda,a}\), the measurement kernel as
\(K_a\), and the exposure factor as \(\mathcal E_a\).
The resulting observed intensity measure is
\begin{equation}
\Lambda_{\boldsymbol\lambda,a}
=
\mathcal E_a
(K_a)_{\#} \sigma_{\boldsymbol\lambda,a}.
\label{eq:design_count_measure}
\end{equation}
The design variable \(a\) may encode, for example, the incident spectrum,
instrumental resolution, detector response, or acceptance.

\paragraph*{Absolute-intensity information.}
To interpret the total observed count rate as information about the
absolute scale of the scattering measure, the overall experimental
normalization must be known or independently constrained.  This includes
the exposure factor \(\mathcal E_a\) and, when \(K_a\) is
lossy, the corresponding acceptance or detection efficiency.  In a beam
experiment, the exposure may be determined from the monitored incident
fluence. More generally, in
experiments such as cosmic-ray observations, an effective
exposure may instead be established from geometrical acceptance and 
calibrated detection efficiency.  
Absolute intensity is statistically informative even for a parameter that
changes only the total mass of the scattering measure: in a counting
experiment, such a change modifies the total-count statistics.  
More generally, a physical parameter may change
both the total mass and the normalized shape of the scattering measure.
When the absolute normalization is known, both responses contribute to
its statistical distinguishability. 
If the overall normalization is unknown, however, care is required in
interpreting the mass-sensitive component, since it may be confounded
with the unknown scale.

\subsection{Fisher-information-based design criteria}
\label{subsec:design_criteria}

For each design \(a\), let
\(\mathbf F^{(a)}(\boldsymbol{\lambda})\) denote the Fisher information
matrix of the Poisson counting experiment with intensity measure
\(\Lambda_{\boldsymbol\lambda,a}\).  
A local information-based design can then be written as
\begin{equation}
a^*
=
\operatorname*{arg\,max}_{a\in\mathsf A}
\Phi \big[
\mathbf F^{(a)}(\boldsymbol{\lambda})
\big],
\label{eq:design}
\end{equation}
where the scalar objective \(\Phi\) specifies the inference task motivated by the scientific objective.

For a single parameter \(\lambda\), the most straightforward choice is \(
\Phi(\mathbf F)
=
F_{\lambda\lambda}\). 
In most practical experiments, the parameter of interest cannot be considered
in isolation, and several parameters must be treated simultaneously.
Possible design criteria then include maximizing 
\(\det\mathbf F\) (D-optimality), minimizing a weighted trace of
\(\mathbf F^{-1}\) (A-type criteria), maximizing the smallest information
direction after a physically chosen parameter scaling (E-type criteria).

\subsection{Nuisance parameters and efficient Fisher information}

An important instance of the design objective arises when one parameter
\(\theta\) is regarded as the parameter of interest and the remaining
parameters
\(\boldsymbol{\psi}=(\psi^1,\ldots,\psi^m)\) as nuisance parameters.
An infinitesimal parameter variation defines a tangent direction in the
statistical model, represented here by the corresponding component of the
likelihood score introduced in Sec.~\ref{subsubsec:likelihood_S_l}.
The Fisher matrix gives the inner products between these score
directions. Its off-diagonal elements therefore quantify the overlap
between parameter-induced tangent directions, and hence how well a
variation of \(\theta\) can be distinguished from variations of
\(\boldsymbol{\psi}\).

Writing the Fisher matrix in block form as
\begin{equation}
\mathbf F
=
\begin{pmatrix}
F_{\theta\theta}
&
F_{\theta\boldsymbol{\psi}}
\\
F_{\boldsymbol{\psi}\theta}
&
F_{\boldsymbol{\psi}\boldsymbol{\psi}}
\end{pmatrix},
\end{equation}
the information on \(\theta\) available in the presence of unknown nuisance
parameters is described by the \emph{efficient Fisher information}
~\cite{KaganRao2003, FewsterJupp2013}.
Provided that \(F_{\boldsymbol{\psi}\boldsymbol{\psi}}\) is nonsingular,
it is given by the Schur complement of the nuisance-parameter block,
\begin{equation}
F_{\theta\theta}^{\mathrm{eff}}
=
F_{\theta\theta}
-
F_{\theta\boldsymbol{\psi}}
F_{\boldsymbol{\psi}\boldsymbol{\psi}}^{-1}
F_{\boldsymbol{\psi}\theta}.
\label{eq:fisher_schur_complement}
\end{equation}
For a design intended specifically to estimate \(\theta\) in the presence
of these nuisance parameters, one may take
\(
\Phi(\mathbf F)
=
F_{\theta\theta}^{\mathrm{eff}}
\).

Equivalently, using the multiparameter extension of the likelihood score
introduced in Eq.~\eqref{eq:poisson_likelihood_score_intrinsic}, let
\(S_\theta\) denote the component associated with the parameter of interest
and let \(\mathbf S_{\boldsymbol{\psi}}\) collect the components associated
with the nuisance parameters.
Here,
\(
F_{\theta\boldsymbol{\psi}}\in\mathbb R^{1\times m}\),
\(F_{\boldsymbol{\psi}\boldsymbol{\psi}}\in\mathbb R^{m\times m}\), and
\(\mathbf S_{\boldsymbol{\psi}}\in\mathbb R^{m\times 1}\),
so that the projected nuisance-score contribution is scalar. The
efficient score is therefore
\begin{equation}
S_\theta^{\mathrm{eff}}
=
S_\theta
-
F_{\theta\boldsymbol{\psi}}
F_{\boldsymbol{\psi}\boldsymbol{\psi}}^{-1}
\mathbf S_{\boldsymbol{\psi}}.
\label{eq:efficient_score}
\end{equation}
Its variance gives the efficient Fisher information,
\begin{equation}
F_{\theta\theta}^{\mathrm{eff}}
=
\mathbb E
\left[
\left(
S_\theta^{\mathrm{eff}}
\right)^2
\right].
\end{equation}
Geometrically, the efficient Fisher information is therefore the squared
length of the component of the parameter-of-interest tangent direction
that is orthogonal to the nuisance tangent subspace.

\subsection{Triple-axis spectroscopy: design within a kinematic fiber}
\label{subsec:tas_design}

\subsubsection{Kinematic map and instrumental setting map}
Triple-axis neutron spectroscopy~\cite{Brockhouse1961} provides a useful example in which
the physical scattering outcomes, their kinematic representation, and
the instrumental configuration must be distinguished carefully.
Because the free-neutron dispersion relation couples energy and
momentum, the instrumental resolution in \((\bm{Q},\omega)\) space is
intrinsically multidimensional and can depend strongly on the chosen
kinematic configuration.

For a fixed incident neutron state with momentum \(\bm{k}_i\), let the
outgoing physical outcome be 
\(x=\bm{k}_f\in X=\mathbb R^3_{\bm k_f}\).
The momentum and energy transfers are obtained through the deterministic
kinematic map 
\begin{equation}
T_{\rm kin}: X\longrightarrow\mathcal Q,
\qquad
\bm{k}_f
\longmapsto
\left(
\bm{k}_i-\bm{k}_f,\,
\frac{E(\bm{k}_i)-E(\bm{k}_f)}{\hbar}
\right)
=
(\bm Q,\omega),
\label{eq:tas_kinematic_map}
\end{equation}
where \( E(\bm{k}) = \hbar^2|\bm{k}|^2 /(2m)\) and \(\mathcal Q\) denotes the momentum--energy-transfer space.

Let
\(\sigma_{\boldsymbol\lambda}\)
denote the physical scattering measure on \(X\), parameterized by the
sample parameters \(\boldsymbol\lambda\).
Its representation in momentum--energy-transfer space is the
deterministic pushforward
\(\sigma_{\boldsymbol\lambda}^{Q\omega}
=
(T_{\rm kin})_{\#}
\sigma_{\boldsymbol\lambda}.
\)
Thus \(\sigma_{\boldsymbol\lambda}^{Q\omega}\)
describes the sample-side distribution of momentum and
energy transfers before instrumental resolution is applied.

The instrument configuration introduces a second map into the same
momentum--energy-transfer space.
Let \(a\in\mathsf A\) denote an instrumental setting and define
\begin{equation}
\pi:\mathsf A \longrightarrow\mathcal Q,
\qquad
a\longmapsto
\pi(a)
=
(\bm{Q}_a,\omega_a),
\label{eq:tas_setting_map}
\end{equation}
where \((\bm{Q}_a,\omega_a)\) is the nominal momentum--energy-transfer
point associated with that setting.
The configuration \(a\) may include the incident and final energies,
sample orientation and scattering geometry, collimation and focusing
conditions, and other instrumental parameters that determine the nominal kinematics and instrumental response.

\begin{figure}[htbp]
\centering
\includegraphics[width=11.5cm]{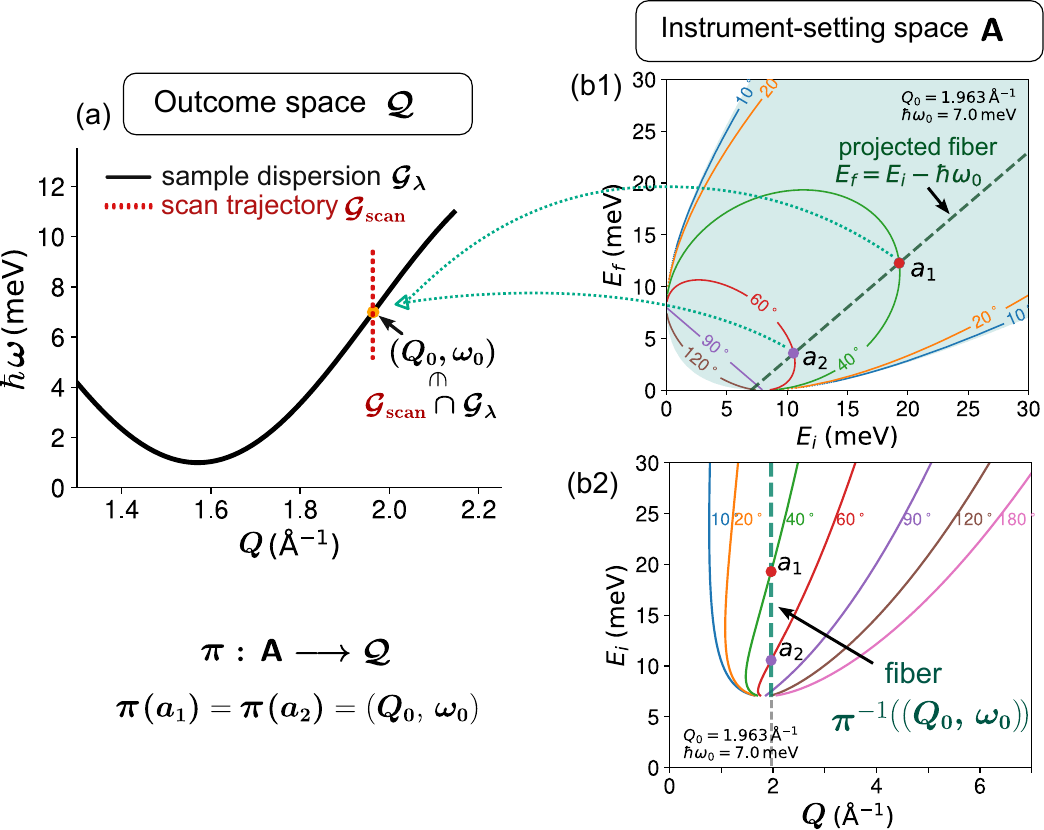}
\caption{
Kinematic map from instrument settings to momentum--energy-transfer
space and the corresponding fiber.
(a) A constant-\(Q\) scan trajectory \(\mathcal G_{\rm scan}\)
intersects the sample dispersion
\(\mathcal G_{\boldsymbol\lambda}\) at the nominal outcome
\((Q_0,\omega_0)\).
(b1) At fixed \(Q_0\), the same outcome is represented in the
\((E_i,E_f)\) setting-space slice by the accessible part of
\(E_f=E_i-\hbar\omega_0\); contours indicate constant scattering angle
\(2\theta_{\mathrm s}\).
(b2) The same fiber
\(\pi^{-1}(\{(Q_0,\omega_0)\})\) is shown in the
\((Q,E_i)\) representation at fixed \(\omega_0\).
Its intersections with the constant-\(2\theta_{\mathrm s}\) curves
identify distinct settings \(a_1\) and \(a_2\) satisfying
\(\pi(a_1)=\pi(a_2)=(Q_0,\omega_0)\).
}
\label{fig:TAS_graph_fiber}
\end{figure}

The two maps therefore play different roles:
\(T_{\rm kin}\) maps individual physical scattering outcomes to their
momentum and energy transfers, whereas \(\pi\) maps an instrumental
configuration to the nominal kinematic point at which the experiment is
performed. 
For a fixed nominal point
\((\bm{Q}_0,\omega_0)\), the set
\begin{equation}
\pi^{-1}
\!\left(
\{(\bm{Q}_0,\omega_0)\}
\right)
=
\left\{
a\in\mathsf A:
\pi(a)
=
(\bm{Q}_0,\omega_0)
\right\}
\label{eq:tas_setting_fiber}
\end{equation}
is the corresponding fiber in configuration space.
Different combinations of incident energy, final energy, and scattering
geometry may therefore realize the same nominal
\((\bm{Q}_0,\omega_0)\). 
Figure~\ref{fig:TAS_graph_fiber} illustrates the relation between the
nominal point \((\bm{Q}_0,\omega_0)\) and its preimage under the setting
map \(\pi\), which contains multiple distinct instrumental settings.
The nominal coordinates therefore specify the kinematic point of the
measurement, but not the complete measurement design.

The distinction between these settings becomes important once the
instrumental resolution is included.
In the conventional Cooper--Nathans description, the distributions of
the incident and final wave vectors induce a multidimensional
resolution distribution in momentum--energy-transfer space
\cite{Cooper:a05676,Popovici:a11747,RN828}.
Within the present framework, this instrumental response is represented
by a setting-dependent resolution kernel
\(K_{{\rm res},a}\). 
The resolution kernel induces the observed measure
per unit exposure,
\(
\nu_{\boldsymbol\lambda,a}
=
(K_{{\rm res},a})_{\#}
\sigma_{\boldsymbol\lambda}^{Q\omega}\). 
The corresponding Poisson intensity measure is then
\begin{equation}
\Lambda_{\boldsymbol\lambda,a}
=
\mathcal E_a\,
\nu_{\boldsymbol\lambda,a} 
=\mathcal E_a\, (K_{{\rm res},a})_{\#}
\sigma_{\boldsymbol\lambda}^{Q\omega},
\label{eq:tas_design_count_measure}
\end{equation}
where \(\mathcal E_a\) contains the exposure and any
setting-dependent, outcome-independent throughput factors.  
Thus the sequence from the physical scattering outcome to the observed
Poisson intensity measure may be written schematically as
\begin{equation}
\sigma_{\boldsymbol\lambda}
\;\xrightarrow{\ (T_{\rm kin})_{\#}\ }\;
\sigma_{\boldsymbol\lambda}^{Q\omega}
\;\xrightarrow{\ (K_{{\rm res},a})_{\#}\ }\;
\nu_{\boldsymbol\lambda,a}
\;\xrightarrow{\ \times\mathcal E_a\ }\;
\Lambda_{\boldsymbol\lambda,a}.
\nonumber
\end{equation}

Importantly, \( \pi(a_1)=\pi(a_2) \) does not imply
\(
K_{{\rm res},a_1}
=
K_{{\rm res},a_2}\). 
Distinct settings within the same kinematic fiber generally differ in
their resolution widths, orientations, correlations, or acceptances, and hence \(
K_{{\rm res},a_1}
\neq
K_{{\rm res},a_2}\). 
Consequently, even for the same intrinsic scattering measure,
\(
\Lambda_{\boldsymbol\lambda,a_1}
\neq
\Lambda_{\boldsymbol\lambda,a_2}
\) in general.
The two configurations therefore constitute distinct statistical
experiments despite sharing the same nominal
\((\mathbf Q_0,\omega_0)\).

\subsubsection{Example: gapped dispersive mode}
This distinction becomes relevant when the scattering structure itself
depends on parameters to be inferred.
For illustration, consider a one-dimensional gapped dispersive mode: 
\(\hbar\omega_{\boldsymbol\lambda}(Q)
=
\Delta
+
2J
\left[
1-\cos(Qd)
\right]\). Here \(d\) is treated as a known structural length scale, while  \(\boldsymbol\lambda=(\Delta,J)\) are taken as the parameters to be inferred. 
The corresponding dispersion graph is
\begin{equation}
\mathcal G_{\boldsymbol\lambda}
=
\left\{
(Q,\omega):
\hbar\omega
=
\Delta
+
2J[1-\cos(Qd)]
\right\}.
\end{equation}
The local deformation of the dispersion in parameter space is given by 
\begin{equation}
\nabla_{\boldsymbol\lambda}
\left(\hbar\omega_{\boldsymbol\lambda}(Q)\right)
=
\begin{pmatrix}
\frac{\partial(\hbar\omega_{\boldsymbol\lambda})}{\partial\Delta} \\
\frac{\partial(\hbar\omega_{\boldsymbol\lambda})}{\partial J}
\end{pmatrix}
=
\begin{pmatrix}
1\\
2[1-\cos(Qd)]
\end{pmatrix}.
\label{eq:tas_parameter_directions}
\end{equation}
Thus a change in \(\Delta\) shifts the mode uniformly along the energy
direction, whereas a change in \(J\) produces a \(Q\)-dependent
deformation.

\begin{figure}[htbp]
\centering
\includegraphics[width=6.5cm]{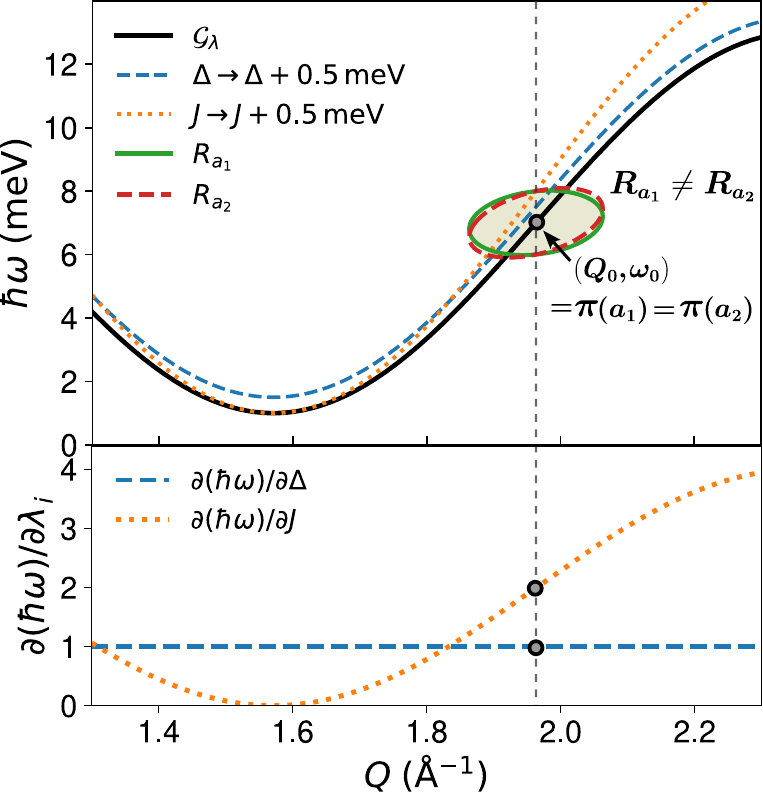}
\caption{
Sample- and instrument-side structures associated with a chosen point
\((Q_0,\omega_0)\) on the illustrative dispersion
\(\hbar\omega_{\boldsymbol\lambda}(Q)
=\Delta+2J[1-\cos(Qd)]\).
(a) Two settings \(a_1\) and \(a_2\) selected from the same kinematic
fiber satisfy
\(\pi(a_1)=\pi(a_2)=(Q_0,\omega_0)\)
but possess different instrumental resolution ellipses \(R_{a1}\neq R_{a2}\), albeit only slightly in this case. 
The reference dispersion and its deformations under changes of
\(\Delta\) and \(J\) are also shown.
(b) The corresponding sample-side parameter directions,
\(\partial(\hbar\omega_{\boldsymbol\lambda})/\partial\Delta=1\)
and
\(\partial(\hbar\omega_{\boldsymbol\lambda})/\partial J
=2[1-\cos(Qd)]\).
The figure separates parameter-dependent deformation of the scattering
structure from setting-dependent instrumental resolution.
}
\label{fig:TAS_resolution}
\end{figure}

In the usual Gaussian approximation, the different resolution kernels
associated with \(a_1\) and \(a_2\) are represented by distinct
resolution ellipsoids, \(R_{a1}\) and \(R_{a2}\), in momentum--energy-transfer space, as depicted
in Fig.~\ref{fig:TAS_resolution}.
A deformation of the sample dispersion caused by \(\Delta\) or \(J\)
therefore changes its overlap with the two resolution kernels
differently.
Consequently, the induced changes in the observed measures,
\(\partial_i\Lambda_{\boldsymbol\lambda,a_1}\) and
\(\partial_i\Lambda_{\boldsymbol\lambda,a_2}\), need not be the same,
even though the two settings correspond to the same nominal momentum
and energy transfer.

The Fisher information matrix
\(\mathbf F^{(a)}(\boldsymbol\lambda)\) provides a direct way to compare
configurations belonging to the same kinematic fiber.
The relevant comparison depends on the physical parameter to be inferred.
For example, if the gap \(\Delta\) is the parameter of interest while
\(J\) is treated as unknown, the appropriate local measure of
distinguishability is the efficient Fisher information
\begin{equation}
F_{\Delta\Delta}^{\mathrm{eff},(a)}
=
F_{\Delta\Delta}^{(a)}
-
F_{\Delta J}^{(a)}
\left(
F_{JJ}^{(a)}
\right)^{-1}
F_{J\Delta}^{(a)}.
\label{eq:tas_gap_efficient_information}
\end{equation}
The corresponding design problem within the kinematic fiber may then be
written as
\begin{equation}
a^*
=
\operatorname*{arg\,max}_{
a\in
\pi^{-1}
(\{(Q_0,\omega_0)\})
}
F_{\Delta\Delta}^{\mathrm{eff},(a)}.
\label{eq:tas_fiber_design}
\end{equation}
This is a local optimization within the kinematic fiber of a fixed
nominal point; a complete design may also optimize the nominal
momentum--energy points and combine multiple measurement settings.

For this simple dispersion, the role of nuisance-parameter elimination is
particularly transparent near the dispersion minimum, where the local
first-order sensitivity to the exchange coupling, 
\(\partial(\hbar\omega_{\boldsymbol\lambda})  / \partial J =
2[1-\cos(Qd)]\) vanishes.  In the idealized local picture, sensitivity to \(\Delta\) is
therefore less confounded with sensitivity to \(J\) at the minimum. 
 For a finite
instrumental resolution, however, the observed measure also contains
contributions from neighboring momentum--energy points, so this local
decoupling need not be exact.  The efficient Fisher information provides
the corresponding criterion directly at the level of the observed
measure.

\subsubsection{Multiple settings and exposure allocation}

Equation~\eqref{eq:tas_fiber_design} describes the optimization within the
fiber of a fixed nominal point. More generally, an experiment may combine
measurements performed at different nominal points, geometries, polarization
channels, or other instrumental settings. Let \(\mathcal E_a\) denote the
exposure allocated to setting \(a\), and let \(\mathbf I^{(\nu_a)}(\boldsymbol\lambda)\) denote the Fisher information
per unit exposure for that setting. For statistically independent
measurements, the Fisher information matrices add,
\begin{equation}
\mathbf F_{\mathrm{tot}}(\boldsymbol\lambda)
=
\sum_{a\in\mathsf A}
\mathbf F^{(a)}(\boldsymbol\lambda)
=
\sum_{a\in\mathsf A}
\mathcal E_a\,
\mathbf I^{(\nu_a)}(\boldsymbol\lambda).
\label{eq:complementary_fisher_additivity}
\end{equation}
Under a fixed total exposure,
\(
\sum_{a\in\mathsf A}\mathcal E_a
=
\mathcal E_{\mathrm{tot}}
\),
the selection of measurement settings and the allocation of exposure among
them form a common experimental-design problem.
Because different settings may be sensitive to different combinations
of parameters, combining them can improve poorly constrained parameter
directions rather than merely increase the total number of detected events.

The example of triple-axis spectroscopy makes the distinction between
kinematic access and measurement design explicit.
The setting map \(\pi\) determines which nominal
\((\bm Q,\omega)\) points are accessible, whereas the observed measure
\(\Lambda_{\bm\lambda,a}\) determines how sensitively a particular setting
probes the local parameter dependence of the scattering measure.
Experimental optimization is therefore a choice not merely of a point in
momentum--energy-transfer space, but of a measurement within the fiber over
that point.

\subsection{Diffuse-sphere model: resolution, throughput, and multiparameter design}
\label{subsec:dimensionless_fringe_design}

We now use the preceding framework to illustrate a familiar experimental
tradeoff between resolution and counting statistics.
The example is deliberately simple: its purpose is not to model a
particular instrument, but to show how the mass--shape decomposition,
kernel contraction, and Fisher-information design criteria enter a
single measurement problem.

Consider a spherical model with characteristic radius \(R\) and a
diffuse surface of width \(w\).
Let \(R_0\) be a nominal radius, let \(q\) denote a signed one-dimensional 
scattering coordinate, and define
\(
 x \coloneqq q R_0,\) 
\(\alpha \coloneqq \ln (R /R_0 ),\) 
\(\rho \coloneqq w /R.\) 
Since \(qR=e^\alpha x\), introduce the dimensionless coordinate
\(z=qR\) and define the diffuse-sphere profile
\begin{equation}
h_\rho(z)
\coloneqq
\left[
\frac{3j_1(z)}{z}
\right]^2
\exp\!\left(-\rho^2 z^2\right),
\label{eq:diffuse_sphere_profile}
\end{equation}
where \(j_1\) is the spherical Bessel function of the first kind and
order one, and the value at \(z=0\) is understood by continuity.
On the fixed reference coordinate \(x=qR_0\), the corresponding profile is
therefore \(
h_{\alpha,\rho}(x)
=
h_\rho\left(e^\alpha x\right).
\)  
The parameter \(\alpha\) changes the characteristic size and hence the
fringe positions on the fixed \(x\) coordinate, whereas \(\rho\)
controls the relative surface width and the damping of higher-order
fringes.

Because the profile is even, it may be extended to the signed coordinate
without changing the physical one-sided form-factor pattern.
The normalized shape is
\begin{equation}
p_{\alpha,\rho}(x)
=
\frac{e^\alpha}{Z_\rho}
h_\rho\!\left(e^\alpha x\right),
\qquad
Z_\rho
=
\int_{-\infty}^{\infty}
h_\rho(z)\,dz.
\label{eq:normalized_fringe_shape}
\end{equation}
The scattering measure can therefore be written directly in
mass--shape form,
\begin{equation}
d\sigma_{\boldsymbol\lambda}(x)
=
r \,
p_{\bm{\theta}}(x)\,dx,
\qquad
\boldsymbol{\lambda}=(r,\boldsymbol{\theta}),
\qquad
\boldsymbol{\theta}=(\alpha,\rho).
\label{eq:fringe_scattering_measure}
\end{equation}
Here \(r=\sigma_{\boldsymbol\lambda}(X)\) is the total scattering
mass, while \(p_{\bm{\theta}}(x)=p_{\alpha,\rho}(x)\) describes its normalized shape.

\subsubsection{Resolution and throughput trade-off}
We next introduce the measurement design when instrumental resolution and
count statistics are coupled. 
Let \(\gamma\) denote the Gaussian resolution width in the dimensionless
recorded coordinate and define the mass-preserving resolution kernel
\( 
K_\gamma(dy\mid x)
=
G_\gamma(y-x)\,dy
\). 
Suppose further that the same instrumental setting accepts only a
fraction \(\eta(\gamma)\) of the scattered events, representing the
instrumental throughput. Finer resolution (smaller \(\gamma\)) may 
be accompanied by lower throughput (smaller \(\eta\)).
Resolution and throughput can then be combined into the
sub-probability kernel
\begin{equation}
\widetilde K_\gamma(dy\mid x)
=
\eta(\gamma)\,
K_\gamma(dy\mid x).
\label{eq:fringe_effective_kernel}
\end{equation}
The observed Poisson intensity measure is therefore
\(
\Lambda_{\boldsymbol\lambda;\gamma}
=
\mathcal E\,
\big(\widetilde K_\gamma\big)_\#
\sigma_{\boldsymbol\lambda},
\) 
or, in density form,
\begin{equation}
d\Lambda_{\boldsymbol\lambda;\gamma}(y)
=
\mathcal E\,r\,
\eta(\gamma)\,
\left[
G_\gamma*p_{\boldsymbol\theta}
\right](y)\,dy.
\label{eq:fringe_observed_measure}
\end{equation}

Since \(K_\gamma\) is mass preserving, the normalized detected shape is simply 
\begin{equation}
p_{\boldsymbol\theta}(y;\gamma)
\coloneqq
\left[
G_\gamma*p_{\boldsymbol\theta}
\right](y),
\label{eq:p_y_theta_gamma}
\end{equation}
whereas the total mass of the observed Poisson measure is
\(
M(\boldsymbol\lambda;\gamma)
\coloneqq
\Lambda_{\boldsymbol\lambda;\gamma}(Y)
=
\mathcal E\,r\,\eta(\gamma)
\). 
Thus the instrumental setting affects the observed measure in two
distinct ways: the resolution kernel \(K_\gamma\) modifies its normalized
shape, while the throughput factor \(\eta(\gamma)\) modifies its total mass.

The mass--shape decomposition of Fisher information expressed in
Eq.~\eqref{eq:poisson_mass_shape_Fisher_matrix} now applies directly:
\begin{equation}
\mathbf F(\boldsymbol\lambda;\gamma)
=
\frac{
\nabla_{\boldsymbol\lambda} M(\boldsymbol\lambda;\gamma)\,
\nabla_{\boldsymbol\lambda} M(\boldsymbol\lambda;\gamma)^{\mathsf T}
}{
M(\boldsymbol\lambda;\gamma)
}
+
M(\boldsymbol\lambda;\gamma)
\begin{pmatrix}
0 & \mathbf 0 \\
\mathbf 0 & \mathbf I^{(P)}(\boldsymbol\theta;\gamma)
\end{pmatrix},
\label{eq:fringe_mass_shape_fisher}
\end{equation}
where
\(
\nabla_{\boldsymbol\lambda}
=
(\partial_r,\partial_\alpha,\partial_\rho)^{\mathsf T}
\),
and \(\mathbf I^{(P)}(\boldsymbol\theta;\gamma)\) is the per-event
Fisher information matrix of the normalized detected shape
\(p_{\boldsymbol\theta}(y;\gamma)\). 
For the parameterization
\(\boldsymbol\lambda=(r,\alpha,\rho)=(r,\boldsymbol\theta)\), 
the Fisher matrix takes the block form
\begin{equation}
\mathbf F(\boldsymbol\lambda;\gamma)
=
M(\boldsymbol\lambda;\gamma)
\begin{pmatrix}
1/r^2 & \mathbf 0 \\
\mathbf 0 & \mathbf I^{(P)}(\boldsymbol\theta;\gamma)
\end{pmatrix}
=
M(\boldsymbol\lambda;\gamma)
\begin{pmatrix}
1/r^{2} & 0 & 0 \\
0 & I_{\alpha\alpha}^{(P)}(\boldsymbol\theta;\gamma)
  & I_{\alpha\rho}^{(P)}(\boldsymbol\theta;\gamma) \\
0 & I_{\rho\alpha}^{(P)}(\boldsymbol\theta;\gamma)
  & I_{\rho\rho}^{(P)}(\boldsymbol\theta;\gamma)
\end{pmatrix}.
\label{eq:fringe_fisher_block}
\end{equation}

The information on \(\alpha\) and \(\rho\) is carried
by the block for the shape parameters.  
Its magnitude, however, is proportional to
the expected number of detected events; 
\begin{equation}
\mathbf F_{\boldsymbol\theta\boldsymbol\theta}
(\boldsymbol\lambda;\gamma)
=
M(\boldsymbol\lambda;\gamma)\, 
\mathbf I^{(P)}(\boldsymbol\theta;\gamma)
=
\mathcal E  r\,
\eta(\gamma)
\mathbf I^{(P)}(\boldsymbol\theta;\gamma).
\label{eq:fringe_information_factorization}
\end{equation}
Thus \(\mathbf I^{(P)}(\boldsymbol\theta;\gamma)\) quantifies the information carried by each
detected event through the normalized shape, while
\(M(\boldsymbol\lambda;\gamma)\) determines how much of that information
is accumulated during the exposure.

For the total scattering mass \(r\) itself, the information
is carried by the mass term, 
\begin{equation}
F_{rr}(\boldsymbol\lambda;\gamma)
=
\frac{M(\boldsymbol\lambda;\gamma)}{r^2}.
\end{equation}
For the logarithmic mass coordinate \(\xi=\ln r\), this reduces simply to \(F_{\xi\xi}=M(\boldsymbol\lambda;\gamma)\).
Thus, in logarithmic coordinates, the Fisher information for a relative change in the total scattering mass is simply the expected number of detected events, as expected for Poisson counting statistics.

The factorization Eq.~\eqref{eq:fringe_information_factorization} 
makes the resolution--throughput competition
particularly transparent.
Broadening the resolution width \(\gamma\) introduces additional
parameter-independent blurring and therefore cannot increase the
per-event shape information, by the contraction result of
Sec.~\ref{sec:information_contraction}.

Increasing the instrumental acceptance width generally increases the
throughput \(\eta(\gamma)\) and hence the expected number of detected
events at the expense of resolution.
The experimental design therefore involves a trade-off between
per-event shape information and counting statistics. 
Accordingly, after factoring out the exposure \(\mathcal E\) and the
total scattering mass \(r\), the information available for the shape
parameters is
\begin{equation}
\frac{
\mathbf F_{\boldsymbol\theta\boldsymbol\theta}
(\boldsymbol\lambda;\gamma)
}{
\mathcal E r
}
=
\eta(\gamma)\,
\mathbf I^{(P)}(\boldsymbol\theta;\gamma).
\label{eq:fringe_information_per_exposure_scale}
\end{equation}
Thus the information efficiency of an instrumental setting is determined
by the product of its throughput and the per-event shape information. 
For the shape parameters, the exposure \(\mathcal E\) and the total
 mass \(r\) enter only as scale factors:
they increase the number of detected events without changing the
information carried by each event.

\begin{figure}[t]
  \centering
  \includegraphics[width=7.5cm]{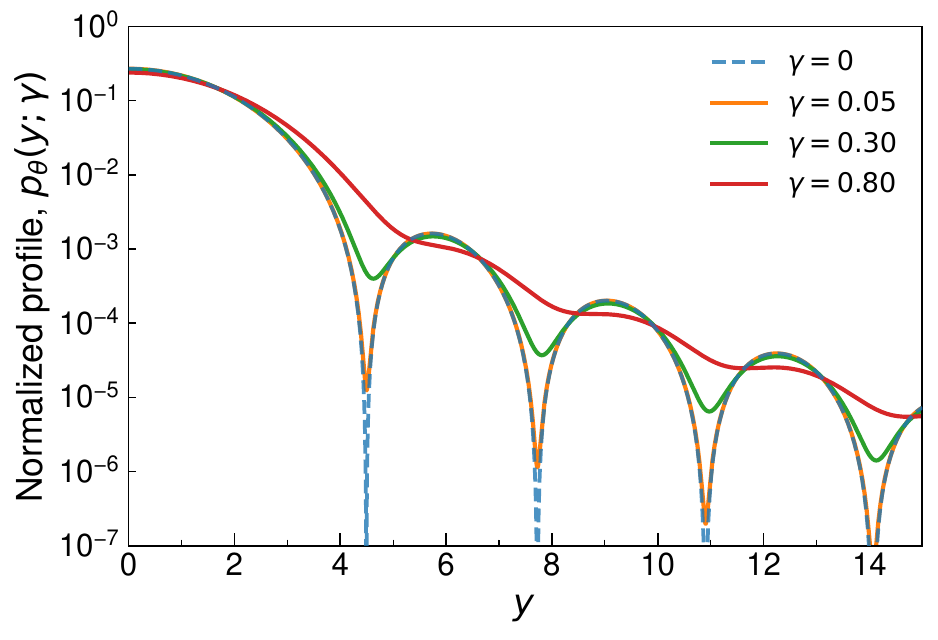}
  \caption{
  Resolution broadening for the dimensionless diffuse-sphere model of
  Eq.~\eqref{eq:p_y_theta_gamma}.
  The normalized profile at the reference size
  \(\alpha=0\) (\(R=R_0\)) and \(\rho=0.08\) is compared with profiles
  obtained after convolution with Gaussian resolutions of width
  \(\gamma=0\), \(0.05\), \(0.30\), and \(0.80\).
  Increasing \(\gamma\) progressively suppresses the higher-order fringes.
  }
  \label{fig:fringe_profiles}
\end{figure}

To illustrate how the resolution kernel modifies the observable shape,
Fig.~\ref{fig:fringe_profiles} shows numerically convolved profiles for
several values of \(\gamma\).
As \(\gamma\) increases, the form-factor minima are progressively filled
and the higher-order fringes are suppressed.
These changes reduce the distinction between nearby profiles produced by
small variations of \(\alpha\) and \(\rho\), providing a direct visual
counterpart to the loss of per-event shape information
\(\mathbf I^{(P)}(\boldsymbol\theta;\gamma)\).

\subsubsection{Criteria of optimality: D-, A-, E-, and c-optimal}
The Fisher matrix itself describes the local statistical
distinguishability provided by a given instrumental setting.
To compare different settings by a single design objective, however,
one must specify which aspect of this multiparameter distinguishability
is scientifically relevant. This leads to the standard notion of an
optimality criterion \cite{Elfving1952, Kiefer1959, Pukelsheim2006, Atkinson2007} in experimental design. 
Such criteria map the Fisher information matrix to a scalar quantity,
emphasizing, for example, the overall uncertainty volume
(D-optimality), the average parameter variance (A-optimality), the
least-constrained parameter direction (E-optimality), or the precision
of a particular parameter combination (\(c\)-optimality).

Let \(\Phi\) denote this scalar design criterion. The preferred
instrumental design can then be written as
\begin{equation}
\gamma^*
=
\operatorname*{arg\,max}_{\gamma}
\Phi\!\left[
\mathbf F_{\boldsymbol\theta\boldsymbol\theta}
(\boldsymbol\lambda;\gamma)
\right].
\label{eq:fringe_design_optimization}
\end{equation}
Different choices of \(\Phi\) correspond to different scientific
objectives and include the optimal criteria.

For joint estimation of
\(\boldsymbol\theta=(\alpha,\rho)\), D-optimality corresponds to
\begin{equation}
\Phi_D
=
\det
\mathbf F_{\boldsymbol\theta\boldsymbol\theta},
\end{equation}
which favors a small overall Fisher ellipse, whereas E-optimality
corresponds to
\begin{equation}
\Phi_E
=
\lambda_{\min}
\left(
\mathbf F_{\boldsymbol\theta\boldsymbol\theta}
\right),
\end{equation}
which favors improvement of the least-constrained local parameter
direction.
Another standard choice, A-optimality, minimizes
\(
\operatorname{Tr}
\mathbf F_{\boldsymbol\theta\boldsymbol\theta}^{-1}
\),
and therefore emphasizes the average parameter variance.
The dimensionless parameterization used here fixes the relative scaling
of \(\alpha\) and \(\rho\) when comparing such matrix-based criteria.

If \(\rho\) is regarded as fixed and only the radius parameter
\(\alpha\) is of interest, a natural criterion is simply
\( \Phi =
F_{\alpha\alpha}\). 
If instead \(\alpha\) is the parameter of interest while \(\rho\) must
be estimated simultaneously, a natural parameter-specific design
criterion is the efficient Fisher information. 
This is precisely the local \(c\)-optimal criterion.  
The scalar objective may be written as 
\begin{equation}
\Phi_c(\boldsymbol\lambda;\gamma)
=
\frac{1}{
\mathbf c^{\mathsf T}
F^{-1}(\boldsymbol\lambda;\gamma)
\mathbf c}.
\end{equation}
For the parameter of interest \(\alpha\), taking
\(\mathbf c=\mathbf e_\alpha\), so that
\(\mathbf c^{\mathsf T}\boldsymbol\lambda=\alpha\), gives
\begin{equation}
\Phi_c(\boldsymbol\lambda;\gamma)
=
\frac{1}{
\mathbf e_\alpha^{\mathsf T}
F^{-1}(\boldsymbol\lambda;\gamma)
\mathbf e_\alpha}
=
F_{\alpha\alpha}^{\mathrm{eff}}
(\boldsymbol\lambda;\gamma)
=
F_{\alpha\alpha}(\boldsymbol\lambda;\gamma)
-
F_{\alpha\rho}(\boldsymbol\lambda;\gamma)
\left[
F_{\rho\rho}(\boldsymbol\lambda;\gamma)
\right]^{-1}
F_{\rho\alpha}(\boldsymbol\lambda;\gamma).
\end{equation}
Maximizing this quantity is equivalent to minimizing the local variance
bound for \(\alpha\), and corresponds to \(c\)-optimality.

\subsubsection{Numerical simulation and criterion-dependent optima}
To make the competing effects of resolution and counting statistics 
explicit, we introduce a simple phenomenological throughput law
\begin{equation}
\eta(\gamma)
=
\frac{\gamma}{\gamma+\gamma_c},
\qquad
\gamma_c=0.12.
\label{eq:illustrative_throughput}
\end{equation}
The Fisher matrix is numerically evaluated below at the reference size
\(R=R_0\) \((\alpha=0)\) and at \(\rho=0.08\).

\begin{figure}[t]
 \centering
  \includegraphics[width=12cm]{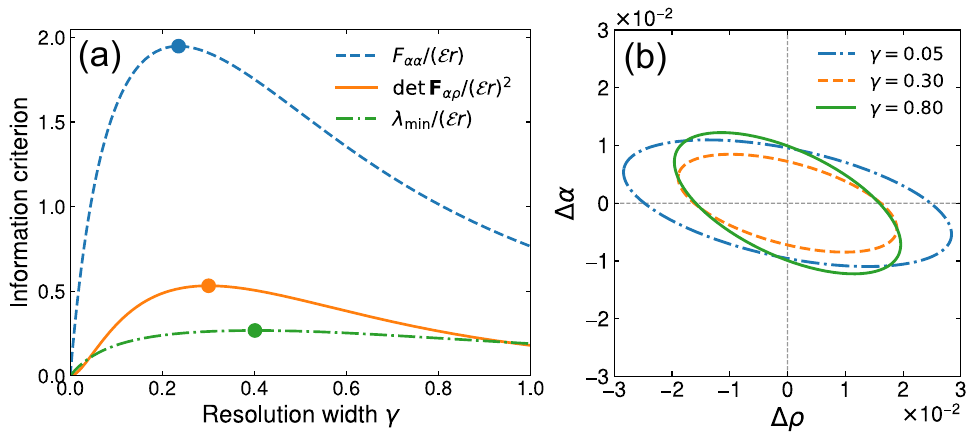}
  \caption{
  (a) Fisher-information design criteria for the diffuse-sphere model with
  the illustrative throughput law
  \(\eta(\gamma)=\gamma/(\gamma+\gamma_c)\),
  \(\gamma_c=0.12\), evaluated at \(\alpha=0\) and \(\rho=0.08\).
  Shown are
  \(F_{\alpha\alpha}/(\mathcal E r)\),
  \(\det\mathbf F_{\boldsymbol\theta\boldsymbol\theta}/(\mathcal E r)^2\),
  and
  \(\lambda_{\min}
  (\mathbf F_{\boldsymbol\theta\boldsymbol\theta})/(\mathcal E r)\),
  corresponding respectively to sensitivity to \(\alpha\) with
  \(\rho\) fixed, D-optimal, and E-optimal joint estimation.
  Their maxima occur at different values of \(\gamma\). 
  (b) Fisher ellipses for the parameter pair \((\alpha,\rho)\) in the
  diffuse-sphere model, shown for
  \(\gamma=0.05\), \(0.30\), and \(0.80\).
  The ellipses correspond to
  \(
  \Delta\boldsymbol\theta^{\mathsf T}
  \mathbf F_{\boldsymbol\theta\boldsymbol\theta}
  \Delta\boldsymbol\theta=\num{1e-4}
  \).  }
  \label{fig:fisher_design_ellipse}
\end{figure}

Figure~\ref{fig:fisher_design_ellipse}(a) compares three different optimal criteria.
For the present model,
\(F_{\alpha\alpha}\) is maximized at
\(\gamma\simeq0.24\),
the D-optimal criterion at
\(\gamma\simeq0.30\),
and the E-optimal criterion at
\(\gamma\simeq0.40\).
The numerical locations of these optima depend on the assumed throughput
law and scattering model, but their non-coincidence illustrates the more
general point that there is no parameter-independent ``best resolution.''

Figure~\ref{fig:fisher_design_ellipse}(b) shows how the instrumental
setting changes the local sensitivity to the two parameters.
As \(\gamma\) varies, both the aspect ratio and orientation of the
Fisher--Rao ellipse change, indicating that the relative sensitivity to
\(\alpha\) and \(\rho\), as well as their local statistical correlation,
depends on the instrumental resolution.

The ellipse area provides a complementary measure of the overall joint
distinguishability.  For a two-parameter Fisher block, the local contour
\( 
\Delta\boldsymbol{\theta}^{\mathsf T}
\mathbf F_{\boldsymbol{\theta}\boldsymbol{\theta}}
\Delta\boldsymbol{\theta}
=
c\)  
has area
\begin{equation}
A (\boldsymbol{\theta};\gamma)
=
\frac{\pi c}
{\sqrt{\det\mathbf F_{\boldsymbol{\theta}\boldsymbol{\theta}}}}.
\end{equation}
Thus the D-optimal criterion in Fig.~\ref{fig:fisher_design_ellipse}(a) is geometrically equivalent to
minimizing the area of the local Fisher--Rao ellipse.  Consistently, the
ellipse at \(\gamma\simeq0.30\) has the smallest area among the three
examples shown in Fig.~\ref{fig:fisher_design_ellipse}(b), matching the maximum of
\(\det\mathbf F_{\boldsymbol{\theta}\boldsymbol{\theta}}\)
in Fig.~\ref{fig:fisher_design_ellipse}(a).
Accordingly, the ellipse contracts from \(\gamma=0.05\) to \(0.30\)
as the gain in throughput dominates, and expands again at
\(\gamma=0.80\) as resolution smearing becomes dominant.

\medskip 
\paragraph*{Joint optimization of instrumentation and data reduction}
Measurement design is not confined to hardware.  The choice of which
outcome coordinates to record and which distinctions to retain in data
reduction also defines part of the measurement process.  By the
contraction result of Sec.~\ref{sec:information_contraction}, such a
reduction preserves Fisher information only when the discarded
distinctions carry no additional local sensitivity to the parameters of
interest; otherwise the lost information cannot be recovered from the
reduced data.

The overall experimental-design pipeline can be summarized as 
\begin{equation}
a
\longrightarrow
\Lambda_{\boldsymbol\lambda,a}
\longrightarrow
\mathbf F^{(a)}(\boldsymbol\lambda)
\longrightarrow
\Phi\!\big[\mathbf F^{(a)}\big],
\nonumber 
\end{equation}
where \(a\) includes both instrumental settings and choices in the
recorded or retained outcome representation,
\(\Lambda_{\boldsymbol\lambda,a}\) is the resulting observed measure,
\(\mathbf F^{(a)}(\boldsymbol\lambda)\) its Fisher information matrix,
and \(\Phi\) specifies the inferential objective.


\section{Connection to quantum measurement}
\label{sec:quantum_origin}

The framework developed in the preceding sections is classical in the
sense that the scattering measure is a finite scalar measure on a space
of experimentally distinguishable outcomes. This does not imply that
the underlying scattering process is classical. Rather, the
measure-theoretic description begins after quantum dynamics has been
converted into classical outcome statistics by a measurement.

In quantum information geometry, the classical Fisher--Rao structure
appears as part of the broader theory of monotone metrics on quantum
states with classical and genuinely quantum 
contributions~\cite{Ciaglia2024}.

The purpose of this section is to locate the finite scattering measure
within this quantum-to-classical hierarchy.  We consider the classical
outcome statistics generated by a specified quantum measurement, while
questions concerning measurement backaction, quantum-state 
distinguishability~\cite{BraunsteinCaves1994,Paris2009}, 
and optimization over quantum measurements are left
outside the present scope.

\subsection{Quantum measurement and outcome sectors}
\label{subsec:quantum_outcome_measures}

A quantum measurement is represented by a positive operator-valued
measure (POVM) on a measurement-outcome space. For a parameterized 
family of quantum states \(\rho_\lambda\), the
Born rule assigns to each specified POVM an ordinary scalar probability
measure on the outcome space. 
The Born rule for a complete POVM produces a normalized probability
measure, whereas the scattering measure considered in this work is a
generally non-normalized finite measure. 
To relate these two descriptions, we regard the scattering outcomes 
as one sector of a complete quantum-measurement outcome space.

Let \(X\) denote the space of scattering outcomes
and  collect all outcomes outside this sector into a complementary (no-scattering)  
outcome \(\cem\):
\begin{equation}
\widetilde X
=
X\sqcup\{\cem\}.
\label{eq:complete_quantum_outcome_space}
\end{equation}
Let \(\widetilde M\) be a normalized POVM on \(\widetilde X\), with
\(
\widetilde M(\widetilde X)
=
\mathbf 1_{\mathcal H}\), where \(\mathbf 1_{\mathcal H}\) denotes the identity operator on the Hilbert space \(\mathcal H\). 
Restricting it to the scattering sector gives
\(M(A)=\widetilde M(A)\) for \(A\in\mathcal X\), and hence
\begin{equation}
M(X)
=\mathbf 1_{\mathcal H} - \widetilde M({\cem}).
\end{equation}

For the quantum-state family \(\rho_\lambda\), the Born rule then
defines a subprobability measure on the scattering-outcome space,
\begin{equation}
\zeta_\lambda(A)
\coloneqq
\operatorname{Tr}
\!\big[
\rho_\lambda M(A)
\big],
\qquad
A\in\mathcal X.
\label{eq:born_scattering_measure}
\end{equation}
Its total mass
\begin{equation}
q_\lambda
\coloneqq
\zeta_\lambda(X)
=
\operatorname{Tr}
\!\big[
\rho_\lambda M(X)
\big]
\leq 1
\label{eq:scattering_probability}
\end{equation}
is the probability that the complete quantum measurement produces an
outcome in the scattering sector.

For \(q_\lambda>0\), define the conditional distribution
\begin{equation}
P_\lambda^M(A)
=
\frac{\zeta_\lambda(A)}
     {q_\lambda},
\qquad
A\in\mathcal X,
\end{equation}
so that
\begin{equation}
\zeta_\lambda
=
q_\lambda P_\lambda^M.
\label{eq:quantum_scattering_factorization}
\end{equation}
The complete Born-rule distribution is therefore
\begin{equation}
\widetilde P_\lambda^M
=
q_\lambda P_\lambda^M
+
(1-q_\lambda)\delta_\cem,
\label{eq:complete_born_measure}
\end{equation}
where \(\delta_\cem\) denotes the Dirac probability measure
concentrated at the complementary outcome \(\cem\).

The finite scattering measure used throughout this work has the
parallel factorization
\(
\sigma_\lambda
=
r_\lambda P_\lambda^M\) with \(
r_\lambda
=
\sigma_\lambda(X)\). 
The mass \(q_\lambda\) is a dimensionless probability associated with
a specified quantum experiment, whereas \(r_\lambda\) is a physical
scattering mass such as a cross section. 
The common structure is an overall mass together with a normalized 
distribution over scattering outcomes.

\paragraph*{Discrete outcome sectors.}
In quantum measurement, outcomes may naturally be resolved into
discrete sectors---for example, energy levels, distinct internal states or different
paths in an interferometer---with continuous variables retained
within each sector.  We therefore introduce a sector
decomposition of the scattering-outcome space: 
\begin{equation}
X
=
\bigsqcup_{c\in\mathcal C}
S_c,
\qquad
\mathcal C=\{1,\ldots,K\},
\label{eq:quantum_sector_partition}
\end{equation}
where \(c\) labels a discrete recorded outcome channel.

Define the conditional weight of sector \(c\), given that the outcome
lies in the scattering sector, by
\begin{equation}
w_{\lambda,c}
\coloneqq
P_\lambda^M(S_c),
\qquad
\sum_{c\in\mathcal C}
w_{\lambda,c}
=
1.
\label{eq:conditional_channel_weight}
\end{equation}
For \(w_{\lambda,c}>0\), let \(P_{\lambda,c}\) denote the
normalized conditional distribution within that sector,
\begin{equation}
P_{\lambda,c}(A)
=
\frac{
P_\lambda^M(A\cap S_c)
}{
w_{\lambda,c}
}.
\label{eq:within_sector_distribution}
\end{equation}
The normalized scattering distribution then decomposes as
\begin{equation}
P_\lambda^M
=
\sum_{c\in\mathcal C}
w_{\lambda,c}P_{\lambda,c}.
\label{eq:sector_decomposition_probability}
\end{equation}

The unconditional probability of sector \(c\) in the complete
measurement is
\(
q_{\lambda,c}
\coloneqq
\zeta_\lambda(S_c)
=
q_\lambda w_{\lambda,c},
\) 
with \(q_\lambda=\sum_c q_{\lambda,c}\).  The complete Born-rule
distribution can therefore be written as
\begin{equation}
\widetilde P_\lambda^M
=
(1-q_\lambda)\delta_{\cem}
+
\sum_{c\in\mathcal C}
q_{\lambda,c}P_{\lambda,c}.
\label{eq:complete_sector_decomposition}
\end{equation}
This separates the recorded outcome into scattering occupancy, the
discrete sector label, and the outcome within that sector.

\subsection{Classical distinguishability of quantum measurement outcomes}
\label{subsec:quantum_classical_fisher}

We now consider the classical distinguishability carried by the
outcomes of a specified quantum measurement.   
Once a POVM \(M\) has been specified, the Born rule
maps the parameterized quantum state \(\rho_{\boldsymbol\lambda}\) 
to the classical statistical model 
\(\widetilde P_{\boldsymbol\lambda}^{M}\), 
to which the Fisher-information framework developed above 
applies directly.

\subsubsection{Hierarchical decomposition of outcome Fisher information}

The sector decomposition introduced above then separates the classical
information into contributions associated with scattering occupancy, the
discrete sector label, and the outcome within each sector.

Applying the mass--shape decomposition  directly to
\(\zeta_\lambda=q_\lambda P_\lambda^M\) gives
\begin{equation}
\mathcal I^{(\zeta)}(\lambda)
=
\frac{\dot q_\lambda^{\,2}}{q_\lambda}
+
q_\lambda
\mathcal I^{(P^M)}(\lambda).
\label{eq:quantum_scattering_sector_FI}
\end{equation}
Because the sectors \(S_c\) are disjoint and the sector label is retained
in the measurement record, the Fisher information of the normalized
scattering distribution further decomposes as
\begin{equation}
\mathcal I^{(P^M)}(\lambda)
=
\sum_{c\in\mathcal C}
\frac{\dot w_{\lambda,c}^{\,2}}{w_{\lambda,c}}
+
\sum_{c\in\mathcal C}
w_{\lambda,c}
\mathcal I^{(P_c)}(\lambda).
\label{eq:categorical_shape_decomposition}
\end{equation}
The first term is the information carried by the discrete sector label,
whereas the second is the average information carried by the conditional
distribution within each sector.
This decomposition is the classical Fisher-information chain rule for a
hierarchical outcome record and is related to the corresponding
decomposition in sequential quantum measurements~\cite{Lu2012}.

Restoring the complementary outcome \(\cem\), the Fisher information of
the complete Born-rule distribution becomes
\begin{align}
\mathcal I^{(\widetilde P^M)}(\lambda)
=
\underbrace{
\frac{\dot q_\lambda^{\,2}}
     {q_\lambda(1-q_\lambda)}
}_{\text{scattering occupancy}}
+
\underbrace{
q_\lambda
\sum_{c\in\mathcal C}
\frac{\dot w_{\lambda,c}^{\,2}}
     {w_{\lambda,c}}
}_{\text{sector label}}
+
\underbrace{
q_\lambda
\sum_{c\in\mathcal C}
w_{\lambda,c}
\mathcal I^{(P_c)}(\lambda)
}_{\text{within-sector shape}}.
\label{eq:complete_hierarchical_FI}
\end{align}
The first term is the Bernoulli process information associated with whether
scattering occurred.  The complete measurement record therefore has a 
hierarchical information structure consisting of three contributions: 
scattering occupancy, the categorical sector label conditional on scattering, and
the conditional within-sector distribution.

\subsubsection{From coherent alternatives to classical outcome sectors}
\label{subsubsec:quantum_to_classical_sectors}

The sector decomposition introduced above applies to distinguishable
measurement outcomes, not directly to coherent quantum alternatives prior
to measurement.  It therefore also identifies the stage at which the
quantum description gives rise to the classical outcome measure used
throughout this work.

For a specified measurement, the Born rule maps the quantum state
\(\rho_\lambda\) to a classical probability measure \(P_\lambda^M\).
It admits the sector decomposition,
\(
P_\lambda^M
=
\sum_c
w_{\lambda,c}P_{\lambda,c}.
\) 
The corresponding categorical contribution in
Eq.~\eqref{eq:complete_hierarchical_FI} is present only when the
measurement record resolves the sector label and the relative sector
weights \(w_{\lambda,c}\) depend on the parameter \(\lambda\).
If the sector weights are parameter independent, this contribution
vanishes even when the sectors are experimentally resolved.

Before such a classical outcome record is defined, however, coherent
quantum alternatives cannot in general be interpreted as separate
sectors.  In an interferometer, for example, the two paths remain coherent
alternatives as long as their amplitudes contribute jointly to the
interference pattern.  Acquiring which-path information does not merely
refine an existing classical partition: it changes the measurement so that
the path label becomes a distinguishable outcome and may thereby modify
the interference structure.

At the level of a quantum instrument, a resolved outcome label may carry
more than classical statistical information: it also labels the
corresponding conditional transformation of the quantum state.  This
connects the present discussion to established theories of quantum
instruments\cite{Davies1970, Ozawa1984} and logical 
reversibility~\cite{UedaImotoNagaoka1996}.  
Here our concern is specifically the
classical outcome measure generated by a specified measurement, for which
the mass--shape and sector decompositions developed above apply.

\subsubsection{Rare-event limit and Poisson counting}
\label{subsubsec:quantum_poisson_limit}

The preceding discussion concerns the classical outcome distribution
generated by a specified quantum measurement.  The subsequent passage
from repeated trials to Poisson counting is entirely classical.

Suppose first that \(n\) incident probes are independently subjected
to the same fixed measurement \(M\), producing
\(
Y_i \overset{\mathrm{i.i.d.}}{\sim} \widetilde P_\lambda^M,\) 
\(
i=1,\ldots,n.
\)
Here \(n\) is the prescribed number of trials. 
By additivity of Fisher information for independent trials, 
the Fisher information of the complete \(n\)-trial record is
\begin{equation}
F^{(n)}(\lambda)
= 
n\,
\mathcal I^{(\widetilde P^M)}(\lambda).
\label{eq:ntrial_outcome_FI}
\end{equation}
Writing
\(q_\lambda=\sum_c q_{\lambda,c}\), the Fisher information of the
complete single-trial outcome distribution may be written as
\begin{equation}
\mathcal I^{(\widetilde P^M)}(\lambda)
=
\frac{\dot q_\lambda^{\,2}}{1-q_\lambda}
+
\sum_{c\in\mathcal C}
\frac{\dot q_{\lambda,c}^{\,2}}{q_{\lambda,c}}
+
\sum_{c\in\mathcal C}
q_{\lambda,c}\,
\mathcal I^{(P_c)}(\lambda).
\end{equation}
Hence, for \(n\) independent trials,
\begin{align}
F^{(n)}(\lambda)
=
n\mathcal I^{(\widetilde P^M)}(\lambda)
=
\frac{\dot M_\lambda^{\,2}}
     {n(1-M_\lambda/n)}
+
\sum_{c\in\mathcal C}
\frac{\dot M_{\lambda,c}^{\,2}}{M_{\lambda,c}}
+
\sum_{c\in\mathcal C}
M_{\lambda,c}\,
\mathcal I^{(P_c)}(\lambda),
\label{eq:ntrial_fisher_sector}
\end{align}
where
\(
M_{\lambda,c}=nq_{\lambda,c}\), 
\(
M_\lambda=\sum_{c\in\mathcal C}M_{\lambda,c}
=nq_\lambda.
\)

To connect this repeated-trial description with the Poisson observation
model introduced in Sec.~\ref{subsec:poisson_scattering_likelihood}, 
we embed the finite-\(n\) model in a
sequence of increasingly rare single-trial scattering experiments.
Accordingly, we write the single-trial scattering probabilities as
\(q_{\lambda,c}^{(n)}\) and \(q_\lambda^{(n)}\).

For each \(n\), let \(\Xi_{i}^{(n)}\) denote the random counting measure generated
by the \(i\)-th single-probe trial:
\begin{equation}
\Xi_{i}^{(n)}
=
\begin{cases}
0,
& \text{if no scattering event occurs},\\[1mm]
\delta_{X_i^{(n)}},
& \text{if a scattering event occurs at }X_i^{(n)}\in X .
\end{cases}
\label{eq:single_trial_random_measure}
\end{equation}
If the resolved sectors are \(c\in\mathcal C\), the mean scattering
measure generated by a single trial is the sub-probability measure
\begin{equation}
\sigma_\lambda^{(n)}
=
\sum_{c\in\mathcal C}
q_{\lambda,c}^{(n)}P_{\lambda,c},
\label{eq:single_trial_scattering_measure}
\end{equation}
so that, for \(A\in\mathcal X\),
\begin{equation}
\mathbb E_\lambda
\left[
\Xi_{i}^{(n)}(A)
\right]
=
\sigma_\lambda^{(n)}(A).
\end{equation}
Its total mass
\begin{equation}
q_\lambda^{(n)}
=
\sigma_\lambda^{(n)}(X)
=
\sum_{c\in\mathcal C}q_{\lambda,c}^{(n)}
\le 1
\end{equation}
is the probability that a single trial produces a scattering event.

The complete scattering record from the \(n\) trials is the
superposition
\begin{equation}
\mathcal N_\lambda^{(n)}
=
\sum_{i=1}^{n}
\Xi_{i}^{(n)},
\label{eq:n_trial_point_process}
\end{equation}
which is itself a random counting measure, with mean measure
\begin{equation}
\mathbb E_\lambda
\left[
\mathcal N_\lambda^{(n)}(A)
\right]
=
n\,\sigma_\lambda^{(n)}(A).
\end{equation}

The rare-event limit is the regime in which the number of trials \(n\) 
increases while the scattering probability in each individual trial
becomes correspondingly small, so that the expected total event measure
approaches a finite nonzero limit.  At the measure level, we express
this scaling as
\begin{equation}
n\sigma_\lambda^{(n)}
\longrightarrow
\Lambda_\lambda,
\qquad
0<\Lambda_\lambda(X)<\infty ,
\label{eq:rare_event_measure_scaling}
\end{equation}
where the convergence is understood in a sense sufficient for the
integrals entering the Laplace functional below to converge.  In
particular,
\begin{equation}
nq_\lambda^{(n)}
\longrightarrow
\Lambda_\lambda(X)
=
M_\lambda,
\label{eq:rare_event_total_mass_scaling}
\end{equation}
and therefore \(q_\lambda^{(n)}\to0\). 

The limiting point process can be identified through its Laplace
functional~\cite{Baccelli2009}.  For any nonnegative test function
\(h\), independence of the trials gives
\begin{align}
\mathbb E_\lambda
\left[
\exp\left(
-\int_X h(x)\,
\mathcal N_\lambda^{(n)}(dx)
\right)
\right]
&=
\left[
1-
\int_X
\left(1-e^{-h(x)}\right)
\sigma_\lambda^{(n)}(dx)
\right]^n
\nonumber\\
&\longrightarrow
\exp\left[
-\int_X
\left(1-e^{-h(x)}\right)
\Lambda_\lambda(dx)
\right].
\label{eq:rare_event_laplace_limit}
\end{align}
The limiting expression is the Laplace functional of a Poisson point
process with intensity measure \(\Lambda_\lambda\).  Hence
\begin{equation}
\mathcal N_\lambda^{(n)}
\Longrightarrow
\mathcal N_\lambda,
\qquad
\mathcal N_\lambda
\sim
\operatorname{PPP}(\Lambda_\lambda),
\label{eq:rare_event_ppp_limit}
\end{equation}
where \(\Longrightarrow\) denotes convergence in distribution of the
random counting measures.  Thus the Poisson observation model of
Sec.~\ref{subsec:poisson_scattering_likelihood} is recovered from the
repeated-trial description.

For the sector decomposition above, if
\(
nq_{\lambda,c}^{(n)}
\longrightarrow
M_{\lambda,c},
\) 
then
\(
\Lambda_\lambda
=
\sum_{c\in\mathcal C}
M_{\lambda,c}P_{\lambda,c},\) and 
\(M_\lambda
=
\sum_{c\in\mathcal C}M_{\lambda,c}.
\)
The restrictions of the limiting Poisson process to disjoint resolved
sectors are therefore independent Poisson processes with mean counts
\(M_{\lambda,c}\).

To pass to the Fisher-information limit, assume the corresponding regular 
convergence of the parameter derivatives,
\begin{equation}
n\dot q_{\lambda,c}^{(n)}
\longrightarrow
\dot M_{\lambda,c},
\end{equation}
for sectors with positive limiting intensity.  Writing
\(
M_{\lambda,c}^{(n)}
=
nq_{\lambda,c}^{(n)},\) and  
\(
M_\lambda^{(n)}
=
\sum_c M_{\lambda,c}^{(n)},
\) 
the Fisher-information contribution from the complementary
no-scattering outcome becomes
\begin{equation}
\frac{
\bigl(\dot M_\lambda^{(n)}\bigr)^2
}{
n\left(1-M_\lambda^{(n)}/n\right)
}
\longrightarrow
0,
\end{equation}
whereas the scattering contributions converge to
\begin{equation}
F^{(n)}(\lambda)
\longrightarrow
\sum_{c\in\mathcal C}
\left[
\frac{\dot M_{\lambda,c}^{\,2}}{M_{\lambda,c}}
+
M_{\lambda,c}
\mathcal I^{(P_c)}(\lambda)
\right]
=
F(\lambda).
\label{eq:rare_event_fisher_limit}
\end{equation}
In the rare-event limit, writing
\(M_{\lambda,c}=M_\lambda w_{\lambda,c}\), the full Fisher information
can be decomposed as
\begin{equation}
F(\lambda)
=
\frac{\dot M_\lambda^{\,2}}{M_\lambda}
+
M_\lambda
\sum_{c\in C}
\frac{\dot w_{\lambda,c}^{\,2}}{w_{\lambda,c}}
+
M_\lambda
\sum_{c\in C}
w_{\lambda,c} I^{(P_c)}(\lambda).
\end{equation}
The disappearance of the separate no-scattering contribution in the
rare-event limit does not make the complementary outcome superfluous.
At finite trial number, the complete scattering-occupancy information
contains contributions from both scattering and no-scattering outcomes.
The rare-event limit shows that, as scattering becomes rare, the
separate contribution from the no-scattering outcome vanishes, whereas
the information carried by the number of scattering events remains
finite and converges to the Poisson mass term
\(\dot M_\lambda^{\,2}/M_\lambda\).

Outside the rare-event regime, the finite-trial description remains
the appropriate statistical model, and the no-scattering outcome can
retain a nonvanishing contribution to the Fisher information.  The
Poisson description should therefore be understood as a particular
limit of the more general repeated-trial measurement model, rather than
as a generic consequence of counting measurement outcomes.

In the notation of Sec.~\ref{subsec:poisson_scattering_likelihood}, 
the limiting intensity measure is 
\(
n \sigma_{\lambda}^{(n)} \longrightarrow \Lambda_\lambda
=
\mathcal E\,\sigma_\lambda .
\) 
Thus the rare-event limit recovers the Poisson point-process
observation model and its Fisher information, determined by the finite
intensity measure \(\Lambda_\lambda\) through its total mass and
normalized outcome distribution. 

\section{Summary}
\label{sec:summary}

We have formulated scattering experiments in terms of finite
measures on spaces of physically distinguishable outcomes.  
In this formulation, 
the scattering measure is the underlying representation-independent
object, while differential cross sections are its density
representations with respect to chosen reference measures.  
Instrumental response and data reduction enter as 
transformations of measures:
deterministic operations as pushforwards and stochastic processes such
as finite resolution, channel mixing, and inefficiency as probability
or sub-probability kernels.  This separates the physical scattering
content from its coordinate representation and from the operations by
which it is recorded.

For a parameterized family of finite scattering measures, this
description leads to an intrinsic Fisher geometry.  The score
and Fisher metric can be defined at the measure level without choosing
a density representation.  Because the measure is finite rather than
normalized, its total mass remains part of the statistical model, and
the Fisher information decomposes exactly into \emph{mass} and
\emph{shape} contributions.  These represent two distinct sources of
parameter distinguishability: the absolute scale of the scattering
intensity and its normalized distribution.  Under Poisson counting
statistics, the same intrinsic metric is realized as the Fisher
information of the observed point process per unit exposure.

Measurement and experimental design are thereby placed within the same
information-processing framework.  For a parameter-independent
probability kernel, the output score is the conditional expectation of
the input score given the recorded outcome, and the information loss is
determined by the conditional variation of the unresolved score.
Resolution, retained coordinates, channel mixing, instrumental settings,
exposure allocation, and data reduction can therefore be compared
according to the parameter-sensitive distinctions they preserve.  
Thus, the design problem is to optimize the Fisher information of the observed
measure for the inferential objective of interest, including the effects
of nuisance parameters.

Finally, the measure-theoretic framework clarifies the place of the
scattering measure within the hierarchy from quantum measurement to
classical outcome statistics. 
Through the Born rule, a quantum measurement
induces probabilities for distinguishable outcome sectors; retaining
the scattering sector together with its occurrence weight gives the
finite measure that enters the classical description developed here.
Instrumental response and data reduction then act on this outcome
measure through classical transformations.  
This separates limitations associated with the quantum measurement from
information subsequently lost through the apparatus or data reduction.

Although developed here specifically for scattering phenomena, 
the same structure applies more
broadly whenever a physical process generates a finite measure on
distinguishable outcomes that is subsequently transformed into recorded
data.  From this perspective, the central experimental question is not
merely how a distribution is represented, but which
parameter-sensitive distinctions survive the complete experimental 
chain.

\begin{acknowledgments}
This study was inspired by neutron scattering experiments performed using iNSE under the JRR-3 General User Program managed by ISSP, the University of Tokyo (Proposal No. 26405), as well as those at J-PARC MLF approved by the NSPAC of IMSS, KEK (Proposal No. 2024S07). 
This study was partly supported by JSPS KAKENHI Grant No. JP24K08298, and by the Precise Measurement Technology Promotion Foundation (PMTP-F). 
\end{acknowledgments}

\section*{Conflict of Interest}

The authors have no conflicts to disclose

\section*{Data Availability Statement}

The data that support the findings of this study are available from the 
corresponding author upon reasonable request.

\bibliography{mybib}


\appendix

\section{Commonly used notation}
\label{app:notation}
\begingroup

\small  

\setlength{\tabcolsep}{5pt}
\renewcommand{\arraystretch}{1.04} 

\noindent

\begin{tabular*}{0.99\linewidth}{
    @{}
    >{\raggedright\arraybackslash}p{6em}
    @{\extracolsep{\fill}}
    >{\raggedright\arraybackslash}p{0.69\linewidth}
    >{\raggedright\arraybackslash}p{5.5em}
    @{}
}
\toprule
Symbol & Description & First defined \\
\midrule

$(X,\mathcal X)$
&
Measurable space of physical scattering outcomes
&
Sec.~\ref{subsec:event_space_scattering_measure}
\\

$x\in X$
&
Individual physical scattering outcome
&
Sec.~\ref{subsec:event_space_scattering_measure}
\\

$A\in\mathcal X$
&
Measurable event, i.e., a measurable set of outcomes
&
Sec.~\ref{subsec:event_space_scattering_measure}
\\

$\sigma$
&
Finite scattering measure
&
Sec.~\ref{subsec:event_space_scattering_measure}
\\

$r$
&
Total mass of the scattering measure, 
$r_\lambda=\sigma(X)$
&
Eq.~\eqref{eq:scattering_measure_total_mass}
\\

$\mu$
&
Reference measure used to represent a scattering measure by a density
&
Sec.~\ref{subsec:reference_measure_density}
\\

$s$
&
Scattering density,
$s=d\sigma/d\mu$
&
Eq.~\eqref{eq:RN-density_sigma}
\\

$P_\lambda$
&
Normalized scattering-outcome probability measure,
$P_\lambda=\sigma_\lambda/r_\lambda$
&
Eq.~\eqref{eq:measure_mass_shape_decomposition}
\\

$p_\lambda$
&
Density of $P_\lambda$ with respect to the reference measure
&
Eq.~\eqref{subsubsec:factorized_response}
\\

$\mathcal M_+(X,\mathcal X)$
&
Space of finite positive measures on $(X,\mathcal X)$
&
Eq.~\eqref{eq:parameterized_measure_model}
\\

$(Y,\mathcal Y)$
&
Measurable space of recorded outcomes
&
Sec.~\ref{subsec:recorded_outcome_spaces}
\\

$T$
&
Measurable deterministic map between outcome spaces
&
Sec.~\ref{subsec:deterministic_transformations}
\\

$T_\#$
&
Pushforward map induced by $T$ on finite measures
&
Eq.~\eqref{eq:pushforward_definition}
\\

$T^*$
&
Pullback of functions or observables,
$T^*f=f\circ T$
&
Sec.~\ref{subsec:deterministic_transformations}
\\

$K(dy\mid x)$
&
Probability or sub-probability kernel from an input outcome $x$
to a recorded outcome $y$
&
Sec.~\ref{subsec:physical_to_recorded_outcomes}
\\

$K_\#$
&
Forward measurement map induced by a kernel on finite measures
&
Eq.~\eqref{eq:measure_level_map_kernel}
\\

$K^*$
&
Backward action of a kernel on functions or observables
&
Eq.~\eqref{eq:kernel_backward_action}
\\

$\nu_\lambda$
&
Finite measure induced from $\sigma_\lambda$ by a measurement
transformation
&
Sec.~\ref{subsec:recorded_outcome_spaces}
\\

$\lambda,\ \boldsymbol\lambda$
&
Scalar or vector physical parameter of the scattering-measure family
&
Sec.~\ref{subsec:parameterized_scattering_measures}
\\

$\dot{\sigma}_\lambda$
&
Tangent measure; an overdot denotes differentiation with respect to
$\lambda$
&
Eq.~\eqref{eq:tangent_measure_definition}
\\

$u_\lambda$
&
Intrinsic score,
$u_\lambda=d\dot{\sigma}_\lambda/d\sigma_\lambda$
&
Eq.~\eqref{eq:intrinsic_measure_score}
\\

$\mathcal D(\sigma\Vert\tau)$
&
Generalized Kullback--Leibler divergence between finite positive measures
&
Eq.~\eqref{eq:extended_KL_measure}
\\

$g_\sigma$
&
Fisher--Rao metric on tangent directions at the finite measure $\sigma$
&
Sec.~\ref{subsec:intrinsic_fisher_scattering_measure}
\\

$\mathcal I^{(\sigma)}$
&
Intrinsic Fisher information of a finite-measure family
&
Eq.~\eqref{eq:intrinsic_fisher_scattering_measure}

\\

$\mathbf I^{(\sigma)}$ 
&
Multiparameter matrix form of $\mathcal I^{(\sigma)}$
&
Eq.~\eqref{eq:intrinsic_fisher_matrix}
\\

$u^{(r)},\ u^{(P)}$
&
Mass and normalized-shape components of the intrinsic score
&
Eq.~\eqref{eq:intrinsic_mass_shape_FI_decomposition}
\\

$\mathcal I^{(P)},\ \mathbf I^{(P)}$
&
Fisher information of the normalized shape $P_\lambda$, and its multiparameter matrix form
&
Eq.~\eqref{eq:shape_fisher_information}
\\

$ds_{\mathrm F}^{2}$
&
Infinitesimal Fisher line element
&
Eq.~\eqref{eq:fisher_line_element_single}
\\

$\mathcal E$
&
Parameter-independent experimental exposure
&
Eq.~\eqref{eq:ideal_count_intensity}
\\

$\Lambda_\lambda$
&
Poisson intensity measure,
$\Lambda_\lambda=\mathcal E\sigma_\lambda$
&
Eq.~\eqref{eq:ideal_count_intensity}
\\

$\calN$
&
Random counting measure of the observed point process
&
Eq.~\eqref{eq:counting_measure}
\\

$\mathrm{PPP}$
&
Poisson point process
&
Sec.~\ref{subsec:poisson_scattering_likelihood}
\\

$\mathbb{P}_\Lambda$
&
Probability law of the Poisson point process with intensity measure \(\Lambda\)
&
Sec.~\ref{subsec:poisson_scattering_likelihood}
\\

$M_\lambda$
&
Expected total event count,
$M_\lambda=\Lambda_\lambda(X)=\mathcal E r_\lambda$
&
Sec.~\ref{subsec:poisson_scattering_likelihood}
\\

$S_\lambda(\calN)$
&
Likelihood score of the Poisson counting experiment
&
Eq.~\eqref{eq:poisson_likelihood_score_intrinsic}
\\

$F$
&
Fisher information of the specified statistical observation model
&
Eq.~\eqref{eq:poisson_fisher_intrinsic}
\\

$\mathbf F$
&
Fisher information matrix of the specified statistical observation model
&
Eq.~\eqref{eq:poisson_fisher_matrix_intrinsic}
\\

$v_\lambda$
&
Intrinsic score of the transformed measure
$\nu_\lambda=K_\#\sigma_\lambda$
&
Sec.~\ref{sec:information_contraction}
\\

$\mathbb E_\lambda[\cdot\mid Y]$
&
Conditional expectation with respect to the joint distribution induced
by $\sigma_\lambda$ and the measurement kernel
&
Eq.~\eqref{eq:conditional_score}
\\

$\mathbf u_{\boldsymbol\lambda},
\mathbf v_{\boldsymbol\lambda}$
&
Input and output score vectors for a multiparameter family
&
Eq.~\eqref{eq:vector_intrinsic_score}
\\

$\mathsf A,\ a$
&
Set of measurement designs and an individual design,
$a\in\mathsf A$
&
Sec.~\ref{subsec:design_kernels}
\\

$\Lambda_{\boldsymbol\lambda,a}$
&
Observed Poisson intensity measure for design $a$,
$\Lambda_{\boldsymbol\lambda,a}
=\mathcal E_a(K_a)_\#\sigma_{\boldsymbol\lambda,a}$
&
Eq.~\eqref{eq:design_count_measure}
\\

$\Phi$
&
Scalar design criterion applied to a Fisher information matrix
&
Eq.~\eqref{eq:design}
\\

$F_{\theta\theta}^{\mathrm{eff}}$
&
Efficient Fisher information for a parameter of interest $\theta$
&
Eq.~\eqref{eq:fisher_schur_complement}
\\

$\rho_\lambda$
&
Parameterized quantum state
&
Sec.~\ref{subsec:quantum_outcome_measures}
\\

$\widetilde X=X\sqcup\{\cem\}$
&
Complete quantum-measurement outcome space, including the complementary
(no-scattering) outcome $\cem$
&
Eq.~\eqref{eq:complete_quantum_outcome_space}
\\

$\widetilde M,\ M$
&
Complete POVM and its restriction to the scattering-outcome sector
&
Sec.~\ref{subsec:quantum_outcome_measures}
\\

$\zeta_\lambda$
&
Born-rule subprobability measure on the scattering-outcome space
&
Eq.~\eqref{eq:born_scattering_measure}
\\

$q_\lambda$
&
Probability that the quantum measurement produces an outcome in the
scattering sector,
$q_\lambda=\zeta_\lambda(X)$
&
Eq.~\eqref{eq:scattering_probability}
\\

$P_\lambda^M$
&
Normalized scattering-outcome distribution conditional on scattering
for the specified quantum measurement $M$
&
Eq.~\eqref{eq:quantum_scattering_factorization}
\\

$w_{\lambda,c}$
&
Conditional weight of outcome sector $c$
&
Eq.~\eqref{eq:conditional_channel_weight}
\\

\bottomrule
\end{tabular*}

\endgroup

\section{Composition of kernels across outcome spaces}
\label{app:kernel_composition}

Successive measurement and data-reduction stages may act between
different outcome spaces.  This Appendix gives the explicit composition
rule for the kernel-induced transformations introduced in
Sec.~\ref{sec:kernel}, together with the corresponding backward action
on observables.

Two elementary examples make the composition rule explicit.
First, consider two successive Gaussian-resolution kernels of the form
introduced in Eq.~\eqref{eq:Gaussian_detector_resolution_kernel}, with
standard deviations \(\gamma_1\) and \(\gamma_2\).  Their composition is
again a Gaussian-resolution kernel with a broadened standard deviation:
\begin{equation}
K_{\gamma_2}\circ K_{\gamma_1}
=
K_{\sqrt{\gamma_1^2+\gamma_2^2}} .
\label{eq:gaussian_kernel_composition}
\end{equation}
As a discrete counterpart, consider two successive symmetric
two-channel mixing kernels of Eq.~\eqref{eq:two_channel_kernel},
\(K_{\varepsilon_1}\) and \(K_{\varepsilon_2}\).  Since kernel
composition reduces to matrix multiplication for discrete outcome
spaces,
\begin{equation}
K_{\varepsilon_2}K_{\varepsilon_1}
=
K_{\varepsilon_{\mathrm{eff}}}, \qquad
\varepsilon_{\mathrm{eff}}
=
\varepsilon_1+\varepsilon_2
-2\varepsilon_1\varepsilon_2 .
\label{eq:two_channel_kernel_composition}
\end{equation}

These examples involve two kernels of the same type, but no such
restriction is required.  Continuous and discrete outcome spaces may be
combined, and kernels representing different physical or data-reduction
operations may be composed in exactly the same way, provided that the
output space of one stage is the input space of the next.

For a general two-stage transformation, let
\[
K_1:(X,\mathcal X)\to(Y,\mathcal Y),
\qquad
K_2:(Y,\mathcal Y)\to(Z,\mathcal Z)
\]
be kernels between measurable outcome spaces.  Their composition is the
kernel from \(X\) to \(Z\) defined by
\begin{equation}
(K_2\circ K_1)(C\mid x)
=
\int_Y
K_2(C\mid y)\,K_1(dy\mid x),
\qquad
C\in\mathcal Z.
\label{eq:kernel_composition_two_stage}
\end{equation}
Thus the intermediate outcome \(y\) is integrated over, yielding a
single effective kernel for the two successive stages.

The corresponding action on observables proceeds in the reverse order.
For a bounded measurable observable \(f:Z\to\mathbb R\), the second kernel first gives
\(
(K_2^*f)(y)
=
\int_Z f(z)\,K_2(dz\mid y).
\) 
Applying \(K_1^*\) then gives
\begin{align}
(K_1^*K_2^*f)(x)
&=
\int_Y (K_2^*f)(y)\,K_1(dy\mid x)
\nonumber\\
&=
\int_Y
\left[
\int_Z f(z)\,K_2(dz\mid y)
\right]
K_1(dy\mid x).
\label{eq:backward_kernel_composition}
\end{align}
The same iterated integral is obtained by applying the composed kernel
directly:
\begin{align}
\bigl((K_2\circ K_1)^*f\bigr)(x)
&=
\int_Z
f(z)\,(K_2\circ K_1)(dz\mid x)
\nonumber\\
&=
\int_Y
\int_Z
f(z)\,K_2(dz\mid y)\,K_1(dy\mid x).
\end{align}
Hence
\begin{equation}
(K_2\circ K_1)^*
=
K_1^*K_2^*.
\label{eq:adjoint_kernel_composition}
\end{equation}
Forward transformations of measures therefore compose in the order of
the experimental stages, whereas their dual action on observables
composes in the reverse order.

For probability kernels, the inner integral
\[
(K_2^*f)(y)
=
\int_Z f(z)\,K_2(dz\mid y)
\]
may be interpreted as the conditional mean of the final observable for
a fixed intermediate outcome \(y\).  The outer integral then averages
this conditional mean over the preceding transition from \(x\) to
\(y\).  In this sense, Eq.~\eqref{eq:backward_kernel_composition} is the
kernel form of the tower property of conditional expectation.

Given a kernel \(K\) from a measurable outcome space
\((X,\mathcal X)\) to another measurable space \((Y,\mathcal Y)\),
a measure \(\sigma\) on \(X\) induces the measure 
\(\nu = K_{\#}\sigma\) on \(Y\) according to 
Eq.~\eqref{eq:detector_output_measure2}. 
In an actual experiment, several such transformations may occur successively between different
outcome spaces.  If
\begin{equation}
    X
    \xrightarrow{K_1}
    Y_1
    \xrightarrow{K_2}
    Y_2
    \xrightarrow{K_3}
    \cdots
    \xrightarrow{K_{n-1}}
    Y_{n-1}
    \xrightarrow{K_n}
    Y_{n} ,
\label{eq:kernel_sequence_outcome_spaces}
\end{equation}
then, setting
\(
\nu^{(0)} \equiv  \sigma
\), 
the measures propagate forward according to
\begin{equation}
    \nu^{(j)}
    =
    (K_j)_\#\nu^{(j-1)}
    \quad
    j=1,\, \ldots,\,n. 
\end{equation}
Consequently, the measure on the final outcome space is
\begin{equation}
    \nu^{(n)}
    =
    (K_n)_\#\nu^{(n-1)}
    =
    (K_n)_\#(K_{n-1})_\#  \cdots (K_1)_\# \nu^{(0)}
    =
    (K_n\circ\cdots\circ K_2\circ K_1)_\#
    \sigma .
\label{eq:composed_kernel_general}
\end{equation}

The same chain acts on observables in the opposite direction. For a
measurable function \(f\) on the final outcome space \(Y_n\), set
\(
f^{(n)}:=f
\)
and define recursively
\begin{equation}
    f^{(j-1)}
    =
    K_j^*f^{(j)},
    \quad
    j=n,\ldots,1.
\label{eq:kernel_backward_recursion}
\end{equation}
Hence
\begin{equation}
    f^{(0)}
    =
    K_1^* f^{(1)}
    =
    K_1^*K_2^*\cdots K_n^*f
    =
    (K_n\circ\cdots\circ K_1)^*f .
\label{eq:kernel_backward_composition}
\end{equation}
For successive measurement stages, kernels compose in the forward
direction, while observables are propagated backward in the reverse
order: 
\(
(K_n\circ\cdots\circ K_1)^*
=
K_1^*\cdots K_n^*
\).  
For probability kernels, this is the kernel form of iterated
conditional expectation.

The function \(f^{(0)}: X \to \mathbb{R}\) in 
Eq.~\eqref{eq:kernel_backward_composition}  may  be interpreted as 
an effective observable on the physical outcome space \(X\), 
incorporating the entire chain of measurement and data-reduction 
operations. 
For a fixed physical outcome \(x\in X\), \(f^{(0)}(x)\) gives 
the expected value of the final observable \(f\) after propagation through the complete chain.

As a simple illustration, consider a finite-resolution measurement described by a kernel \(K\) from \(X\) to  \(Y\), followed by selection of a measurable region \(S\in\mathcal Y\). Taking the final observable to be the indicator of the selected region, \(f=\mathbf 1_S\), its backward image on the physical outcome space is
\begin{equation}
f^{(0)}(x)
=
K^*\mathbf 1_S(x)
=
\int_Y \mathbf 1_S(y)\, K(dy\mid x)
=
K(S\mid x).
\end{equation}
Thus, \(f^{(0)}(x)\) gives the probability that a physical outcome \(x\) contributes to the selected data. In the deterministic limit, \(K(dy\mid x)=\delta_{T(x)}(dy)\), this becomes \(
f^{(0)}(x)
=
\mathbf 1_S(T(x))
=
\mathbf 1_{T^{-1}(S)}(x) 
\), 
recovering the ordinary preimage of the selected region. A finite-resolution kernel therefore replaces the sharp \(0\)-or-\(1\) preimage indicator by a probabilistic acceptance function on the physical outcome space.

At every intermediate stage, the forward and backward recursions are
linked by the same duality,
\begin{equation}
    \big\langle
    f^{(j)},\nu^{(j)}
    \big\rangle
    =
    \big\langle
    f^{(j-1)},\nu^{(j-1)}
    \big\rangle ,
    \quad
    j=1,\, \ldots,\, n,
\label{eq:kernel_stagewise_duality}
\end{equation}
and therefore
\begin{equation}
    \big\langle f, \nu^{(n)}\big\rangle
    =
    \big\langle f^{(0)},\sigma \big\rangle.
\label{eq:kernel_composed_duality}
\end{equation}
The action of the entire experimental chain may be represented either by propagating the measure forward or by propagating the final observable backward to the physical outcome space. This is the compositional form of the
forward--backward duality described above.

An arbitrarily long chain of transformations can be regarded,
at the measure level, as a single induced map.
When the intermediate stages need not be distinguished explicitly,
we use \(\nu\) generically for an induced outcome measure. 
Recording and data processing are important examples of this general
structure, but they do not constitute mathematically distinct types of
transformation.

Finally, composition also makes transparent the behavior of the total
mass.
If every kernel in Eq.~\eqref{eq:kernel_sequence_outcome_spaces} is a 
probability kernel,  
\begin{equation}
    \sigma(X)
    =
    \nu^{(1)}(Y_1)
    =
    \cdots
    =
    \nu^{(n-1)}(Y_{n-1})
    =
    \nu^{(n)}(Y_n).
\label{eq:mass_preservation_kernel_sequence}
\end{equation}
In realistic experiments, some stages are typically represented by
sub-probability kernels, owing to finite acceptance, inefficiency, or
event selection. The total mass cannot increase along such a sequence, 
\begin{equation}
    \sigma(X)
    \geq
    \nu^{(1)}(Y_1)
    \geq
    \cdots
    \geq
    \nu^{(n-1)}(Y_{n-1})
    \geq
    \nu^{(n)}(Y_n).
\label{eq:mass_decrease_kernel_sequence}
\end{equation}

This monotonicity of the total mass should be distinguished from the
behavior of information that is useful for inferring the parameter of interest.
Even when the total mass is preserved, a transformation may reduce distinctions 
between different parameter values, as in resolution broadening or channel mixing. 
Mass preservation and preservation of parameter-relevant information are
therefore distinct properties of transformations between outcome measures.
The latter is quantified by Fisher information and its contraction under
measurement kernels in Secs.~\ref{sec:fisher_scattering_measure}
and~\ref{sec:information_contraction}.

\section{Concrete examples of induced measures and experimental reductions}
\label{app:concrete_reductions}

The measure-level formulation of Sec.~\ref{sec:kernel} includes many
operations that are routinely performed in scattering experiments.
This Appendix gives several concrete examples that make explicit how
the outcome space, the measure, and its density representation change
under such operations. The examples also illustrate the distinction
between an invertible change of coordinates and a genuine reduction of
experimentally distinguishable outcomes.

\subsection{Coordinate transformation: energy and time of flight}
\label{app:tof_coordinate}

An invertible coordinate transformation changes the representation of
the scattering measure without discarding physical outcomes. Consider
a nonrelativistic particle of mass \(m\) traveling over a fixed flight
path \(L\). If the physical outcome is described by its kinetic energy
\(E>0\), the corresponding ideal time of flight is
\begin{equation}
t=T(E)
=
L\sqrt{\frac{m}{2E}}.
\label{eq:app_tof_energy_map}
\end{equation}
The measure expressed on the time-of-flight space is the pushforward
\(T_{\#}\sigma\).

The same transformation acts in the opposite direction on observables.
For example, let
\begin{equation}
f(t)
=
\mathbf{1}_{[t_1,t_2]}(t)
\end{equation}
select a time-of-flight interval. Its pullback to energy space is
\begin{equation}
(T^*f)(E)
=
f(T(E))
=
\mathbf{1}_{[E_2,E_1]}(E),
\qquad
E_i=\frac{mL^2}{2t_i^2}.
\end{equation}
Consequently,
\begin{align}
\int
\mathbf{1}_{[t_1,t_2]}(t)\,
(T_{\#}\sigma)(dt)
&=
(T_{\#}\sigma)([t_1,t_2])
\nonumber\\
&=
\sigma([E_2,E_1])
=
\int
\mathbf{1}_{[E_2,E_1]}(E)\,
\sigma(dE).
\label{eq:app_tof_duality}
\end{align}
This is a simple realization of the pushforward--pullback duality:
measures move forward under \(T_{\#}\), whereas observables move
backward under \(T^*\). No information is removed by the ideal
invertible transformation; only the coordinate representation changes.

A slightly less trivial example is provided by an energy-dependent
weighting.  
Conversely, consider an observable defined on the energy space,
\begin{equation}
g(E)=\frac{C}{\sqrt{E}}.
\end{equation}
The corresponding observable on the time-of-flight space can be written
as its pullback along the inverse coordinate map: 
\begin{equation}
f(t)
=
\bigl[(T^{-1})^*g\bigr](t)
=
g(T^{-1}(t))
= C\sqrt{\frac{2t^2}{mL^2}}
= C\sqrt{\frac{2}{m}}\,\frac{t}{L}.
\end{equation}

Accordingly,
\begin{equation}
\int g(E)\,\sigma(dE)
=
\int
\bigl[(T^{-1})^*g\bigr](t)\,
(T_{\#}\sigma)(dt).
\end{equation} 
Since \(T\) maps energy to time of flight, \(T^{*}\) pulls 
observables on the time-of-flight space
back to the energy space, whereas an observable defined on the energy
space is transferred to the time-of-flight space by the pullback
\((T^{-1})^{*}\) of the inverse coordinate map. 
In the present example,
the weighting proportional to \(1/\sqrt{E}\) in energy space therefore
corresponds to an observable proportional to \(t\) in time-of-flight
space.

\subsection{Binning}
\label{app:binning}

Binning is a deterministic map from a possibly continuous outcome space
to a discrete set of bin labels. Let
\begin{equation}
X
=
A_1\sqcup A_2\sqcup\cdots\sqcup A_M
\end{equation}
be a measurable partition of the outcome space and define
\begin{equation}
T_{\mathrm{bin}}:X\longrightarrow\{1,\ldots,M\},
\qquad
T_{\mathrm{bin}}(x)=j
\quad
\text{for }x\in A_j.
\end{equation}
The corresponding binned measure is
\begin{equation}
\nu^{\mathrm{bin}}
=
(T_{\mathrm{bin}})_{\#}\sigma.
\end{equation}
For a singleton \(\{j\} \),
\begin{equation}
\nu^{\mathrm{bin}}(\{j\})
=
\sigma(A_j).
\label{eq:app_bin_measure}
\end{equation}
If the scattering measure has a density representation \(
\sigma(dx)=s(x)\,\mu(dx), \)  then
\begin{equation}
\nu^{\mathrm{bin}}(\{j\})
=
\int_{A_j}s(x)\,\mu(dx).
\label{eq:app_bin_density_integral}
\end{equation}
Thus, a histogram bin represents the measure assigned to a finite
region of outcome space, rather than the value of a density at a
representative point.

The information-contraction result gives an immediate interpretation
of finite binning. For a parameterized recorded distribution
\(P_{\lambda}\), let
\begin{equation}
p_j(\lambda)
=
P_{\lambda}(A_j).
\end{equation}
Its Fisher information after binning is
\begin{equation}
F_{\mathrm{bin}}(\lambda)
=
\sum_{j=1}^{M}
\frac{
\left[\partial_{\lambda}p_j(\lambda)\right]^2
}{
p_j(\lambda)
},
\end{equation}
and satisfies
\begin{equation}
F_{\mathrm{bin}}(\lambda)
\leq
F_Y(\lambda),
\end{equation}
where \(F_Y\) is the Fisher information available in the continuously
resolved record.

If a partition
\(\mathcal{P}_{\mathrm{fine}}\) refines
\(\mathcal{P}_{\mathrm{coarse}}\), the coarser histogram is itself a
deterministic pushforward of the finer one. Hence
\begin{equation}
F^{(\mathcal{P}_{\mathrm{coarse}})}
\leq
F^{(\mathcal{P}_{\mathrm{fine}})}
\leq
F_Y.
\label{eq:app_partition_refinement}
\end{equation}
Under the usual regularity conditions, increasingly fine partitions
approach the Fisher information of the continuously resolved record.

\subsection{Projection and marginalization}
\label{app:projection_marginalization}

A common reduction retains one recorded coordinate while discarding
another. Suppose that the outcome space is
\begin{equation}
(X,\mathcal{X})
=
(U\times V,\mathcal{U}\otimes\mathcal{V}),
\end{equation}
with individual outcomes \(x=(u,v)\). Retaining only \(u\) corresponds
at the outcome level to the projection
\begin{equation}
\pi_U:U\times V\longrightarrow U,
\qquad
\pi_U(u,v)=u.
\end{equation}
The induced measure on \(U\) is the pushforward
\begin{equation}
\sigma_U
=
(\pi_U)_{\#}\sigma,
\end{equation}
so that, for \(B\in\mathcal{U}\),
\begin{equation}
\sigma_U(B)
=
\sigma(B\times V).
\label{eq:app_marginal_measure}
\end{equation}

Suppose now that \(\sigma\) has density \(s(u,v)\) with respect to the
product reference measure \(\mu_U\otimes\mu_V\).  Since \(s\) is a
Radon--Nikodym density, it is nonnegative and measurable.  Tonelli's
theorem applies and gives
\begin{align}
\sigma_U(B)
&=
\int_{B\times V}
s(u,v)\,
(\mu_U\otimes\mu_V)(dudv)
\nonumber \\
&=
\int_B
\left[
\int_V
s(u,v)\,\mu_V(dv)
\right]
\mu_U(du).
\label{eq:app_marginal_tonelli}
\end{align}
In particular, Tonelli's theorem ensures that
\(
u
\longmapsto
\int_V s(u,v)\,\mu_V(dv)
\) 
is a measurable nonnegative function on \(U\).
Because \(\sigma\) is finite, this function is finite for
\(\mu_U\)-almost every \(u\).

Equation~\eqref{eq:app_marginal_tonelli} identifies the
quantity in brackets as the Radon--Nikodym 
density of \(\sigma_U\) with respect to \(\mu_U\), conventionally called the
marginal density:
\begin{equation}
s_U(u)
=
\int_V
s(u,v)\,\mu_V(dv).
\label{eq:app_marginal_density}
\end{equation}
Here, \(\mu_V\) is the reference measure on \(V\); for an ordinary
continuous coordinate, \(\mu_V(dv)=dv\).  Thus \(s(u,v)\) is the density
of the original measure on \(U\times V\), whereas \(s_U(u)\) is the
density of its marginal measure on \(U\).  For each retained value of
\(u\), the marginal density accumulates the contributions from all
values of the discarded coordinate \(v\).

For example, a two-dimensional detector image may be projected onto 
the horizontal coordinate by summing over the vertical coordinate. 
The same operation can thus be 
described at three different levels: projection at the outcome level,
pushforward or marginalization at the measure level, and integration
over the discarded coordinate at the density level.  Keeping these
levels distinct avoids identifying a particular density operation with
the underlying transformation of experimental outcomes.

\subsection{Radial reduction and the induced reference measure}
\label{app:radial_reduction}

Radial reduction gives a particularly useful example because the
geometrical factor appearing in the reduced density is inherited from
the reference measure rather than from the physics itself.

First consider a measure on three-dimensional space 
\(\mathbb{R}^3\), and define the radial map
\begin{equation}
T:\mathbb{R}^3 \longrightarrow[0,\infty),
\qquad
T(\bm k)=|\bm k|=k.
\label{eq:app_radial_map}
\end{equation}
The radially reduced measure is \(
\nu
=
T_{\#}\sigma\), or, 
\(
\nu(B) =
\sigma\!\left(T^{-1}(B)\right), 
\)  for a measurable set \(B\subset[0,\infty)\). 
The fiber over a fixed \(k\) is the sphere
\begin{equation}
T^{-1}(\{k\})
=
\left\{
\bm k:\ |\bm k|=k
\right\},
\end{equation}
so radial reduction identifies all points on the same sphere and
discards their directional distinction.

Suppose \(\sigma\) admits a density representation, 
\begin{equation}
\sigma(d\bm k)
=
s(\bm k)\,d^3\bm k.
\end{equation}
Using \(
\bm k=k\hat{\bm k}\) and
\(d^3\bm k
=
k^2\,dk\,d\Omega,
\) 
one obtains
\begin{align}
\nu(B)
&=
\int_{T^{-1}(B)}
s(\bm k)\,d^3\bm k
=
\int_B
\left[
k^2
\int_{S^2}
s(k\hat{\bm n})\,d\Omega
\right]
dk.
\end{align}
Hence
\begin{equation}
\frac{d\nu}{dk}
=
k^2
\int_{S^2}
s(k\hat{\bm n})\,d\Omega.
\label{eq:app_general_radial_density}
\end{equation}
For a spherically symmetric density,
\(s(k\hat{\bm n})=s(k)\), this becomes
\begin{equation}
\frac{d\nu}{dk}
=
4\pi k^2 s(k).
\label{eq:app_radial_density_3d}
\end{equation}

The factor \(4\pi k^2\) is geometrical. More precisely, 
the entire factor is the Radon--Nikodym density of the 
pushed-forward reference measure with respect to \(dk\):
\begin{equation}
(T_{\#}d^3\bm k)(dk)
=
4\pi k^2\,dk.
\label{eq:app_radial_reference_3d}
\end{equation}
Radial reduction thus makes particularly transparent that the geometrical factor 
\(4\pi k^2\) taken as a whole, belongs to the induced measure rather 
than to the  physical content.

More generally, for the radial map
\begin{equation}
T:\mathbb R^n\to[0,\infty),
\qquad
T(\bm k)=|\bm k|=k,
\end{equation}
the reference measure \(d^n\bm k\) is pushed forward as 
\(
T_{\#}(d^n\bm k)
=
\omega_{n-1} k^{n-1}\,dk,
\) 
where
\(
\omega_{n-1}
=
\frac{2\pi^{n/2}}{\Gamma(n/2)}
\) 
is the surface area of the unit \((n-1)\)-sphere.

\subsection{Auxiliary recorded coordinates and event classification}
\label{app:auxiliary_coordinate}

Experimental event records often contain coordinates that are not 
themselves primary physical observables of interest, but are 
retained to classify  events, or distinguish different categories 
of recorded signals. 
Examples include pulse-shape discriminants, coincidence or veto tags, 
and other classification variables. Such a mark need not be 
intrinsically devoid of physical information; rather, its role is 
parameter- and model-dependent. For a specified event class and 
parameter family, it may carry no additional information about 
the parameter of interest once the primary event coordinates are known.

Let an event contain a coordinate \(x\) of physical interest together
with an auxiliary mark \(h\).
Suppose that, for the event class under consideration,
\begin{equation}
p_{\lambda}(x,h)
=
p_{\lambda}(x)\,g(h),
\label{eq:app_auxiliary_factorization}
\end{equation}
where \(g(h)\) is independent of \(\lambda\). Then
\begin{equation}
\frac{\partial_{\lambda}  p_{\lambda}(x,h)}{p_{\lambda}(x,h)} 
=
\frac{\partial_{\lambda}  p_{\lambda}(x)}{p_{\lambda}(x)},
\end{equation}
and the detailed value of \(h\) carries no additional Fisher
information about \(\lambda\). Marginalizing over \(h\) is therefore
information preserving for this parameter family.

This does not imply that \(h\) is experimentally useless. Pulse height,
for example, may be essential for distinguishing the desired event
class from background or electronic noise. A cut on \(h\), however,
should be distinguished from simply omitting \(h\) from the retained
coordinates. If a threshold rejects true events, it changes the total
mass of the observed finite measure and may remove information carried
by the event rate as well as by the distribution of retained events.

More generally, the relevance of a recorded coordinate is determined
not by whether it appears in the raw event record, but by whether it
resolves parameter-sensitive distinctions. This is the same criterion
that appears geometrically in the score-projection relation and
operationally in the measurement-design problem.

\section{Explicit examples of Fisher-information contraction}
\label{app:contraction_examples}

This Appendix gives explicit calculations for the examples summarized
in Sec.~\ref{subsec:contraction_examples}.  We restrict attention here
to normalized distributions, so that the Fisher information refers to
shape information per recorded event.

\subsection{Symmetric mixing of two discrete channels}
\label{app:discrete_channel_mixing}

Consider the two-channel distribution
\begin{equation}
\mathbf p_\lambda
=
\begin{pmatrix}
\lambda\\
1-\lambda
\end{pmatrix},
\qquad
0<\lambda<1.
\label{eq:two_channel_input_distribution_app}
\end{equation}
The two channels may, for example, represent spin-up and spin-down
measurement outcomes.
Its per-event Fisher information is
\begin{align}
\mathcal I^{(p)}(\lambda)
&=
\sum_i
\frac{1}{p_i(\lambda)}
\left[
\frac{\partial p_i(\lambda)}{\partial\lambda}
\right]^2
=
\frac{1}{\lambda}
+
\frac{1}{1-\lambda}
=
\frac{1}{\lambda(1-\lambda)}.
\label{eq:two_channel_input_FI_app}
\end{align}

Applying the symmetric mixing matrix
\begin{equation}
K_\varepsilon
=
\begin{pmatrix}
1-\varepsilon & \varepsilon\\
\varepsilon & 1-\varepsilon
\end{pmatrix},
\qquad
0\leq\varepsilon\leq\frac12,
\end{equation}
gives
\begin{equation}
\mathbf q_\lambda
=
K_\varepsilon\mathbf p_\lambda
=
\begin{pmatrix}
q_\lambda\\
1-q_\lambda
\end{pmatrix},
\qquad
q_\lambda
=
\varepsilon+(1-2\varepsilon)\lambda.
\label{eq:two_channel_output_app}
\end{equation}
Since
\(
\partial_\lambda q_\lambda=1-2\varepsilon
\),
the Fisher information after mixing is
\begin{align}
\mathcal I^{(q)}(\lambda)
&=
(1-2\varepsilon)^2
\left[
\frac{1}{q_\lambda}
+
\frac{1}{1-q_\lambda}
\right]
=
\frac{(1-2\varepsilon)^2}
     {q_\lambda(1-q_\lambda)}.
\label{eq:two_channel_output_FI_app}
\end{align}
The explicit expressions give
\begin{equation}
\mathcal I^{(p)}(\lambda)
-
\mathcal I^{(q)}(\lambda)
=
\frac{\varepsilon(1-\varepsilon)}
{\lambda(1-\lambda)\,
 q_\lambda(1-q_\lambda)}
\geq0.
\label{eq:two_channel_loss_relation_app}
\end{equation}
By the conditional-variance identity of
Eq.~\eqref{eq:FI_loss_conditional_variance}, this difference is exactly
the component of the input score that cannot be inferred from the 
recorded channel.

Relative to the input information,
\begin{align}
R(\lambda,\varepsilon)
\coloneqq 
\frac{
\mathcal I^{(p)}(\lambda)
-
\mathcal I^{(q)}(\lambda)}
{\mathcal I^{(p)}(\lambda)}
=
1-
\frac{\mathcal I^{(q)}(\lambda)}
     {\mathcal I^{(p)}(\lambda)}
=
\frac{\varepsilon(1-\varepsilon)}
     {q_\lambda(1-q_\lambda)}.
\label{eq:relative_information_loss_two_channel_app}
\end{align}
For \(\varepsilon=0\), \(R=0\) and the information is preserved.
For any fixed \(0<\varepsilon<1/2\),
as \(\lambda\to0\) or \(1\), nearly all of the input Fisher information is lost, \(R\to1\).
Thus even weak mixing may produce almost complete relative information loss
near the boundary, where a rare channel carries large score magnitude.

For \(\varepsilon\neq1/2\), \(K_\varepsilon\) is algebraically
invertible,
\[
\mathbf p_\lambda
=
K_\varepsilon^{-1}\mathbf q_\lambda.
\]
This does not contradict Fisher-information contraction.
For \(0<\varepsilon<1/2\), \(K_\varepsilon^{-1}\) is not a stochastic
kernel and amplifies finite-count fluctuations.
Moreover,
\(
\det K_\varepsilon
=
1-2\varepsilon
\to 0\) 
  \((\varepsilon \to 1/2)\), 
so the inversion becomes ill-conditioned as the recorded channel loses
its dependence on \(\lambda\).

\subsection{Gaussian detector blur}
\label{app:gaussian_blur}

We next consider a continuous location model.
Let the normalized physical distribution on \(X=\mathbb R\) be
\begin{equation}
p_\lambda(x)
=
\frac{1}{\sqrt{2\pi}\,w}
\exp\!\left[
-\frac{(x-\lambda)^2}{2w^2}
\right],
\label{eq:gaussian_physical_app}
\end{equation}
where \(w\) is the intrinsic width and \(\lambda\) is a location
parameter.
Its score is
\[
\frac{\partial_\lambda  p_\lambda(x)}{p_\lambda(x)}
=
\frac{x-\lambda}{w^2},
\]
and therefore
\begin{equation}
\mathcal I^{(p)}(\lambda)
=
\frac{1}{w^2}.
\label{eq:gaussian_input_FI_app}
\end{equation}

Let the detector add independent Gaussian blur,
\[
y=x+\xi,
\qquad
\xi\sim\mathcal N(0,\gamma^2),
\]
with resolution kernel
\begin{equation}
k_\gamma(y\mid x)
=
\frac{1}{\sqrt{2\pi}\gamma}
\exp\!\left[
-\frac{(y-x)^2}{2\gamma^2}
\right].
\end{equation}
The recorded distribution is the convolution
\begin{align}
q_{\lambda;\gamma}(y)
&=
\int_{\mathbb R}
k_\gamma(y\mid x)\,
p_\lambda(x)\,dx
=
\frac{1}
{\sqrt{2\pi(w^2+\gamma^2)}}
\exp\!\left[
-\frac{(y-\lambda)^2}
       {2(w^2+\gamma^2)}
\right].
\label{eq:gaussian_detector_app}
\end{align}
Hence
\begin{equation}
\mathcal I^{(q)}(\lambda;\gamma)
=
\frac{1}{w^2+\gamma^2}
\leq
\frac{1}{w^2}
=
\mathcal I^{(p)}(\lambda).
\label{eq:gaussian_output_FI_app}
\end{equation}
The retained fraction is therefore
\begin{equation}
\frac{\mathcal I^{(q)}(\lambda;\gamma)}
     {\mathcal I^{(p)}(\lambda)}
=
\frac{1}
     {1+(\gamma/w)^2}.
\label{eq:gaussian_retained_fraction_app}
\end{equation}
The contraction is thus controlled entirely by the dimensionless
resolution ratio \(\gamma/w\).

\subsection{Binary coarse graining}
\label{app:binary_coarse_graining}

Finally, consider a subsequent binary reduction of the continuously
recorded Gaussian outcome
\(y\sim q_{\lambda;\gamma}\).
Define
\begin{equation}
z
=
\begin{cases}
0, & y<t,\\
1, & y\geq t.
\end{cases}
\end{equation}
Let
\[
D^2=w^2+\gamma^2,
\qquad
\beta=\frac{t-\lambda}{D}.
\]
The resulting binary distribution
\(b_{\lambda;\gamma,t}\) has probabilities
\begin{equation}
b_{\lambda;\gamma,t}(0)
=
\Phi_{\rm std}(\beta),
\qquad
b_{\lambda;\gamma,t}(1)
=
1-\Phi_{\rm std}(\beta),
\label{eq:binary_probabilities_app}
\end{equation}
where \(\Phi_{\rm std}\) and \(\phi_{\rm std}\) denote the cumulative
distribution function and density of the standard normal
distribution.

Since
\[
\partial_\lambda
\Phi_{\rm std}(\beta)
=
-\frac{\phi_{\rm std}(\beta)}{D},
\]
the per-event Fisher information for the location parameter is
\begin{equation}
\mathcal I^{(b)}(\lambda;\gamma,t)
=
\frac{
\phi_{\rm std}(\beta)^2
}{
D^2\,
\Phi_{\rm std}(\beta)
[1-\Phi_{\rm std}(\beta)]
}.
\label{eq:binary_location_FI_app}
\end{equation}
For a threshold centered at a nominal value \(\lambda_0\),
\(t=\lambda_0\), evaluated locally at \(\lambda=\lambda_0\),
one has \(\beta=0\),
\(\Phi_{\rm std}(0)=1/2\), and
\(\phi_{\rm std}(0)=1/\sqrt{2\pi}\).
Therefore
\begin{align}
\mathcal I^{(b)}
(\lambda_0;\gamma,t=\lambda_0)
=
\frac{2}{\pi  D^2}=
\frac{2}{\pi} \,
\mathcal I^{(q)}(\lambda_0;\gamma).
\label{eq:binary_centered_location_FI_app}
\end{align}
Thus, despite reducing a continuous coordinate to a single bit, the
centered threshold retains a fraction
\(2/\pi\simeq0.64\) of the local Fisher information for the location
parameter.

The result is strongly parameter dependent.
If instead the Gaussian width \(D\) is treated as the parameter of
interest, the binary Fisher information is
\begin{equation}
\mathcal I_{DD}^{(b)}
=
\frac{
\beta^2\phi_{\rm std}(\beta)^2
}{
D^2\,
\Phi_{\rm std}(\beta)
[1-\Phi_{\rm std}(\beta)]
}.
\label{eq:binary_width_FI_app}
\end{equation}
At the same centered operating point,
\(\beta=0\), and hence
\begin{equation}
\mathcal I_{DD}^{(b)}
=
0.
\end{equation}
A small location shift produces a first-order imbalance between the
two binary outcomes, whereas a symmetric broadening leaves their
probabilities unchanged to first order.

The same coarse-graining operation therefore retains substantial
sensitivity to one parameter direction while annihilating another.
This provides an explicit example of the anisotropic contraction of
the tangent space of a statistical model under data reduction.

\section{Representation invariance}
\label{app:representation_invariance}

\subsection{Fisher metric independence from the choice of reference measure}
\label{secA1}

The intrinsic Fisher information is independent of the reference measure used to
represent the same family of scattering measures by densities, as can be
checked explicitly. Let
\begin{equation}
\widetilde\mu(dx)
=
h(x)\,\mu(dx),
\qquad h(x)>0,
\end{equation}
where \(h\) is independent of \(\lambda\). Then
\begin{equation}
\widetilde s_\lambda(x)
=
\frac{s_\lambda(x)}{h(x)},
\end{equation}
and hence
\begin{align}
\widetilde s_\lambda(x)\,\widetilde\mu(dx)
=
s_\lambda(x)\,\mu(dx), \qquad 
\frac{\partial_\lambda  \widetilde s_\lambda(x)}
{\widetilde s_\lambda(x)}
=
\frac{\partial_\lambda s_\lambda(x) }{s_\lambda(x)}.
\end{align} 
Therefore the Fisher information Eq.~\eqref{eq:intrinsic_fisher_density_representation} 
is unchanged.  
The invariance of the intrinsic expression in
Eq.~\eqref{eq:intrinsic_fisher_scattering_measure} is even more direct, since it
is defined solely by \(\sigma_\lambda\) and its parameter derivative, 
without requiring any choice of reference measure.

\subsection{Explicit KL divergence independence of density representation }
If two measures \(\sigma\) and \(\tau\) admit densities \(s\) and \(t\), respectively,  with respect to a
common reference measure \(\mu\),
\begin{equation}
d\sigma=s\,d\mu,
\qquad
d\tau=t\,d\mu,
\end{equation}
the generalized KL divergence of Eq.~\ref{eq:extended_KL_measure} is then
\begin{equation}
\mathcal D(\sigma\Vert\tau)
=
\int_X
\left[
s(x)\ln\frac{s(x)}{t(x)}
-s(x)+t(x)
\right]
d\mu(x).
\label{eq:extended_KL_density}
\end{equation}
Although Eq.~\eqref{eq:extended_KL_density} is written in terms of
densities, the divergence itself is a property of the measures and
does not depend on the choice of density representation. 
This invariance can be seen explicitly under an invertible change of
outcome coordinates.  If \(y=T(x)\), both densities acquire the same
Jacobian factor \(J_{T^{-1}}(y) \equiv 
|\det\left(\partial x / \partial y \right)|\), so that their 
ratio is unchanged:
\begin{equation}
\frac{s_Y(y)}{t_Y(y)}
=
\frac{
s_X(x) J_{T^{-1}}(y)
}{
t_X(x) J_{T^{-1}}(y)
}
=
\frac{s_X(x)}{t_X(x)}.
\label{eq:KL_jacobian_cancellation}
\end{equation}
Together with
\(s_Y(y)\,dy=s_X(x)\,dx\) and
\(t_Y(y)\,dy=t_X(x)\,dx\), this leaves
\(\mathcal D(\sigma\Vert\tau)\) unchanged.
Thus the divergence characterizes the difference between the measures
rather than between particular coordinate representations of their
densities.

\end{document}